\documentclass[preprint,3p,english,authoryear]{elsarticle}
\usepackage{lmodern}

\usepackage[T1]{fontenc}
\usepackage[latin9]{inputenc}
\usepackage{xcolor}
\usepackage{babel}
\usepackage{array}
\usepackage{rotfloat}
\usepackage{booktabs}
\usepackage{mathtools}
\usepackage{multirow}
\usepackage{amsmath}
\usepackage{amssymb}
\usepackage{stackrel}
\usepackage{graphicx}
\usepackage{setspace}
\usepackage[unicode=true, hidelinks=true]{hyperref}
\usepackage{natbib} 
\usepackage[normalem]{ulem}
\usepackage[title]{appendix}
\makeatletter

\providecommand{\tabularnewline}{\\}

\newtheorem{prop}{Proposition}

\makeatother

\begin{document}

\newgeometry{top=1.25in,left=1in,right=1in,bottom=1.25in}

\title{A multifactor regime-switching model for inter-trade durations in
the limit order market \tnoteref{t1}}
\tnotetext[t1]{All co-authors agree with the contents of the manuscript. We thank to the National Science Foundation, DMS-1612501 for financial support. Declarations of interest: none.}

\author[HNU]{Zhicheng Li\fnref{fn1}}
\ead{zhicheng.li@hnu.edu.cn}

\author[SBU]{Haipeng Xing\fnref{fn1}}
\ead{xing@ams.sunysb.edu}

\author[CUHK]{Xinyun Chen\corref{cor1}}
\ead{chenxinyun@cuhk.edu.cn}

\address[HNU]{Center of Economics, Finance and Management Studies, Hunan University, China}
\address[SBU]{Department of Applied Math, Stony Brook University, United States}
\address[CUHK]{Institute of Data and Decision Analytics,The Chinese University of Hong Kong, Shenzhen, China }

\fntext[fn1]{Zhicheng Li and Haipeng Xing are equally contributed as the first author.}
\cortext[cor1]{Corresponding author: Xinyun Chen, Institute of Data and Decision Analytics, The Chinese University of Hong Kong, Shenzhen, 
2001 Longxiang Road, Longgang District, Shenzhen, China, 86-0755-23517025.}

\begin{doublespacing}

\begin{abstract}
This paper studies inter-trade durations in the NASDAQ limit
order market and finds that inter-trade durations in ultra-high frequency
have two modes. One mode is to the order of approximately $10^{-4}$ seconds, and
the other is to the order of $10^0$ seconds. This phenomenon and other empirical evidence suggest that there
are two regimes associated with the dynamics of inter-trade durations, and the regime switchings are
driven by the changes of high-frequency traders (HFTs) between providing and taking liquidity.
To find how the two modes depend on information in the limit order book (LOB), we propose a two-state multifactor
regime-switching (MF-RSD) model for inter-trade durations, in which the probabilities transition matrices
are time-varying and depend on some lagged LOB factors. 
The MF-RSD model has good in-sample fitness and the superior out-of-sample performance,
compared with some benchmark duration models.
Our findings of the effects of LOB factors on the inter-trade durations 
help to understand more about the high-frequency market microstructure.\newline


\end{abstract}

\maketitle

\section{Introduction}
In the last two decades, there has been a rapid development of algorithmic trading and high-frequency trading in financial markets. The evolution of the market from human involvement to computer control, from operating in time frames of minutes to time scales of microseconds, has changed the market in a fundamental way (\cite{o2015high}). In this paper, we document a bimodal phenomenon that is prevalent in the distributions of inter-trade duration for stocks listed in the NASDAQ limit order market. The two modes are significantly different in their scales: one is of the order of $10^{-4}$ seconds, and the other is of the order of $10^0$ seconds. To the best of our knowledge, our paper is the first to document such a phenomenon in high-frequency duration data. This phenomenon, together with other empirical evidence, suggests that there are two regimes associated with the dynamics of inter-trade duration, and the regime switchings, 
therefore, reflects the dynamics of the market microstructure in a high-frequency world.

To better understand the bimodal phenomenon, we propose a two-state
multifactor regime-switching duration (MF-RSD) model with transition
probability matrices depending on the information in the limit order book.
We study the properties of the model, including the ergodicity, bimodality,
and exogeneity, and use the expectation-maximization (EM)
algorithm to make an inference on the model parameters. Simulation studies
validate our estimation method and show that the MR-RSD model can successfully generate
the bimodal distribution of inter-trade durations. Then, we use the proposed model to 
study the inter-trade durations in the NASDAQ limit order market. 
The results shows that the MF-RSD model not only fits the data well and but also has a good 
out-of-sample performance in predicting the inter-trade durations. We use 5000 in-sample data 
to estimate the model parameters and then do 1-step-ahead forecasting of the next inter-trade duration
for each single stock. A benchmark comparison with some classical duration models, 
such as the autoregressive conditional duration (ACD) model by \cite{engle1998autoregressive} 
and the Markov-switching multifractal inter-trade duration (MSMD) model by \cite{chen2013markov}, 
shows that the our-of-sample performance of 
our MF-RSD model is better than that of the other two in predicting the arrival time of the next trade.

We have analyzed the underlying regimes that are estimated from the MF-RSD model and find that 
these two regimes are related to the endogenous liquidity provision and consumption of HFTs.
Specifically, we use the permanent price change and the realized price spread to represent
the profit of taking and providing liquidity, and see their relationship between the underlying regimes. 
Our results show that the profit of taking liquidity is large when the market in the
short-duration regime and the HFTs are probably using aggressive trading strategies, while in the 
long-duration regime when the HFTs mainly act as the passive market makers, the profit of providing liquidity is 
relatively large. Moreover, we regress the short-term price change on the regime levels and the regime-switching
probabilities and find that a large price change happens in a short time after the short-duration regime, and 
there will be a stronger price movement if the market just switches from the long-duration regime to the short-duration regime.
This result suggests that the HFTs change the trading strategy to taking liquidity because they foresee that 
a large price change is going to occur subsequently.

We also study some economic hypotheses in the market microstructure theory 
by applying the MF-RSD model to 25 Nasdaq stocks in a whole month. In particular, 
we investigate the sensitivity of the transition probabilities of inter-trade durations in the MF-RSD model to various LOB
variables and summarize which types of market information the traders or their algorithms learn
from and react to in the LOB. Most of the results are consistent with the existing findings in the literature. For instance,
1) order book depth imbalance is an indicator for price movement and a large depth imbalance
would lead HFTs to actively trade for earning a profit, which causes short inter-trade durations;
2) HFTs are more likely to provide liquidity as market makers when price spread is large. 
At that time, market orders are initiated by non-HFTs and the inter-trade duration would be more likely in the long-duration regime; and 3) a large trade volume reflects a strong trading willingness on one side of the market, which is
a signal of informed trading. Therefore, HFTs would actively follow up and trade when they observe the volume of the last trade is very large.

Besides, several new findings are obtained. First,
 the impact of order book depth imbalance on the regime switchings of inter-trade durations is significant 
(positive for inter-trade durations switching from the long-duration regime 
to the short duration regime and negative for the opposite direction) mostly
for the stocks whose price spread is tight, while it is not obvious for the stocks whose spread is slack. 
This is because the depth imbalance between best ask and best bid 
is only informative for price movement when the price spread between them is small. 
Second, the effect of price spread is asymmetric. A large price spread tends to keep inter-trade duration
stay in the long-duration regime if the preceding inter-trade duration is long. 
However, a large price spread is not a significant signal 
which leads the inter-trade duration to switch into the long-duration regime from the short-duration regime.
Third, an increase in price spread, which is contemporaneously associated with a change in mid-price,
is shown to have a significant effect on the regime switching of inter-trade durations, i.e., it is significantly negative
for inter-trade durations switching from the long-duration regime to the short-duration regime and significantly
positive for inter-trade durations switching from the short-duration regime to the long one.

The remainder of the paper is organized as follows. Section 2 presents the related literature. 
Section 3 discusses the data we use and provides the empirical facts of inter-trade durations, in particular, the bimodal distribution
of inter-trade durations. In Section 4 we introduce our MF-RSD model and
discuss the model's implied properties. The estimation method and a simulation study are also given in Section 4.
In Section 5 we implement the empirical analysis of the NASDAQ limit order market and summarize the main findings.
Section 6 gives our concluding remarks.

\section{Related literature}
Over the past twenty years, many financial duration models have been proposed. 
Most of them aim to capture the empirical properties of inter-trade durations.
One pioneering work is the ACD model by \cite{engle1998autoregressive},
which expresses the conditional expectation of duration as a linear function of past
durations and past conditional expectations of durations, and successfully explains the observed clustering effect of inter-trade durations. 
After that, many varieties of ACD extensions were developed, such as the logarithmic ACD model by \cite{bauwens2000logarithmic}, the threshold ACD model by \cite{zhang2001nonlinear}, the Markov-switching ACD model by \cite{hujer2002markov} and the stochastic conditional duration (SCD) model by \cite{bauwens2004stochastic}.  These models can generate a rich collection of nonlinear dynamics that fit different duration series.
Recently, \cite{chen2013markov} proposed the MSMD model, in which the stochastic intensity can be decomposed into multiple
fractals and each of them follows a distinct hidden Markov process, capture the long memory property of inter-trade durations 
that has been discussed in both the theoretical and applied literature \citep{jasiak1999persistence,diebold2001long,DEO20103715}.

Another strand of papers considers other variables' effects on the dynamics of inter-trade durations 
by incorporating them into the modeling, which shed light on our work. 
For instance, \cite{engle2000econometrics} and \cite{bauwens2003asymmetric} have jointly modeled 
the inter-trade durations and other events of interest, such as the price process. The MF-RSD model in our paper 
has a similar spirit as the regime-switching model in \cite{diebold1994regime}, and we link the regimes of inter-trade durations with the LOB
variables via time-varying transition probabilities. The method of making the Markov transition matrix time-varying and dependent on some variables is also employed in \cite{kim2008estimation}, \cite{kang2014estimation}, and \cite{chang2017new}, while they further considered and addressed a potential endogeneity problem in which the innovations in determining regime switching are possibly correlated with the observed time series.

Our work is also built on some works which study information learning in financial markets. In traditional market microstructure theory, traders learn information about security from market data, especially the trade-related information \citep{kyle1985continuous, glosten1985bid,hasbrouck1991measuring}. However, in the high-frequency world, the basic unit of market information becomes limit orders rather than trades \citep{hautsch2012market, o2015high, lo2015resiliency}. Many papers have studied the informativeness in the LOB both empirically and theoretically. \cite{harris2005information} showed that the limit orders in the NYSE are informative about price changes and that NYSE specialists would use this information to provide liquidity. \cite{cont2014price} investigated the price impact of order book events using NYSE TAQ data and found that price changes are driven by the order book imbalance. \cite{lipton2013trade} and \cite{cartea2018enhancing} studied the informativeness of order book imbalance in detail and found that it is a good predictor for the arrival of trades. Recently, \cite{aliyev2018learning} build a market microstructure model in which the liquidity providers not only learn about the fundamental value of the asset but also learn about the extent of informed tradings through order flow imbalance. The model theoretically explains why order imbalance sometimes destabilizes markets and provides more understanding on the dynamic process of how markets digest order imbalance.

One more strand of research related to our paper focuses on HFTs' trading strategies and the influences of LOB factors on them. Some studies \citep{menkveld2013high, hagstromer2013diversity, LI2005533} mentioned that a significant proportion of high-frequency traders (HFTs) employ market-making strategies and, consequently, are very sensitive to the transient fluctuation in market liquidity. On the other hand, some papers \citep{brogaard2014high, HFreview} pointed out that HFTs would also use aggressive directional strategies when they anticipate the direction of order flow or price movement in the short run. Furthermore, several studies have shown that 
HFTs switch the trading strategies between supplying liquidity and consuming liquidity according to different LOB statuses. 
For example, \cite{carrion2013very} found that HFTs are good at timing and hence provide liquidity when the price spread
is large and take liquidity when the spread is small. \cite{goldstein2017high} showed that HFTs condition their strategy on order book imbalance, i.e., they supply liquidity on the thick side of the order book and demand liquidity on the thin side. \cite{van2019high} found that HFTs trade with large institutional orders, i.e., they initially lean against these orders but eventually change direction and take a position in the same direction as the most informed institutional orders. Recently,
\cite{foster2019microliquidity} used a new market microstructure model to explore a fragile liquidity when HFTs switch from
liquidity provision to liquidity consumption on the basis of unexpected information signals.

\section{Data and empirical properties}
\subsection{Data}
Our data are downloaded from LOBSTER (\href{https://lobsterdata.com/}{https://lobsterdata.com/}),
which provides high-quality LOB data for all NASDAQ stocks since June
2007,  based on NASDAQ's Historical TotalView-ITCH data. LOBSTER simultaneously generates two files.
One is a `message' file, which contains indicators for the type of event causing an update of the LOB in the requested price range, 
with decimal precision in nanoseconds ($10^{-9}$ second).
The other file is an `order-book' file that records the evolution of the LOB up to the requested number of levels at the time when the `message' file is updated.
Through the `message' file, we can easily calculate the time length between two consecutive trades, i.e., the inter-trade duration. 
And from the `order-book' file, we are able to construct various LOB variables.

The sample period of our study is January 2013, which contains
21 trading days, and we have randomly selected 25 stocks from the NASDAQ-100 index
with various market capitalizations. A simple description of these selected
stocks is presented in Table \ref{table:25stocks} in the Appendices. Because the frequency of our
data is ultra-high, and we are particularly interested in the intra-day
dynamics of inter-trade durations, we will analyze each stock
in each trading day separately. To give a clear illustration,
we summarize our results for the 25 stocks but provide
detailed results for the Microsoft Corporation (MSFT) in the rest of our paper.

\subsection{Empirical properties}

\begin{figure}[!h]
\centerline{\includegraphics[scale=0.3]{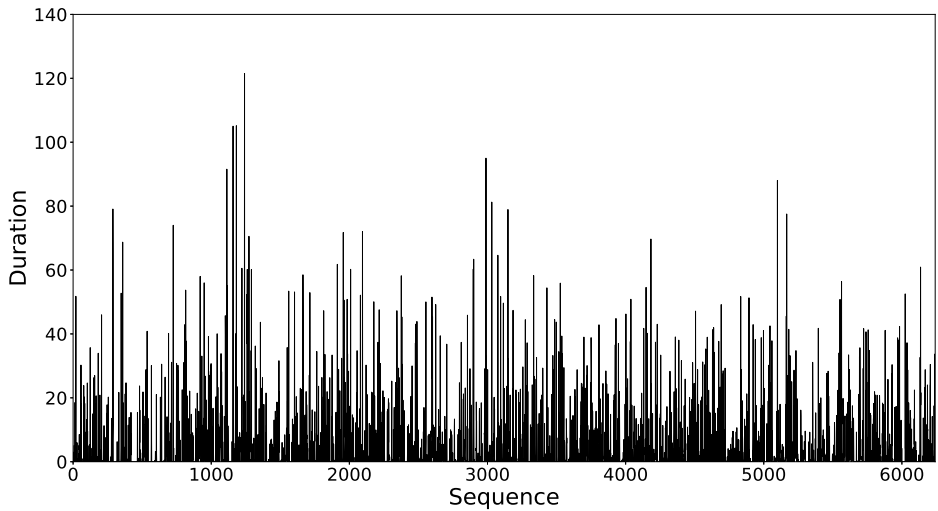}}\caption{The series of MSFT inter-trade durations on January 2nd, 2013. The horizontal axis is the chronological order of the durations, and the vertical axis records the duration length. The inter-trade durations are adjusted for calendar effects.}
\label{fig:SeriesMSFT}
\end{figure}

\begin{figure}[!h]
\centerline{\includegraphics[scale=0.5]{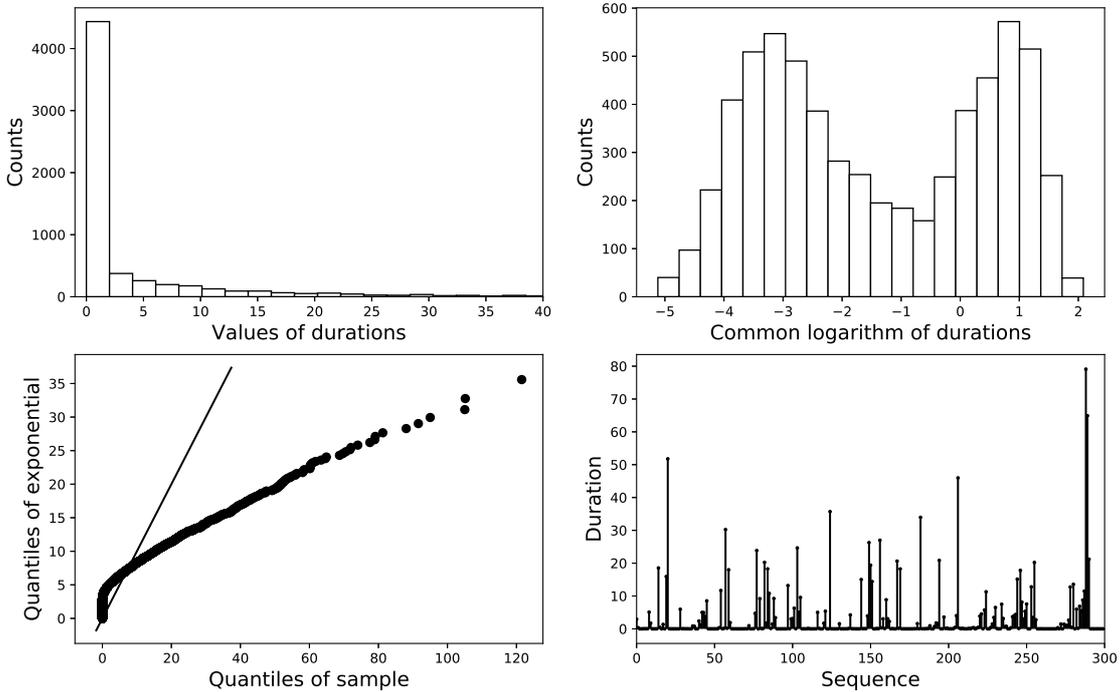}}
\caption{Some empirical facts of MSFT inter-trade durations on January 2nd, 2013. Top left: histogram of the inter-trade durations. Top right: histogram of the common logarithms of inter-trade durations. Bottom left: the exponential duration Q-Q plot for the sample durations. Bottom right: the first 300 inter-trade durations.}
\label{fig:durationFacts}
\end{figure}

In Figure \ref{fig:SeriesMSFT}, we first plot the series of MSFT inter-trade durations on January 2nd, 2013.\footnote{As the inter-trade duration series exhibits a significant intra-day calendar effect, which will be elucidated in Section V, the data that we use throughout the paper are the adjusted inter-trade durations that have been adjusted by the intra-day calendar effect.}
There are more than 6000 inter-trade durations on that day. Some of them are extremely short and in the order of $10^{-6}$ seconds, and some of them are very long and in the order of 100 seconds. We perform a histogram of these inter-trade durations in the first subplot of Figure \ref{fig:durationFacts}, which shows a heavy tailedness and that the majority is less than seconds. In fact, the inter-trade durations that are less than 1 second account for 65\% of the total share. Hence, in the second subplot of Figure \ref{fig:durationFacts}, we scrutinize the inter-trade durations by checking the histogram of their common logarithms and find a bimodal distribution with two distinct modes. Moreover, the inter-trade durations of the remaining 24 stocks also
exhibit similar bimodal distributions, and we will give a detailed discussion in the next subsection.

The bottom left panel of Figure \ref{fig:durationFacts}
presents an exponential duration quantile-quantile (Q-Q) plot for the MSFT inter-trade durations,
which indicates a nonexponential duration distribution with a long right tail
and that the distribution is over-dispersed
(i.e., the standard deviation exceeds the mean), in contrast
to the equality that would be obtained if the trade events follow a classical point process and 
the durations were exponentially distributed \citep{haipeng2016durations}. 
In particular, the mean of MSFT inter-trade durations is 4.07 seconds, while the standard deviation is 9.54 seconds.
Another empirical property of inter-trade durations is the clustering effect,
which has been discussed in \cite{engle1998autoregressive}, 
\cite{jasiak1999persistence}, and \citet[Section 11.2.3]{laixing2008}.
Here, we illustrate this effect by showing a random sample of 300 consecutive
inter-trade durations in Figure \ref{fig:durationFacts},
from which we can see that the clustering effect appears in a manner in which
short (or long) inter-trade durations are followed by short (or long) inter-trade durations.

\subsection{Bimodal distribution}
\begin{figure}[!h]
\centerline{\includegraphics[scale=0.3]{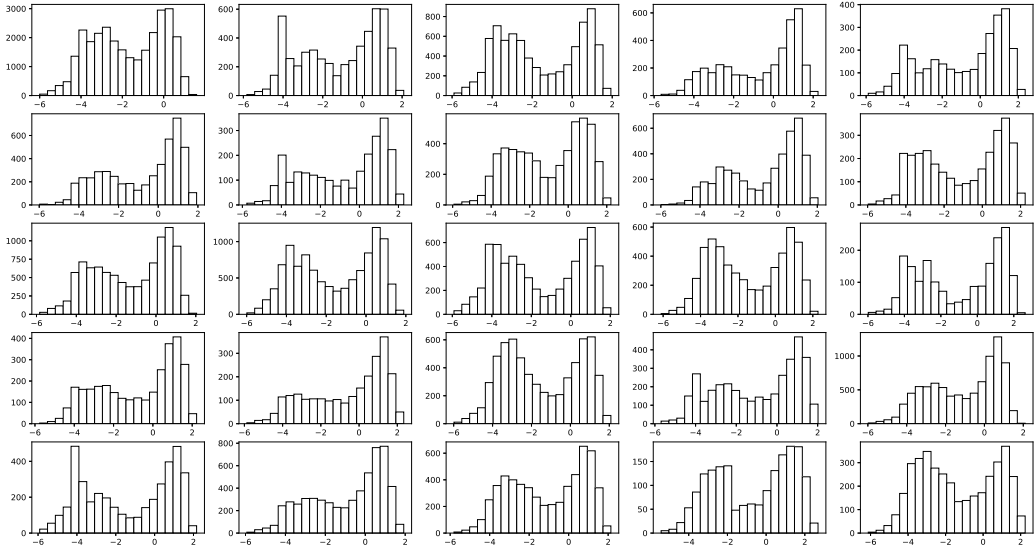}}
\caption{Histograms of the common logarithms of the inter-trade durations for 25 NASDAQ stocks on January 2nd, 2013. 
The order from left to right and from top to bottom is consistent with their alphabetical order, which is presented in Table \ref{table:25stocks}.}
\label{fig:bimodal}
\end{figure}

\begin{figure}[!h]
\centerline{\includegraphics[scale=0.46]{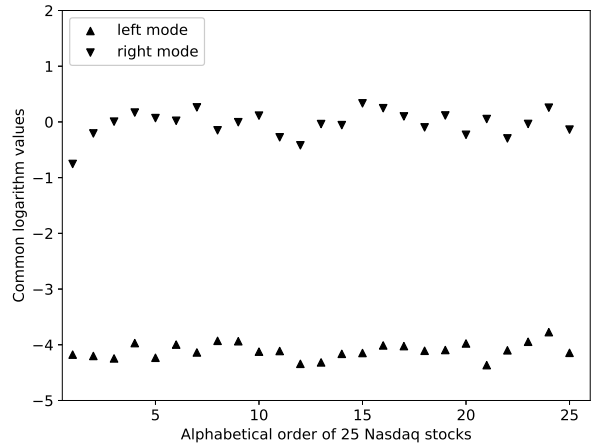}}
\caption{Estimated two modes of the distribution of the inter-trade durations for 25 NASDAQ stocks.}
\label{fig:twomodes}
\end{figure}

We have investigated the distributions of inter-trade durations for all 25 stocks in our sample.
Figure \ref{fig:bimodal} shows the histograms of their common logarithms,\footnote{It plots the histograms for the calendar-effect-adjusted inter-trade durations on the same trading day, i.e., January 2nd, 2013. In the Appendices, we also provide the histograms for the raw inter-trade durations and the aggregated inter-trade durations in one month. They all show bimodal distribution.} and they all exhibit a similar bimodal pattern.
To locate the two modes, we fit these empirical inter-trade duration distributions by a mixture of inverse Gaussian distributions. 
The results are plotted in Figure \ref{fig:twomodes}, from which we can see that the left modes are 
approximately $10^{-4}$ s, and the right modes are located between $10^{-1}$s and $10^{1}$ s.
To the best of our knowledge, our paper is the first to document such a phenomenon in high-frequency duration data.
This newly found bimodal distribution of inter-trade durations, together with the clustering effect we have shown,
suggest that there may be two regimes associated with
the dynamics of inter-trade durations, corresponding to the quick and slow trading periods.

Based on \cite{hasbrouck2013low}, which shows that the high-frequency
trading firms have effective-latency in milliseconds by using NASDAQ data in the year 2007,
we think that the clustering of very short inter-trade durations (most of the time below $10^{-1}$ s) is caused by HFTs.
Moreover, according to some literature mentioned in Section 2 \citep{brogaard2014high,carrion2013very},
we conjecture that the regime switchings between the long inter-trade durations and the short inter-trade durations
are driven by HFTs' behaviors of changing trading strategies for profitability. 
When the market is relatively stable, HFTs mainly use market making strategies to earn profit from the price 
spread between the best ask and the best bid.\footnote{Sometimes HTFs' profit is supplemented by a rebate from exchange for resting liquidity.} 
At that time, trades are mostly initiated by non-HFTs and the time interval between two trades would be relatively long.
While when HFTs receive some signal that probably indicates a short-term price movement, they
may switch to aggressive trading strategies and trade in the direction of price movement to gain profit.
Consequently, the market enters a regime in which HFTs consume liquidity and the inter-trade durations become very short.

\section{Multifactor Regime-Switching Duration Model}

\subsection{Model specification}
Given the empirical properties and the bimodal distribution of inter-trade durations, we
propose the following multifactor regime-switching duration (MF-RSD) model with two hidden states.
Let $t_{i}$ be the calendar time for the $i$-th trade in a trading day for one particular stock, where $t_{0}<t_{1}<\cdots<t_{n}$
and $t_{0}$ is the opening time for the market. The $i$-th inter-trade
duration is defined as
\[
y_{i}=t_{i}-t_{i-1},
\]
and there are two regimes for the inter-trade durations.  Under each regime, $y_i$ is an i.i.d.
random variable drawn from a prespecified distribution, i.e.,
\begin{equation}
y_{i}\sim\begin{cases}
f_{1}(y; \boldsymbol{\theta_{1}})\qquad\textrm{if}\,s_{i}=1,\\
f_{2}(y; \boldsymbol{\theta_{2}})\qquad\textrm{if}\,s_{i}=2.
\end{cases}
\end{equation}

\noindent $s_i$ is a two-state Markov chain
with a probability transition matrix
$$
\mathbb{P}_{i}=\left(\begin{array}{cc}
1- P_{12,i} & P_{12,i}\\
P_{21,i} & 1-P_{21,i}
\end{array}\right),
$$

\noindent in which $P_{jk,i} = P(s_{i}=k | s_{i-1}=j)$ is
the transition probability from state $j$ to state $k$ (where $j,k=1,2$ ) for the
$i$-th trade. 

We think the transition probabilities in our model are time-varying and depend on some exogenous variables.
Denote ${\bf x}_{i-1}$ the vector that contains some lagged variables which have been realized 
on or before the $(i-1)$-th trade,
and we concretely assume that the $P_{jk,i}$ satisfies the following logistic forms,
\begin{equation}
\log\frac{P_{12,i}}{1-P_{12,i}}=\boldsymbol{\beta}_{12}^{T}\boldsymbol{\mathbf{x}}_{i-1}, \qquad\log\frac{P_{21,i}}{1-P_{21,i}}=\boldsymbol{\beta}_{21}^{T}\boldsymbol{\mathbf{x}}_{i-1}.
\end{equation}
Hence, multiple factors can be incorporated into the model to consider their effects to the regime switchings
of inter-trade durations. In our study, we mainly focus on the impact of LOB factors on the dynamics of inter-trade durations.
Furthermore, considering the range of inter-trade durations and the simplicity of estimation,
we assume that the distribution $f_k(y)$ $(k=1,2)$
is inverse Gaussian (IVG). Specifically, assume that
conditional on state $k$, the
probability density function for $y_{i}$ is as follows:
\begin{equation}
f(y_{i}|s_{i}=k)=\left(\frac{\lambda_{k}}{2\pi y_{i}^{3}}\right)^{\frac{1}{2}}\exp\left(-\frac{\lambda_{k}(y_{i}-\mu_{k})^{2}}{2\mu_{k}^{2}y_{i}}\right)\qquad k=1,2,
\label{eq:densityFun}
\end{equation}
where $\mu_k$ and $\lambda_k$ are the respective mean and shape parameters for two IVGs.
Suppose that $s_i=1$ corresponds to the short-duration regime, and we should have $\mu_1<\mu_2$.

Compared with other inter-trade duration models, our MF-RSD model
has several advantages in modeling high-frequency duration data.
First, it uses a parsimonious structure to directly capture the observed bimodal distribution and clustering of inter-trade durations,
which helps us to understand the dynamics of market more intuitively.
Second, by introducing the covariates that potentially determine the evolution of inter-trade durations, 
we can improve the model's predictive power by using the values of covariates 
that are realized before the $i-$th inter-trade.
Finally, given the identified relationships between the covariates and the arrival time for trades,
the MF-RSD model can also be used to test some economic hypotheses on the limit order market.

\subsection{Implied properties}
Here we study some implied properties of the MF-RSD model. We begin
by showing the stationarity, ergodicity, and over-dispersion.
Then, we present the conditions under which the MF-RSD model can generate the bimodality.
Finally, we discuss the rationality of assuming that the covariates that determine the regime switchings of inter-trade durations are exogenous.

\subsubsection{Stationarity, ergodicity, and over-dispersion}
If we have the factor series $\mathbf{x}_i$ is a stationary sequence in a compact set, by the Theorem 5.5 of \cite{orey1991markov}, 
we can conclude that the underlying regime state $s_i$ for inter-trade duration is stationary and ergodic in the sense that 
$\lim_{i\to\infty}P(s_{i}=k|s_{0}=j)\to\pi_{k}$ for $j,k\in\{1,2\}$,
with $\pi_{k}\geq0$ and $\sum_{k}\pi_{k}=1$. In particular, $\pi=(\pi_{1},\pi_{2})'$
satisfies that 
$\pi P=\pi$, $P=E[\mathbb{P}(\mathbf{x}_i)].$

Moreover, let $s^{*}$ follows the stationary distribution of $s_{i}$, then we can write the stationary distribution of 
$y^{*}$ as $y(s^{*})$. Denote the random variable $y(k)\sim f(\boldsymbol{\theta_{k}})$ and define the dispersion $d_{k}=\frac{\text{Var}(y(k))}{E[y(k)]^{2}}$.
Suppose that $E[y(k)]$ are not equal for $k=1,2$. Then we have the dispersion of $y^{*}$ that, 
$d=\frac{\text{Var}(y^{*})}{E[y^{*}]^{2}}>\min_{k}d_{k}.$ Because
\begin{align*}
\text{Var}(y^{*}) & =\text{Var}(E[y^{*}|s^{*}])+E[\text{Var}(y^{*}|s^{*})]>E[\text{Var}(y^{*}|s^{*})]\\
 & =E[d_{s^{*}}E[y^{*}|s^{*}]^{2}]\geq(\min_{k}d_{k})\cdot E[E[y^{*}|s^{*}]^{2}]\geq(\min_{k}d_{k})\cdot E[y^{*}]^{2}.
\end{align*}
The last inequality follows the Jensen's Inequality. Hence, if the individual $f(\boldsymbol{\theta_{k}})$
already has over-dispersion, the stationary distribution of $y^{*}$ must be over-dispersed.

\subsubsection{Bimodality}
A bimodal distribution mostly arises as a mixture of two different unimodal distributions; however,
certain requirements need to be met. According to the Theorem 2 in \cite{dovsla2009conditions} which 
studies a mixture of two general unimodal densities, we have derived the following proposition for a mixture of two IVGs:

\begin{prop}
Let $p$ be the mixture ratio of two IVGs, i.e., $g(y)=p\cdot f(y;\mu_1,\lambda_1)+(1-p)\cdot f(y;\mu_2,\lambda_2), p\in(0,1)$.
Define the modes of two IVGs as $M_1$ and $M_2$ ($M_1<M_2$), and
$N_k=\mu_k\big[ -\big(1+\frac{9\mu_k^{2}}{4\lambda_k^2}\big)^{\frac{1}{2}}-\frac{3\mu_k}{2\lambda_k}\big]$ for $k=1,2$.
We further define a function of $y$ as
\begin{equation}
R(y)=\frac{\mu_1^2\lambda_2-\mu_2^2\lambda_1}{2\mu_1^2\mu_2^2}-\frac{\lambda_2-\lambda_1}{2y^2}-\frac{1}{y-N_2}
+\frac{1}{y-M_1}+\frac{1}{y-N_1}+\frac{1}{M_2-y}.
\label{eq:Ryfunction}
\end{equation}
Then, $g$ is bimodal if and only if the followings are satisfied:
\begin{enumerate}
\item Between the interval $(M_1,M_2)$, there is a $y^{\star}$ that makes $R(y)<0$.
\item There exist roots $y_1$ and $y_2$ of equation R(y)=0 such that $M_1<y_1<y_2<M_2$. Define
$$
\frac{1}{p_k}=1+\left(\frac{\lambda_1}{\lambda_2}\right)^{\frac{3}{2}}\left(\frac{\mu_2}{\mu_1}\right)^{2}
\frac{(y_k-M_1)(y_k-N_1)}{(M_2-y_k)(y_k-N_2)}
\exp\left[\left(\frac{\lambda_2}{2\mu_2^2}-\frac{\lambda_1}{2\mu_1^2}\right)y_k +\frac{\lambda_2-\lambda_1}{2y_k} + C\right],
$$
where $C= 2\mu_1\mu_2(\lambda_1\mu_2-\lambda_2\mu_1)$
and $k=1,2$. The mixture ratio $p$ should be in the range $(p_1,p_2)$.
\end{enumerate}
\end{prop}

The proof for Proposition 1 is given in the Appendices. In our MF-RSD model, the above conditions are easy to be satisfied.
The stationary probability for $s_i=1$ is the mixture ratio $p$. The IVG under the short-duration regime has a smaller mode $M_1$.
Specifically, we can assume $\mu_1<\mu_2$ and $\lambda_1<\lambda_2$ \footnote{They are consistent with our empirical results}.
Hence, when $y$ is between $M_1$ and $M_2$ (it is certainly greater than $N_1$ and $N_2$, which are negative),
$R(y)$ in \eqref{eq:Ryfunction} is very likely to be negative, especially when $\mu_1<<\mu_2$. Moreover, as in \eqref{eq:Ryfunction}, $\lim_{y\to M_1}R(y)=+\infty$ and $\lim_{y\to M_2}R(y)=+\infty$, $R(y)$ will first decrease to a negative value from $M_1$ to $y^{\star}$ 
and increase back to positive infinity when $y$
approaches $M_2$. Therefore, there must be two roots for the equation $R(y)=0$ in the interval $(M_1,M_2)$, which are $y_1$
and $y_2$. The mixture ratio $p$, which is the stationary probability for $s_i=1$, should be in the range $(p_1,p_2)$.

\subsubsection{Exogeneity}
In our MF-RSD model, we assume that the covariates that drive the dynamics of inter-trade durations are exogenous.
This assumption is reasonable in the following two senses.

First, according to \cite{hautsch2011econometrics}, the event probability
of a point process $N(t)$ depends on information that is (at least
instantaneously) known prior to $t$. If variables
are known and constant prior to an event $i$ , we call them time-invariant
covariates, and they can be considered weakly exogenous. Conversely,
the \textit{time-varying} covariate process $\boldsymbol{x}(t)$ can continuously change between two
consecutive events. To be weakly exogenous for $y_{i}$,
the process $x(t)$ must be c\`adl\`ag with
$\boldsymbol{x}_{\breve{N}(t)+1}=\boldsymbol{x}(t_{i-1})$
for all $t$ with $t_{i-1}<t\leq t_{i}$.
In our case, the duration event $y_{i}$ is defined as
$y_{i}\coloneqq t_{i}-t_{i-1},$
and $P_{jk,i}=P(s_{i}=k|s_{i-1}=j)$
represents the switching probability for the underlying state of event
$i$ changing from $j$ to $k$ (here, $j,k\in\{1,2\}$). Thus, the factors
$\mathbf{x}_{i-1}$ that have been realized on or
before $t_{i-1}$ can be considered weakly exogenous to the occurrence
of event $i$.

Second, if we think the factors $\mathbf{x}_{i-1}$ which represent the LOB information up to $(i-1)$-th trade
are timely revealed to market participants, especially to the HFTs.
Then, their actions are timely made conditional on the market information. As the occurrence of trade is triggered by just
one action of a single trader, the covariates $\mathbf{x}_{i-1}$ that were realized on or before the $(i-1)$-th trade 
can be considered as exogenous to the decision process of the market participant who initiates the $i$-th trade.

\subsection{Estimation}
\subsubsection{EM algorithm}
The direct maximization of the likelihood function of the observations is difficult to implement,
as we cannot observe the underlying states $s_{i}$ and $\{s_{i}\}_{i=1}^{n}$ has $2^{n}$ possible realizations.
Hence, we adopt the standard expectation-maximization (EM) algorithm.
In the E-step, given the information of $\mathbf{x}_{1:n}=\{\mathbf{x}_{i-1}\}_{i=1}^{n}$
and the model's old parameters $\Theta^{(m)}$ ($m=1,2,\ldots$ is the number of iterations),
the expected complete-data log-likelihood is
\begin{align}
&\mathbf{E}_{s_{1:n}}[l(y_{1:n,}s_{1:n}|\mathbf{x}_{1:n},\Theta^{(m)})] \nonumber \\
=&\mathbb{P}(s_{1}=1|\mathcal{F})[\log f(y_{1}|\boldsymbol{\theta_{1}})+\log\rho]+\mathbb{P}(s_{1}=2|\mathcal{F})[\log f(y_{1}|\boldsymbol{\theta_{2}})+\log(1-\rho)]\nonumber \\
 & +\stackrel[i=2]{n}{\sum}\big\{\mathbb{P}(s_{i}=1|\mathcal{F})\log f(y_{i}|\boldsymbol{\theta_{1}})+\mathbb{P}(s_{i}=2|\mathcal{F})\log f(y_{i}|\boldsymbol{\theta_{2}})\label{eq:Ex-loglike}\\
 & +\mathbb{P}(s_{i}=2,s_{i-1}=1|\mathcal{F})\boldsymbol{\beta}_{12}^{T}\mathbf{x}_{i-1}-\mathbb{P}(s_{i-1}=1|\mathcal{F})\log(1+e^{\boldsymbol{\beta}_{12}^{T}\mathbf{x}_{i-1}})\nonumber \\
 & +\mathbb{P}(s_{i}=1,s_{i-1}=2|\mathcal{F})\boldsymbol{\beta}_{21}^{T}\mathbf{x}_{i-1}-\mathbb{P}(s_{i-1}=2|\mathcal{F})\log(1+e^{\boldsymbol{\beta}_{21}^{T}\mathbf{x}_{i-1}})\big\}.\nonumber 
\end{align}
where $\mathcal{F}$ represents information
generated by $\{\mathbf{x}_{i},y_{i}\}_{i=1}^{n}$
and old parameters $\Theta^{(m)}$.
$\rho$ is the conditional stationary probability for state 1, i.e., $\rho=\mathbb{P}(s_{1}=1|\mathcal{F})$.
The conditional posterior distributions $\mathbb{P}(s_{i}=2,s_{i-1}=1|\mathcal{F})$, $\mathbb{P}(s_{i}=1,s_{i-1}=2|\mathcal{F})$, 
$\mathbb{P}(s_{i-1}=1|\mathcal{F})$, and $\mathbb{P}(s_{i-1}=2|\mathcal{F})$ can be calculated using the
the forward-backward algorithm \footnote{\cite{diebold1994regime} has shown the steps to calculate and implement this forward-backward algorithm.}.

Then, in the M-step, we update the model parameters by maximizing the expected log-likelihood \ref{eq:Ex-loglike},
\begin{equation}
\Theta^{(m+1)}=\text{arg}\max_{\Theta}\mathbf{E}_{s_{1:n}}[l(y_{1:n,}s_{1:n}|x_{1:n},\Theta^{(m)})]. 
\label{eq:mstep}
\end{equation}
\noindent We repeat the E-step and M-step until the model parameters converge, i.e.,
\begin{equation*}
||\Theta^{(m+1)}-\Theta^{(m)}|| \leq \text{tolerance}
\end{equation*}

\noindent By plugging the density function \eqref{eq:densityFun} into the expected log-likelihood
function \eqref{eq:Ex-loglike}, the maximization step \eqref{eq:mstep}
yields the following analytical expressions for the model
parameters \footnote{Detailed derivations are shown in the Appendices.}:
\begin{equation}
\widehat{\mu}_{k}=\frac{{\displaystyle \sum_{i=1}^{n}y_{i}\cdot\mathbb{P}(s_{i}=j|\mathcal{F})}}{{\displaystyle \sum_{i=1}^{n}\mathbb{P}(s_{i}=j|\mathcal{F})}}, \quad
\widehat{\lambda}_{k}=\frac{{\displaystyle \mu_{k}^{2}\sum_{i=1}^{n}\mathbb{P}(s_{i}=k|\mathcal{F})}}{{\displaystyle \sum_{i=1}^{n}\frac{(y_{i}-\mu_{k})^{2}}{y_{i}}\cdot\mathbb{P}(s_{i}=k|\mathcal{F})}}, 
\quad k=1,2,
\label{M-parameters}
\end{equation}

\noindent and the regression coefficients $\boldsymbol{\hat{\beta}}_{12}$, $\boldsymbol{\hat{\beta}}_{21}$ satisfy the following equations:
\begin{equation}
\stackrel[i=2]{n}{\sum}\left[\mathbb{P}(s_{i}=2,s_{i-1}=1|\mathcal{F})-\frac{e^{\boldsymbol{\hat{\beta}}_{12}^{T}\mathbf{x}_{i-1}}}{1+e^{\boldsymbol{\hat{\beta}}_{12}^{T}\mathbf{x}_{i-1}}}\cdot\mathbb{P}(s_{i-1}=1|\mathcal{F})\right]\mathbf{x}_{i-1}=\vec{0}
\end{equation}
\begin{equation}
\stackrel[i=2]{n}{\sum}\left[\mathbb{P}(s_{i}=1,s_{i-1}=2|\mathcal{F})-\frac{e^{\boldsymbol{\hat{\beta}}_{21}^{T}\mathbf{x}_{i-1}}}{1+e^{\boldsymbol{\hat{\beta}}_{21}^{T}\mathbf{x}_{i-1}}}\cdot\mathbb{P}(s_{i-1}=2|\mathcal{F})\right]\mathbf{x}_{i-1}=\vec{0}.
\end{equation}

\subsubsection{Estimation consistency}
Given the assumption that the covariates $\mathbf{x}_{i-1}$ are exogenous to the duration process and inter-trade duration $y_i$ is i.i.d
conditional on the underlying state $s_i$, the maximum likelihood (MLE) will obtain consistent estimators,
and they asymptotically converge to the true parameters.
However, we adopt the EM algorithm instead of directly maximizing the likelihood because of the complexity in constructing the log-likelihood function
in the presence of underlying hidden states. Hence, the estimation consistency relies on the convergence of the EM estimator
to the global maximizer of the log-likelihood function.

Define the log-likelihood for the observed data as $L(\Theta)=L(Y|\mathbf{X},\Theta)$,
in which $Y:=y_{1:n}$ and $\mathbf{X}:=\mathbf{x}_{1:n}$. Then, the MLE estimator directly maximizes $L(\Theta)$,
and the EM algorithm iteratively increases the value of $L(\Theta)$ at each step by maximizing
the expected complete-data log-likelihood in \eqref{eq:mstep}, which can be redefined as
\begin{equation*}
\Theta^{(m+1)} = \text{argmax}\,Q(\Theta|\Theta^{(m)})
\end{equation*}
According to \cite{wu1983convergence}, if the unobserved complete-data specification can be described by an exponential family,
the function $Q(\psi | \phi)$ is continuous in both $\psi$ and $\phi$. Then, 
the limits of the EM sequence are the stationary points (local maxima) of the log-likelihood $L(\Theta)$.
This condition obviously holds, as the IVGs belong to the exponential family and the transition probabilities are in logistic form.
Moreover, if $L(\Theta)$ is unimodal, the EM estimator converges to the global maximizer of $L(\Theta)$.

\subsubsection{Standard errors of the estimators}
The variance-covariance matrix of estimators is not a byproduct of the EM algorithm.
To obtain that, we use the supplement EM (SEM) method
proposed by \cite{meng1991using}. The SEM method employs the fact that the rate of convergence of the estimators in the EM is governed
by the fractions of missing information that would increase the variability
of the estimation. In particular, the desired observed variance-covariance
matrix $V$, which is also the inverse of the observed information matrix,
can be expressed as the sum of the covariance matrix of the expected complete-data, 
i.e., $\mathbf{I}_{\text{EC}}^{-1}$, and a variance inflation
part $\Delta V$,
\begin{equation}\label{eq: Covariance Matrix}
V=\mathbf{I}_{\text{EC}}^{-1}+\triangle V, \qquad
\mathbf{I}_{\text{EC}}=\mathbb{E}\left[\mathbf{I}_{\text{C}}(\Theta)\bigg|\mathcal{F}\right]\Bigg|_{\Theta=\hat{\Theta}} 
\end{equation}

\noindent where $\mathbf{I}_{\text{EC}}$ is the expectation of complete-data information matrix $\mathbf{I}_{\text{C}}(\Theta)$
over the conditional distributions evaluated at $\hat{\Theta}$, which can be calculated in the standard EM method.
The variance inflation part $\Delta V$ in \eqref{eq: Covariance Matrix} can be written as
a function of $DM$ as follows:
\begin{equation}
\triangle V=\mathbf{I}_{\text{EC}}^{-1}\cdot DM(I-DM)^{-1},\label{eq: Variance Inflation Part}
\end{equation}
where $I$ is simply the identity matrix and $DM$ is the matrix rate of EM convergence,
whose calculation is shown in detail in \cite{meng1991using}. 
Hence, the standard errors of the estimators are the square roots of the diagonal elements of $V$.

\subsection{Simulation}

\begin{figure}[!h]
\centerline{\includegraphics[scale=0.46]{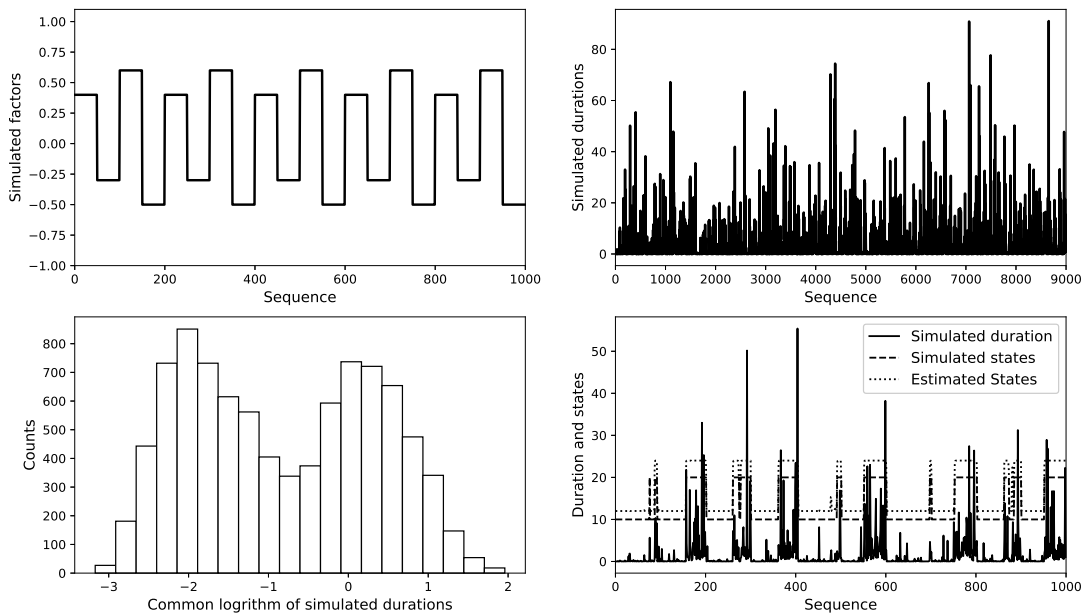}}
\caption{Top left: $\{x_i\}$ $(1\le i \le 1000)$;
Top right: Simulated $\{ y_i \}$.
Bottom left: Histogram of the common logarithm of simulated durations; Bottom right: Comparison of the estimated states and simulated states for the first
1000 points of simulated durations. The estimated states and real
states are scaled differently to see them clearly.}
\label{fig:simStudy}
\end{figure}

We perform a simulation study in this part
to validate our MF-RSD model and the estimation method. 
We provide an example in which the regime-switching probabilities are driven by a one-dimensional factor
$x$, which is piecewise constant and takes alternative
values among $\{0.4, -0.3, 0.6, -0.5\}$, as shown in
Figure \ref{fig:simStudy}. Hence, the regime switching probabilities are
determined as following:
\begin{equation*}
\log\frac{P_{jl,i}}{1-P_{jl,i}}=\beta_{jl,0} + \beta_{jl,1} \cdot x_{i-1},
\end{equation*}
where $1\le j \neq l \le 2$. The model parameters $\mu_k, \lambda_k, \boldsymbol{\beta_{jl}}$ ($k=1,2$) 
are shown in Table \ref{table:SimCompare}.
These values are chosen so that moderate regime-switching
probabilities are generated from the model. We then
simulate $y_i$ $(1\le i \le n)$ for $n=9000$ and
show the results in Figure \ref{fig:simStudy}.
Note that the simulated $y_i$ exhibits a large
variation and that the histogram in the bottom left panel shows a clear bimodal
distribution, which demonstrates that our model
can indeed generate bimodality. We then use the EM algorithm
to estimate the model and show the estimated values in
Table \ref{table:SimCompare}, from which we can see that the estimates
are quite close to the true values with low standard errors, and the confidence
intervals of estimates basically cover the true values. 
We also show the simulated and estimated states $s_i$ $(1\le i \le 1000)$
in the bottom right panel of Figure \ref{fig:simStudy}, from which we find that the difference
between the true and estimated states are quite small, and the underlying states capture the
clustering of long (or short) inter-trade durations.

\begin{table}[!h]
\caption{True and estimated parameters with piecewise
constant $x_i$.}
\medskip{}
\centerline{%
\begin{tabular}{ccccccccc}
\hline 
 & $\mu_{1}$ & $\lambda_{1}$ & $\mu_{2}$ & $\lambda_{2}$ & $\beta_{12,0}$ & $\beta_{12,1}$ & $\beta_{21,0}$ & $\beta_{21,1}$\tabularnewline
\hline 
{\footnotesize{}True values} & {\footnotesize{}0.3} & {\footnotesize{}0.01} & {\footnotesize{}5} & {\footnotesize{}2} & {\footnotesize{}-5} & {\footnotesize{}-7} & {\footnotesize{}-2.6} & {\footnotesize{}6}\tabularnewline
\multirow{2}{*}{\footnotesize{}Estimated values} & {\footnotesize{}0.308} & {\footnotesize{}0.0098} & {\footnotesize{}4.831} & {\footnotesize{}1.984} & {\footnotesize{}-5.022} & {\footnotesize{}-7.384} & {\footnotesize{}-2.538} & {\footnotesize{}5.696} \tabularnewline
& {\footnotesize{}(0.0242)} & {\footnotesize{}(0.000187)} & {\footnotesize{}(0.129)} & {\footnotesize{}(0.0491)} & {\footnotesize{}(0.388)} & {\footnotesize{}(0.996)} & {\footnotesize{}(0.137)} & {\footnotesize{}(0.317)} \tabularnewline
\hline 
\end{tabular}}
\label{table:SimCompare}
\textit{\footnotesize{}Note:}{\footnotesize{} standard errors are shown in parentheses.}
\end{table}

\section{Real data analysis}

\subsection{Data cleaning and descriptive statistics}

In the following, we implement the real data analysis for the inter-trade durations in the NASDAQ limit order market.
The inter-trade durations usually contain the
intra-day calendar effect, i.e., inter-trade durations
tend to be shorter early and late in the day and longer around noon. 
Based on the standard dummy variable procedure (e.g.,\cite{ghysels2004stochastic}),
we find that that the intra-day calendar effect 
is a common phenomena for all stocks in our sample (shown in Section \ref{SC:Calendar effects} in the Appendices.).
Therefore, prior to the analysis, we have adjusted inter-trade durations for the calendar effects.

For each stock in our sample, we aggregate the inter-trade durations in January 2013 and provide the
descriptive statistics in Table \ref{table: durationStatistics}. These statistics include count,
min (the minimum), three quantiles, max (the maximum), mean, standard
deviation, skewness, kurtosis, and over-dispersion (O.D.), which is
the standard deviation divided by the mean. Table \ref{table: durationStatistics} suggests the following:
1) There is a large variation in their counts and maximum values;
2) The minimum values are approximately $10^{-6}$ seconds, and the mean
values are approximately 1-10 seconds; 3) Their median value has a small
variation, with most of them are around or below 0.1 seconds; and 4) They
all exhibit a large dispersion as the values of their over-dispersion
are significantly greater than one.

\begin{table}[!tb]
\caption{Descriptive statistics of inter-trade durations for 25 NASDAQ stocks in January 2013.}
\centerline{%
\begin{tabular}{cccccccccccc}
\hline 
{\footnotesize{}Stock} & {\footnotesize{}Count} & {\footnotesize{}Min} & {\footnotesize{}25\%} & {\footnotesize{}Med.} & {\footnotesize{}75\%} & {\footnotesize{}Max} & {\footnotesize{}Mean} & {\footnotesize{}Std.} & {\footnotesize{}Skew.} & {\footnotesize{}Kurt.} & {\footnotesize{}O.D.}\tabularnewline
\hline 
{\footnotesize{}AAPL} & {\footnotesize{}539369} & {\footnotesize{}4.94E-06} & {\footnotesize{}7.51E-04} & {\footnotesize{}0.03} & {\footnotesize{}0.86} & {\footnotesize{}83.23} & {\footnotesize{}1.04} & {\footnotesize{}2.55} & {\footnotesize{}5.32} & {\footnotesize{}47.70} & {\footnotesize{}2.46}\tabularnewline
{\footnotesize{}ALXN} & {\footnotesize{}88916} & {\footnotesize{}5.28E-06} & {\footnotesize{}1.09E-03} & {\footnotesize{}0.19} & {\footnotesize{}6.05} & {\footnotesize{}286.81} & {\footnotesize{}6.23} & {\footnotesize{}14.03} & {\footnotesize{}4.71} & {\footnotesize{}35.76} & {\footnotesize{}2.25}\tabularnewline
{\footnotesize{}AMZN} & {\footnotesize{}171344} & {\footnotesize{}5.32E-06} & {\footnotesize{}5.03E-04} & {\footnotesize{}0.01} & {\footnotesize{}2.12} & {\footnotesize{}254.97} & {\footnotesize{}3.27} & {\footnotesize{}8.53} & {\footnotesize{}5.44} & {\footnotesize{}50.79} & {\footnotesize{}2.61}\tabularnewline
{\footnotesize{}BIDU} & {\footnotesize{}102830} & {\footnotesize{}5.66E-06} & {\footnotesize{}9.97E-04} & {\footnotesize{}0.14} & {\footnotesize{}5.28} & {\footnotesize{}242.40} & {\footnotesize{}5.48} & {\footnotesize{}12.49} & {\footnotesize{}4.77} & {\footnotesize{}36.05} & {\footnotesize{}2.28}\tabularnewline
{\footnotesize{}BMRN} & {\footnotesize{}47012} & {\footnotesize{}5.57E-06} & {\footnotesize{}6.56E-04} & {\footnotesize{}0.49} & {\footnotesize{}12.66} & {\footnotesize{}437.95} & {\footnotesize{}12.06} & {\footnotesize{}25.07} & {\footnotesize{}3.83} & {\footnotesize{}22.49} & {\footnotesize{}2.08}\tabularnewline
{\footnotesize{}CELG} & {\footnotesize{}138826} & {\footnotesize{}5.00E-06} & {\footnotesize{}1.05E-03} & {\footnotesize{}0.11} & {\footnotesize{}4.18} & {\footnotesize{}213.83} & {\footnotesize{}4.06} & {\footnotesize{}8.66} & {\footnotesize{}4.13} & {\footnotesize{}28.22} & {\footnotesize{}2.13}\tabularnewline
{\footnotesize{}CERN} & {\footnotesize{}44963} & {\footnotesize{}3.84E-06} & {\footnotesize{}1.05E-03} & {\footnotesize{}0.19} & {\footnotesize{}12.46} & {\footnotesize{}706.02} & {\footnotesize{}12.87} & {\footnotesize{}28.38} & {\footnotesize{}4.55} & {\footnotesize{}38.78} & {\footnotesize{}2.21}\tabularnewline
{\footnotesize{}CMCSA} & {\footnotesize{}107143} & {\footnotesize{}5.93E-06} & {\footnotesize{}1.62E-03} & {\footnotesize{}0.16} & {\footnotesize{}4.64} & {\footnotesize{}601.18} & {\footnotesize{}5.29} & {\footnotesize{}12.88} & {\footnotesize{}6.42} & {\footnotesize{}101.10} & {\footnotesize{}2.43}\tabularnewline
{\footnotesize{}COST} & {\footnotesize{}79533} & {\footnotesize{}6.51E-06} & {\footnotesize{}1.87E-03} & {\footnotesize{}0.44} & {\footnotesize{}7.73} & {\footnotesize{}444.95} & {\footnotesize{}6.81} & {\footnotesize{}13.88} & {\footnotesize{}4.84} & {\footnotesize{}56.97} & {\footnotesize{}2.04}\tabularnewline
{\footnotesize{}DISCA} & {\footnotesize{}55283} & {\footnotesize{}5.72E-06} & {\footnotesize{}2.92E-03} & {\footnotesize{}0.82} & {\footnotesize{}11.16} & {\footnotesize{}425.82} & {\footnotesize{}10.06} & {\footnotesize{}20.41} & {\footnotesize{}4.09} & {\footnotesize{}28.31} & {\footnotesize{}2.03}\tabularnewline
{\footnotesize{}EBAY} & {\footnotesize{}169875} & {\footnotesize{}6.39E-06} & {\footnotesize{}3.00E-03} & {\footnotesize{}0.21} & {\footnotesize{}3.52} & {\footnotesize{}363.80} & {\footnotesize{}3.47} & {\footnotesize{}7.61} & {\footnotesize{}5.44} & {\footnotesize{}76.51} & {\footnotesize{}2.20}\tabularnewline
{\footnotesize{}FB} & {\footnotesize{}151244} & {\footnotesize{}4.34E-06} & {\footnotesize{}8.69E-04} & {\footnotesize{}0.08} & {\footnotesize{}3.20} & {\footnotesize{}294.02} & {\footnotesize{}3.92} & {\footnotesize{}9.88} & {\footnotesize{}5.80} & {\footnotesize{}59.27} & {\footnotesize{}2.52}\tabularnewline
{\footnotesize{}GOOG} & {\footnotesize{}135745} & {\footnotesize{}5.50E-06} & {\footnotesize{}7.07E-04} & {\footnotesize{}0.06} & {\footnotesize{}3.75} & {\footnotesize{}235.97} & {\footnotesize{}4.05} & {\footnotesize{}9.16} & {\footnotesize{}4.34} & {\footnotesize{}29.78} & {\footnotesize{}2.26}\tabularnewline
{\footnotesize{}INTC} & {\footnotesize{}124260} & {\footnotesize{}5.16E-06} & {\footnotesize{}1.00E-03} & {\footnotesize{}0.04} & {\footnotesize{}3.20} & {\footnotesize{}361.41} & {\footnotesize{}4.63} & {\footnotesize{}12.23} & {\footnotesize{}5.89} & {\footnotesize{}60.05} & {\footnotesize{}2.64}\tabularnewline
{\footnotesize{}ISRG} & {\footnotesize{}38118} & {\footnotesize{}4.90E-06} & {\footnotesize{}4.43E-04} & {\footnotesize{}0.04} & {\footnotesize{}11.90} & {\footnotesize{}1504.86} & {\footnotesize{}15.02} & {\footnotesize{}37.79} & {\footnotesize{}6.45} & {\footnotesize{}102.77} & {\footnotesize{}2.52}\tabularnewline
{\footnotesize{}KLAC} & {\footnotesize{}70264} & {\footnotesize{}5.17E-06} & {\footnotesize{}3.06E-03} & {\footnotesize{}0.54} & {\footnotesize{}9.21} & {\footnotesize{}600.68} & {\footnotesize{}8.31} & {\footnotesize{}17.46} & {\footnotesize{}5.68} & {\footnotesize{}83.42} & {\footnotesize{}2.10}\tabularnewline
{\footnotesize{}MAR} & {\footnotesize{}39728} & {\footnotesize{}5.29E-06} & {\footnotesize{}3.03E-03} & {\footnotesize{}1.22} & {\footnotesize{}15.60} & {\footnotesize{}2166.25} & {\footnotesize{}14.02} & {\footnotesize{}31.83} & {\footnotesize{}15.83} & {\footnotesize{}795.34} & {\footnotesize{}2.27}\tabularnewline
{\footnotesize{}MSFT} & {\footnotesize{}163839} & {\footnotesize{}5.91E-06} & {\footnotesize{}1.20E-03} & {\footnotesize{}0.05} & {\footnotesize{}2.50} & {\footnotesize{}833.87} & {\footnotesize{}3.72} & {\footnotesize{}10.69} & {\footnotesize{}10.40} & {\footnotesize{}348.43} & {\footnotesize{}2.87}\tabularnewline
{\footnotesize{}NFLX} & {\footnotesize{}182071} & {\footnotesize{}5.05E-06} & {\footnotesize{}5.58E-04} & {\footnotesize{}0.03} & {\footnotesize{}2.41} & {\footnotesize{}242.51} & {\footnotesize{}3.34} & {\footnotesize{}8.47} & {\footnotesize{}5.17} & {\footnotesize{}43.02} & {\footnotesize{}2.54}\tabularnewline
{\footnotesize{}QCOM} & {\footnotesize{}161471} & {\footnotesize{}6.86E-06} & {\footnotesize{}2.97E-03} & {\footnotesize{}0.26} & {\footnotesize{}3.75} & {\footnotesize{}443.04} & {\footnotesize{}3.54} & {\footnotesize{}7.71} & {\footnotesize{}6.08} & {\footnotesize{}115.19} & {\footnotesize{}2.18}\tabularnewline
{\footnotesize{}REGN} & {\footnotesize{}52770} & {\footnotesize{}4.68E-06} & {\footnotesize{}4.76E-04} & {\footnotesize{}0.05} & {\footnotesize{}8.71} & {\footnotesize{}632.70} & {\footnotesize{}10.57} & {\footnotesize{}25.64} & {\footnotesize{}5.30} & {\footnotesize{}50.58} & {\footnotesize{}2.43}\tabularnewline
{\footnotesize{}SBUX} & {\footnotesize{}101250} & {\footnotesize{}5.47E-06} & {\footnotesize{}3.52E-03} & {\footnotesize{}0.57} & {\footnotesize{}6.47} & {\footnotesize{}201.61} & {\footnotesize{}5.51} & {\footnotesize{}10.78} & {\footnotesize{}3.95} & {\footnotesize{}25.89} & {\footnotesize{}1.96}\tabularnewline
{\footnotesize{}TXN} & {\footnotesize{}97907} & {\footnotesize{}5.28E-06} & {\footnotesize{}1.72E-03} & {\footnotesize{}0.27} & {\footnotesize{}5.73} & {\footnotesize{}461.87} & {\footnotesize{}5.87} & {\footnotesize{}13.20} & {\footnotesize{}5.04} & {\footnotesize{}49.99} & {\footnotesize{}2.25}\tabularnewline
{\footnotesize{}VOD} & {\footnotesize{}57699} & {\footnotesize{}5.81E-06} & {\footnotesize{}3.86E-03} & {\footnotesize{}0.78} & {\footnotesize{}10.14} & {\footnotesize{}794.09} & {\footnotesize{}10.81} & {\footnotesize{}25.51} & {\footnotesize{}6.40} & {\footnotesize{}87.76} & {\footnotesize{}2.36}\tabularnewline
{\footnotesize{}YHOO} & {\footnotesize{}91951} & {\footnotesize{}5.24E-06} & {\footnotesize{}1.63E-03} & {\footnotesize{}0.24} & {\footnotesize{}5.92} & {\footnotesize{}460.85} & {\footnotesize{}6.31} & {\footnotesize{}14.89} & {\footnotesize{}5.37} & {\footnotesize{}51.83} & {\footnotesize{}2.36}\tabularnewline
\hline 
\end{tabular}}
\textit{\footnotesize{}Note:}{\footnotesize{} 25\%, Med., and 75\%
are the 25th percentile, 50th percentile (median), and 75th percentile, respectively.
Std., Skew., and Kurt. are the standard deviation, skewness, and kurtosis, respectively.}{\footnotesize\par}
\label{table: durationStatistics}
\end{table}

\subsection{Choice of LOB factors}
To identify the factors that may impact the regime switchings of inter-trade durations, 
we focus the covariates in our model on the LOB factors for the following reasons.
First, many studies advocate that the order flows are determined by the state of the order book \citep{hautsch2012market,huang2015simulating,abergel2015long};
therefore, factors reflecting the state of the LOB are thought to determine
the flow of market orders and hence the inter-trade durations. 
Second, the information flow revealed by the change of LOB contribute a major part of 
information learning for the market participants \citep{harris2005information, cao2009information}, 
especially for the HFTs who take the speed advantage to trade on the information.
Third, a lot of research has found that the algorithm
traders in the limit order market are indeed using LOB factors as signals to set their trading strategies. 
\citep{LI2005533,obizhaeva2013optimal,hendershott2013algorithmic,goldstein2017high}, 
so these factors must play an important role on the arrival time of trades, i.e., the inter-trade durations. 
 
Let $p_a$ and $p_b$ be the prices (dollar prices divided by 10) for the best ask and the best bid, respectively, 
and $v_a^1$ and $v_b^1$ be the depth at the best ask and the best bid, respectively.
We consider the following four LOB factors as the covariates in our MF-RSD model.
\begin{itemize}
\item Depth imbalance ($DI$). $DI=\frac{|v_{a}^{1}-v_{b}^{1}|}{|v_{a}^{1}+v_{b}^{1}|}$.
It is the difference between the depth of the best ask and the
depth of the best bid, divided by the sum of the two depths.
It compares order supplies between the best ask and the best bid, and
a large $DI$ reflects a great imbalance of the two sides in offering liquidity.
Some studies \citep{cont2014price,lipton2013trade,cartea2018enhancing}
suggest that it is an effective predictor for the subsequent price movements and the rate of incoming orders.
\item Price spread (\textit{PS}). $PS=|p_{a}-p_{b}|$, which is the price
difference between the best ask and the best bid. It is commonly considered
as a result of information asymmetry and reveals the stock's overall liquidity \citep{o2003presidential}.
From the perspective of market microstructure,
it measures the transaction cost for the liquidity demander, and normally,
a wider price spread implies weaker trading aggressiveness \citep{ranaldo2004order}.
Moreover, \cite{carrion2013very} and \cite{hendershott2013algorithmic} found that 
high-frequency market makers provide liquidity when
the spread is large while they take liquidity when the spread is small.
\item Trade volume of the last trade, denoted by \textit{TV}. It is the size
of the last trade (volume measured in thousands), which is generally treated as
a momentum variable that reflects the trading magnitude and willingness. Large trades normally convey
more information than small trades \citep{hasbrouck1991measuring} and are also expected
to be accompanied by some small upcoming trades \citep{van2019high}.
\item Price movement of the last trade, denoted by \textit{PM}. It is a dummy
variable that indicates whether the last trade had led to a change in
mid-price, i.e., $\frac{p_{_{a}}+p_{b}}{2}$. We think it might be a momentum variable for 
the price change and we want to see whether market participants treat it as 
an informative signal to identify the price movement in the following moments.
\end{itemize}

\begin{table}[!h]
\caption{Summary statistics for 4 LOB factors, which are depth imbalance ($DI$),
price spread ($PS$), trade volume of the last trade ($TV$), and price
movement of the last trade ($PM$).}

\medskip{}

\centerline{%
\begin{tabular}{c|ccc|ccc|ccc|>{\centering}p{1.6cm}}
\hline 
\multirow{2}{*}{{\footnotesize{}Stock}} & \multicolumn{3}{c|}{{\footnotesize{}$DI$}} & \multicolumn{3}{c|}{{\footnotesize{}$PS$}} & \multicolumn{3}{c|}{{\footnotesize{}$TV$}} & \multicolumn{1}{c}{{\footnotesize{}$PM$}}\tabularnewline
\cline{2-11} 
 & {\footnotesize{}25\%} & {\footnotesize{}Med} & {\footnotesize{}75\%} & {\footnotesize{}25\%} & {\footnotesize{}Med} & {\footnotesize{}75\%} & {\footnotesize{}25\%} & {\footnotesize{}Med} & {\footnotesize{}75\%} & {\footnotesize{} Mean}\tabularnewline
\hline 
{\footnotesize{}AAPL} & {\footnotesize{}0.22} & {\footnotesize{}0.50} & {\footnotesize{}0.78} & {\footnotesize{}0.8} & {\footnotesize{}1.2} & {\footnotesize{}1.7} & {\footnotesize{}0.04} & {\footnotesize{}0.1} & {\footnotesize{}0.12} & {\footnotesize{}0.34}\tabularnewline
{\footnotesize{}ALXN} & {\footnotesize{}0.16} & {\footnotesize{}0.37} & {\footnotesize{}0.71} & {\footnotesize{}0.3} & {\footnotesize{}0.5} & {\footnotesize{}0.7} & {\footnotesize{}0.05} & {\footnotesize{}0.1} & {\footnotesize{}0.1} & {\footnotesize{}0.38}\tabularnewline
{\footnotesize{}AMZN} & {\footnotesize{}0.20} & {\footnotesize{}0.48} & {\footnotesize{}0.77} & {\footnotesize{}0.8} & {\footnotesize{}1.2} & {\footnotesize{}1.7} & {\footnotesize{}0.03} & {\footnotesize{}0.1} & {\footnotesize{}0.1} & {\footnotesize{}0.42}\tabularnewline
{\footnotesize{}BIDU} & {\footnotesize{}0.20} & {\footnotesize{}0.39} & {\footnotesize{}0.71} & {\footnotesize{}0.3} & {\footnotesize{}0.4} & {\footnotesize{}0.7} & {\footnotesize{}0.09} & {\footnotesize{}0.1} & {\footnotesize{}0.15} & {\footnotesize{}0.43}\tabularnewline
{\footnotesize{}BMRN} & {\footnotesize{}0.14} & {\footnotesize{}0.33} & {\footnotesize{}0.58} & {\footnotesize{}0.2} & {\footnotesize{}0.3} & {\footnotesize{}0.5} & {\footnotesize{}0.1} & {\footnotesize{}0.1} & {\footnotesize{}0.1} & {\footnotesize{}0.40}\tabularnewline
{\footnotesize{}CELG} & {\footnotesize{}0.18} & {\footnotesize{}0.39} & {\footnotesize{}0.69} & {\footnotesize{}0.2} & {\footnotesize{}0.4} & {\footnotesize{}0.6} & {\footnotesize{}0.05} & {\footnotesize{}0.1} & {\footnotesize{}0.1} & {\footnotesize{}0.40}\tabularnewline
{\footnotesize{}CERN} & {\footnotesize{}0.20} & {\footnotesize{}0.43} & {\footnotesize{}0.75} & {\footnotesize{}0.3} & {\footnotesize{}0.5} & {\footnotesize{}0.7} & {\footnotesize{}0.03} & {\footnotesize{}0.1} & {\footnotesize{}0.1} & {\footnotesize{}0.38}\tabularnewline
{\footnotesize{}CMCSA} & {\footnotesize{}0.22} & {\footnotesize{}0.45} & {\footnotesize{}0.70} & {\footnotesize{}0.1} & {\footnotesize{}0.1} & {\footnotesize{}0.1} & {\footnotesize{}0.1} & {\footnotesize{}0.2} & {\footnotesize{}0.6} & {\footnotesize{}0.08}\tabularnewline
{\footnotesize{}COST} & {\footnotesize{}0.20} & {\footnotesize{}0.40} & {\footnotesize{}0.66} & {\footnotesize{}0.2} & {\footnotesize{}0.2} & {\footnotesize{}0.3} & {\footnotesize{}0.05} & {\footnotesize{}0.1} & {\footnotesize{}0.12} & {\footnotesize{}0.36}\tabularnewline
{\footnotesize{}DISCA} & {\footnotesize{}0.20} & {\footnotesize{}0.39} & {\footnotesize{}0.67} & {\footnotesize{}0.2} & {\footnotesize{}0.2} & {\footnotesize{}0.4} & {\footnotesize{}0.08} & {\footnotesize{}0.1} & {\footnotesize{}0.11} & {\footnotesize{}0.36}\tabularnewline
{\footnotesize{}EBAY} & {\footnotesize{}0.23} & {\footnotesize{}0.48} & {\footnotesize{}0.72} & {\footnotesize{}0.1} & {\footnotesize{}0.1} & {\footnotesize{}0.2} & {\footnotesize{}0.1} & {\footnotesize{}0.17} & {\footnotesize{}0.34} & {\footnotesize{}0.18}\tabularnewline
{\footnotesize{}FB} & {\footnotesize{}0.24} & {\footnotesize{}0.48} & {\footnotesize{}0.74} & {\footnotesize{}0.1} & {\footnotesize{}0.1} & {\footnotesize{}0.1} & {\footnotesize{}0.1} & {\footnotesize{}0.3} & {\footnotesize{}0.7} & {\footnotesize{}0.08}\tabularnewline
{\footnotesize{}GOOG} & {\footnotesize{}0.19} & {\footnotesize{}0.49} & {\footnotesize{}0.79} & {\footnotesize{}1.7} & {\footnotesize{}2.6} & {\footnotesize{}3.7} & {\footnotesize{}0.02} & {\footnotesize{}0.09} & {\footnotesize{}0.1} & {\footnotesize{}0.40}\tabularnewline
{\footnotesize{}INTC} & {\footnotesize{}0.22} & {\footnotesize{}0.46} & {\footnotesize{}0.74} & {\footnotesize{}0.1} & {\footnotesize{}0.1} & {\footnotesize{}0.1} & {\footnotesize{}0.1} & {\footnotesize{}0.3} & {\footnotesize{}1.2} & {\footnotesize{}0.03}\tabularnewline
{\footnotesize{}ISRG} & {\footnotesize{}0.20} & {\footnotesize{}0.58} & {\footnotesize{}0.94} & {\footnotesize{}3.5} & {\footnotesize{}4.9} & {\footnotesize{}6.3} & {\footnotesize{}0.006} & {\footnotesize{}0.05} & {\footnotesize{}0.1} & {\footnotesize{}0.41}\tabularnewline
{\footnotesize{}KLAC} & {\footnotesize{}0.20} & {\footnotesize{}0.42} & {\footnotesize{}0.66} & {\footnotesize{}0.1} & {\footnotesize{}0.2} & {\footnotesize{}0.2} & {\footnotesize{}0.1} & {\footnotesize{}0.1} & {\footnotesize{}0.2} & {\footnotesize{}0.30}\tabularnewline
{\footnotesize{}MAR} & {\footnotesize{}0.20} & {\footnotesize{}0.41} & {\footnotesize{}0.64} & {\footnotesize{}0.1} & {\footnotesize{}0.1} & {\footnotesize{}0.2} & {\footnotesize{}0.1} & {\footnotesize{}0.1} & {\footnotesize{}0.2} & {\footnotesize{}0.29}\tabularnewline
{\footnotesize{}MSFT} & {\footnotesize{}0.24} & {\footnotesize{}0.49} & {\footnotesize{}0.77} & {\footnotesize{}0.1} & {\footnotesize{}0.1} & {\footnotesize{}0.1} & {\footnotesize{}0.1} & {\footnotesize{}0.32} & {\footnotesize{}1.12} & {\footnotesize{}0.03}\tabularnewline
{\footnotesize{}NFLX} & {\footnotesize{}0.14} & {\footnotesize{}0.38} & {\footnotesize{}0.68} & {\footnotesize{}0.9} & {\footnotesize{}1.4} & {\footnotesize{}2} & {\footnotesize{}0.07} & {\footnotesize{}0.1} & {\footnotesize{}0.11} & {\footnotesize{}0.45}\tabularnewline
{\footnotesize{}QCOM} & {\footnotesize{}0.23} & {\footnotesize{}0.46} & {\footnotesize{}0.70} & {\footnotesize{}0.1} & {\footnotesize{}0.1} & {\footnotesize{}0.2} & {\footnotesize{}0.1} & {\footnotesize{}0.2} & {\footnotesize{}0.4} & {\footnotesize{}0.15}\tabularnewline
{\footnotesize{}REGN} & {\footnotesize{}0.05} & {\footnotesize{}0.33} & {\footnotesize{}0.67} & {\footnotesize{}1.4} & {\footnotesize{}2} & {\footnotesize{}2.9} & {\footnotesize{}0.04} & {\footnotesize{}0.1} & {\footnotesize{}0.1} & {\footnotesize{}0.45}\tabularnewline
{\footnotesize{}SBUX} & {\footnotesize{}0.21} & {\footnotesize{}0.44} & {\footnotesize{}0.68} & {\footnotesize{}0.1} & {\footnotesize{}0.1} & {\footnotesize{}0.2} & {\footnotesize{}0.1} & {\footnotesize{}0.1} & {\footnotesize{}0.26} & {\footnotesize{}0.22}\tabularnewline
{\footnotesize{}TXN} & {\footnotesize{}0.24} & {\footnotesize{}0.50} & {\footnotesize{}0.74} & {\footnotesize{}0.1} & {\footnotesize{}0.1} & {\footnotesize{}0.1} & {\footnotesize{}0.1} & {\footnotesize{}0.2} & {\footnotesize{}0.5} & {\footnotesize{}0.11}\tabularnewline
{\footnotesize{}VOD} & {\footnotesize{}0.31} & {\footnotesize{}0.60} & {\footnotesize{}0.84} & {\footnotesize{}0.1} & {\footnotesize{}0.1} & {\footnotesize{}0.1} & {\footnotesize{}0.1} & {\footnotesize{}0.28} & {\footnotesize{}0.7} & {\footnotesize{}0.09}\tabularnewline
{\footnotesize{}YHOO} & {\footnotesize{}0.26} & {\footnotesize{}0.51} & {\footnotesize{}0.75} & {\footnotesize{}0.1} & {\footnotesize{}0.1} & {\footnotesize{}0.1} & {\footnotesize{}0.1} & {\footnotesize{}0.3} & {\footnotesize{}0.9} & {\footnotesize{}0.05}\tabularnewline
\hline 
\end{tabular}}
\textit{\footnotesize{}Note:}{\footnotesize{} 25\%, Med., and 75\%
are the 25th percentile, 50th percentile (median), and 75th percentile, respectively.}{\footnotesize\par}
\label{table:factorStatistics}
\end{table}

In Table \ref{table:factorStatistics}, we show some summary statistics of these LOB factors.
The minimum and maximum values of $DI$ are 0 and 1, and the minimum values of $PS$ and $TV$ are 
always $0.1$ (1 cent) and $0.001$ (1 share), respectively; we do not report these values in the table.
Since the factor $PM$ is a dummy variable, we only report its mean, or equivalently the ratio of the times when it equals to 1.
From Table \ref{table:factorStatistics}, we find that the percentiles of $DI$ are very similar
across stocks. The factors $TV$ and $PM$ do have some
differences from stock to stock, while the most divergent factor is $PS$.
For some stocks, such as CMCSA, FB, INTC, MSFT, TXN, VOD and YHOO, 
the factor $PS$ are most time fixed at 0.1 (the minimum tick size). 
But for some other stocks, such as GOOG, ISRG, and REGN,
$PS$ varies much and its values can be very large.

\subsection{In-sample regression and model prediction}
We first implement the in-sample analysis for the inter-trade durations of MSFT on January 2, 2013,
according to the following regression models for the transition probabilities:
\begin{equation}\label{model.4lob}
\begin{aligned}
\log\frac{P_{12}}{1-P_{12}} & =\beta_{12,0}+\beta_{12,1}DI+\beta_{12,2}PS+\beta_{12,3}TV+\beta_{12,4}PM,\\
\log\frac{P_{21}}{1-P_{21}} & =\beta_{21,0}+\beta_{21,1}DI+\beta_{21,3}PS+\beta_{21,4}TV+\beta_{21,5}PM.
\end{aligned}
\end{equation}
The model is estimated using the EM procedure as described in Section 4 and the results are reported in Table \ref{table:MSFTregression}.
In addition to the full model - Model 6, we also estimate the submodels (Models 1-5).
The distribution parameters estimated from the full model are very close to those from the submodels,
and the estimated factor coefficients in the full model are 
having similar signs and significance levels as those in the submodels,
suggesting that there is no strong multicollinearity in the LOB factors.
In the rest of the empirical study, we shall only report the results using the full model (M6).

\begin{table}[!h]
\caption{Estimation results for MSFT on January 2, 2013, in 6 regression models. }
\medskip{}
\hspace{1.4cm}
\begin{tabular}{>{\centering}b{1.5cm}>{\centering}p{1.4cm}>{\centering}p{1.4cm}>{\centering}p{1.4cm}>{\centering}p{1.4cm}>{\centering}p{1.4cm}>{\centering}p{1.4cm}}
\hline 
{\footnotesize{}Parameters} & {\footnotesize{}M1} & {\footnotesize{}M2} & {\footnotesize{}M3} & {\footnotesize{}M4} & {\footnotesize{}M5} & {\footnotesize{}M6}\tabularnewline
\hline 
{\footnotesize{}$\hat{\mu}_{1}$} & {\scriptsize{}0.131}{\scriptsize\par}
{\scriptsize{}(.0632){*}} & {\scriptsize{}0.125}{\scriptsize\par}

{\scriptsize{}(.0602){*}} & {\scriptsize{}0.132}{\scriptsize\par}

{\scriptsize{}(.0637){*}} & {\scriptsize{}0.145}{\scriptsize\par}

{\scriptsize{}(.0690){*}} & {\scriptsize{}0.131}{\scriptsize\par}

{\scriptsize{}(.0632){*}} & {\scriptsize{}0.142}{\scriptsize\par}

{\scriptsize{}(.0670){*}{*}}\tabularnewline
{\footnotesize{}$\hat{\lambda}_{1}$} & {\scriptsize{}0.000234}{\scriptsize\par}

{\scriptsize{}(5.45e-6){*}{*}} & {\scriptsize{}0.000234}{\scriptsize\par}

{\scriptsize{}(5.45e-6){*}{*}} & {\scriptsize{}0.000234}{\scriptsize\par}

{\scriptsize{}(5.45e-6){*}{*}} & {\scriptsize{}0.000234}{\scriptsize\par}

{\scriptsize{}(5.44e-6){*}{*}} & {\scriptsize{}0.000234}{\scriptsize\par}

{\scriptsize{}(5.45e-6{*}{*}} & {\scriptsize{}0.000234}{\scriptsize\par}

{\scriptsize{}(5.44e-6){*}{*}}\tabularnewline
{\footnotesize{}$\hat{\mu}_{2}$} & {\scriptsize{}10.066}{\scriptsize\par}

{\scriptsize{}(.418){*}{*}} & {\scriptsize{}10.070}{\scriptsize\par}

{\scriptsize{}(.419){*}{*}} & {\scriptsize{}10.062}{\scriptsize\par}

{\scriptsize{}(.418){*}{*}} & {\scriptsize{}10.025}{\scriptsize\par}

{\scriptsize{}(.423){*}{*}} & {\scriptsize{}10.065}{\scriptsize\par}

{\scriptsize{}(.418){*}{*}} & {\scriptsize{}10.024}{\scriptsize\par}

{\scriptsize{}(.422){*}{*}}\tabularnewline
{\footnotesize{}$\hat{\lambda}_{2}$} & {\scriptsize{}2.411}{\scriptsize\par}

{\scriptsize{}(.109){*}{*}} & {\scriptsize{}2.407}{\scriptsize\par}

{\scriptsize{}(.109){*}{*}} & {\scriptsize{}2.403}{\scriptsize\par}

{\scriptsize{}(.109){*}{*}} & {\scriptsize{}2.324}{\scriptsize\par}

{\scriptsize{}(.106){*}{*}} & {\scriptsize{}2.409}{\scriptsize\par}

{\scriptsize{}(.109){*}{*}} & {\scriptsize{}2.318}{\scriptsize\par}

{\scriptsize{}(.105){*}{*}}\tabularnewline
{\footnotesize{}$\hat{\beta}_{12,0}$ } & {\scriptsize{}-1.061}{\scriptsize\par}

{\scriptsize{}(.0393){*}{*}} & {\scriptsize{}-0.424}{\scriptsize\par}

{\scriptsize{}(.0723){*}{*}} & {\scriptsize{}-0.781}{\scriptsize\par}

{\scriptsize{}(.120){*}{*}} & {\scriptsize{}-0.991}{\scriptsize\par}

{\scriptsize{}(.0459){*}{*}} & {\scriptsize{}-1.064}{\scriptsize\par}

{\scriptsize{}(.0407){*}{*}} & {\scriptsize{}0.881}{\scriptsize\par}

{\scriptsize{}(.177){*}{*}}\tabularnewline
{\footnotesize{}$\hat{\beta}_{12,1}$} & {\scriptsize{}\_} & {\scriptsize{}-1.249}{\scriptsize\par}

{\scriptsize{}(.128){*}{*}} & {\scriptsize{}\_} & {\scriptsize{}\_} & {\scriptsize{}\_} & {\scriptsize{}-1.654}{\scriptsize\par}

{\scriptsize{}(.138)}\tabularnewline
{\footnotesize{}$\hat{\beta}_{12,2}$ } & {\scriptsize{}\_} & {\scriptsize{}\_} & {\scriptsize{}-2.237}{\scriptsize\par}

{\scriptsize{}(.913){*}} & {\scriptsize{}\_} & {\scriptsize{}\_} & {\scriptsize{}-8.145}{\scriptsize\par}

{\scriptsize{}(1.118){*}{*}}\tabularnewline
{\footnotesize{}$\hat{\beta}_{12,3}$} & {\scriptsize{}\_} & {\scriptsize{}\_} & {\scriptsize{}\_} & {\scriptsize{}-0.0944}{\scriptsize\par}

{\scriptsize{}(.0368){*}} & {\scriptsize{}\_} & {\scriptsize{}-0.107}{\scriptsize\par}

{\scriptsize{}(.0350){*}{*}}\tabularnewline
{\footnotesize{}$\hat{\beta}_{12,4}$ } & {\scriptsize{}\_} & {\scriptsize{}\_} & {\scriptsize{}\_} & {\scriptsize{}\_} & {\scriptsize{}0.0479}{\scriptsize\par}

{\scriptsize{}(.156)} & {\scriptsize{}0.157}{\scriptsize\par}

{\scriptsize{}(.179)}\tabularnewline
{\footnotesize{}$\hat{\beta}_{21,0}$} & {\scriptsize{}-0.442}{\scriptsize\par}

{\scriptsize{}(.0433){*}{*}} & {\scriptsize{}-0.925}{\scriptsize\par}

{\scriptsize{}(.0915){*}{*}} & {\scriptsize{}0.345}{\scriptsize\par}

{\scriptsize{}(.328)} & {\scriptsize{}-0.886}{\scriptsize\par}

{\scriptsize{}(.0569){*}{*}} & {\scriptsize{}-0.443}{\scriptsize\par}

{\scriptsize{}(.0432){*}{*}} & {\scriptsize{}-1.018}{\scriptsize\par}

{\scriptsize{}(.417){*}}\tabularnewline
{\footnotesize{}$\hat{\beta}_{21,1}$} & {\scriptsize{}\_} & {\scriptsize{}0.947}{\scriptsize\par}

{\scriptsize{}(.158){*}{*}} & {\scriptsize{}\_} & {\scriptsize{}\_} & {\scriptsize{}\_} & {\scriptsize{}1.139}{\scriptsize\par}

{\scriptsize{}(.168){*}{*}}\tabularnewline
{\footnotesize{}$\hat{\beta}_{21,2}$} & {\scriptsize{}\_} & {\scriptsize{}\_} & {\scriptsize{}-7.685}{\scriptsize\par}

{\scriptsize{}(3.187){*}} & {\scriptsize{}\_} & {\scriptsize{}\_} & {\scriptsize{}-4.539}{\scriptsize\par}

{\scriptsize{}(3.831)}\tabularnewline
{\footnotesize{}$\hat{\beta}_{21,3}$} & {\scriptsize{}\_} & {\scriptsize{}\_} & {\scriptsize{}\_} & {\scriptsize{}0.342}{\scriptsize\par}

{\scriptsize{}(.0320){*}{*}} & {\scriptsize{}\_} & {\scriptsize{}0.341}{\scriptsize\par}

{\scriptsize{}(.0305){*}{*}}\tabularnewline
{\footnotesize{}$\hat{\beta}_{21,4}$} & {\scriptsize{}\_} & {\scriptsize{}\_} & {\scriptsize{}\_} & {\scriptsize{}\_} & {\scriptsize{}0.345}{\scriptsize\par}

{\scriptsize{}(.692)} & {\scriptsize{}1.445}{\scriptsize\par}

{\scriptsize{}(.810)}\tabularnewline
\hline 
\end{tabular}
\label{table:MSFTregression}

\textit{\scriptsize{}Notes:}{\scriptsize{} M1 is a model without any
LOB factor. M2-M5 are models having 1 LOB factor, with $DI$, $PS$,
$TV$ and $PM$, respectively. M6 is a model with all 4 LOB factors.
The robust standard errors are shown in parentheses. $*p<0.05$ and $**p<0.01$.}{\scriptsize\par}
\end{table}

From Table \ref{table:MSFTregression}, we can see that the estimated mean of the long-duration regime is much larger than
that of the short-duration regimes, i.e., $\hat{\mu}_2>>\hat{\mu}_1$.
Based on the estimated values for the parameters of two IVGs, we also calculate the modes of the two distributions. 
They are around $8\times10^{-5}$ seconds and $1$ second, 
which are consistent with the bimodal distribution of inter-trade durations discussed in Section 3. 
Moreover, we have noticed the following results for the LOB factors. 
First, the estimated coefficients for the $DI$ is significantly positive for ${P}_{12}$ ($\hat{\beta}_{12,1}>0$) 
and significantly negative for ${P}_{21}$ ($\hat{\beta}_{21,1}<0$). Second, 
the $PS$ is shown to be both negative for ${P}_{12}$ and ${P}_{21}$ while its coefficient is only significant in ${P}_{12}$.
Third, $\hat{\beta}_{12,3}$ is significantly less than 0 and $\hat{\beta}_{21,3}$ is significantly greater than 0, 
suggesting that the $TV$ is a significant factor for the regime switchings of inter-trade durations. 
Last, the factor $PM$ seems to be insignificant in this case because 
both $\hat{\beta}_{12,4}$ and $\hat{\beta}_{21,4}$ are not significantly different than 0. 
Nonetheless, the LOB factors fairly impact the dynamics of inter-trade duration via regime-switching probabilities, 
and we expect that our MF-RSD model can enhance the prediction of inter-trade durations by considering their effects. Moreover, we'd like to understand more about 
the economics implication of the effects of LOB factors on the regime switchings of 
inter-trade durations, which will be discussed in detail in Section \ref{SC: LOB factors}.

Given the estimated values of model parameters, we examine the model prediction power by 
first considering a 1-step-ahead out-of-sample prediction for the underlying states.
The 1-step-ahead prediction of $s_i$ is given by
\begin{equation}
\begin{aligned}
\mathrm{E}_i(s_{i+1}) & =1\cdot\left[\mathbb{P}(s_i=1|\mathcal{F})\hat{P}_{11,i+1}+\mathbb{P}(s_i=2|\mathcal{F})\hat{P}_{21,i+1}\right]\\
 & +2\cdot\left[\mathbb{P}(s_i=1|\mathcal{F})\hat{P}_{12,i+1}+\mathbb{P}(s_i=2|\mathcal{F})\hat{P}_{22,i+1}\right],
\end{aligned}
\end{equation}

\noindent and the transition probabilities at $i$ are
given by
\begin{equation}
\begin{aligned}
\log\frac{\hat{P}_{12,i+1}}{1-P_{12,i+1}} & =\hat{\beta}_{12,0}+\hat{\beta}_{12,1}DI_i+\hat{\beta}_{12,2}PS_i
+\hat{\beta}_{12,3}TV_i +\hat{\beta}_{12,4}PM_i\\
\log\frac{\hat{P}_{21,i+1}}{1-P_{21,i+1}} & =\hat{\beta}_{21,0}+\hat{\beta}_{21,1}DI_i+\hat{\beta}_{21,2}PS_i
+\hat{\beta}_{21,3}TV_i+\hat{\beta}_{21,4}PM_i.
\end{aligned}
\label{eq:estimatedP}
\end{equation}
\noindent which can be calculated given the information up to $i-$th trade.

\begin{figure}[!h]
\centerline{\includegraphics[scale=0.36]{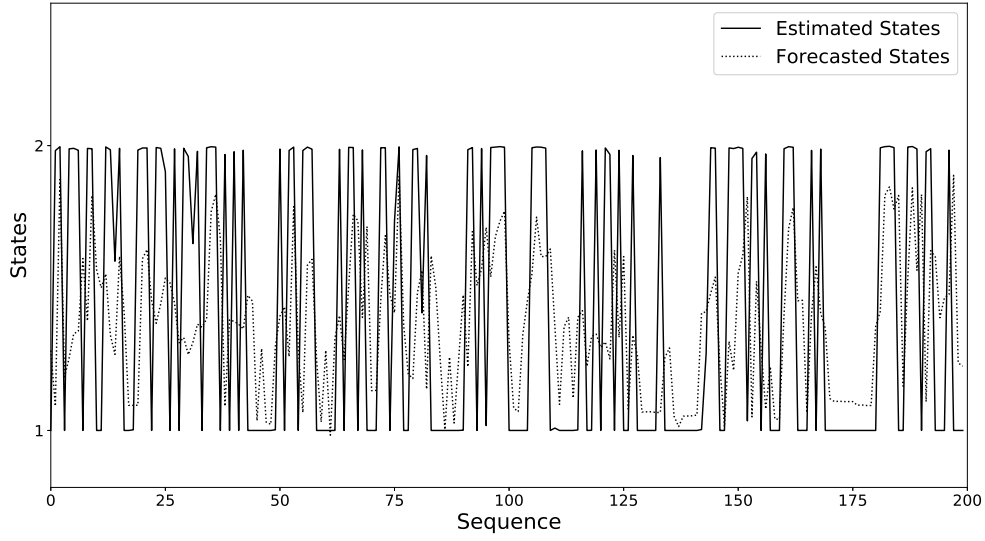}}
\caption{1-step-ahead out-of-sample forecasted versus estimated regime states.}
\label{fig:forecastStates}
\end{figure}

We use the MSFT inter-trade duration series on January 2, 2013, to demonstrate
the above prediction. The model parameter is estimated using the first 5000 samples,
and 1-step-ahead forecasting is made for the next 200 durations.
Figure \ref{fig:forecastStates} shows the comparison between the estimated states and the 1-step-ahead forecasts.
The estimated states are the underlying states inferred from the real data, and the forecasted states
are 1-step-ahead out-of-sample prediction.\footnote{The states jump between two values, 
i.e., state 1 (the short-duration regime) and state 2 (the long-duration regime)} From the result, we can see
that the forecasted states almost capture the variation tendency of
the underlying estimated states.

Furthermore, the 1-step-ahead prediction for the inter-trade duration is expressed as
\begin{equation}
\begin{aligned}
\mathrm{E}_{n}(y_{n+1}) & =\hat{\mu}_{1}\cdot\left[\mathbb{P}(s_n=1|\mathcal{F})\hat{P}_{11,n+1}+\mathbb{P}(s_n=2|\mathcal{F})\hat{P}_{21,n+1}\right]\\
 & +\hat{\mu_{2}}\cdot\left[\mathbb{P}(s_n=1|\mathcal{F})\hat{P}_{12,n+1}+\mathbb{P}(s_n=2|\mathcal{F})\hat{P}_{22,n+1}\right].
\end{aligned}
\end{equation}
To see the predictive performance of our model, we
compare our results in predicting the next inter-trade duration with those of two classical
duration models, i.e., the ACD model by \cite{engle1998autoregressive} and the MSMD model by \cite{chen2013markov}.
The ACD model we considered has the following form:
$$
y_{i} =\varphi_{i}\varepsilon_{i}, \qquad
\varphi_{i} =\omega+\alpha y_{i-1}+\beta\varphi_{i-1},
$$

\noindent where $\varepsilon_{i}$ follows an inverse
Gaussian distribution. The MSMD model we considered
has five levels, and the following form:
\begin{align*}
y_{i} & =\frac{\varepsilon_{i}}{\lambda_{i}}=\varphi_{i}\varepsilon_{i}, \quad
\varphi_{i} =\varphi\prod_{k=1}^{5}M_{k,i}, \\
M_{k,i} &=\begin{cases}
M & \textrm{w.p. }1-(1-\gamma_{5})^{b^{k-5}}\\
M_{k,i-1} & \textrm{w.p. }(1-\gamma_{5})^{b^{k-5}}
\end{cases}\quad k=1\ldots5, \quad
M =\begin{cases}
m_{0} & \textrm{w.p. }1/2\\
1-m_{0} & \textrm{w.p. }1/2
\end{cases}
\end{align*}

\noindent in which $\varepsilon_{i}$ also follows an inverse
Gaussian distribution.

Table \ref{table:benchmark} shows the comparison among
our model, the ACD(1,1) model and the MSMD(5) model.
We use the BIC as an indicator to compare their in-sample fitness.
From the results, we find that the in-sample fitness
of our MF-RSD model is slightly worse than that of the ACD(1,1) model but better than that of the MSMD(5) model.
This is probably because the ACD model has the most parsimonious structure and the least number of parameters.
For the out-of-sample prediction power, we use the RMSE between 
the real duration and the predicted duration as a measurement. 
Nonetheless, from the results in Table \ref{table:benchmark},
our MF-RSD model performs significantly better than the other two models.

\begin{table}[!h]
\caption{The evaluation of in-sample fitness and out-of-sample forecast for
25 NASDAQ stocks on January 2, 2013, with a benchmark comparison to
the ACD and MSMD models. }
\medskip{}
\centerline{%
\begin{tabular}{c|c|ccc|ccc|ccc}
\hline 
 &  & \multicolumn{3}{c|}{Log-likelihood} & \multicolumn{3}{c|}{BIC} & \multicolumn{3}{c}{RMSE}\tabularnewline
\cline{3-11} 
Stock & $n$ & \multicolumn{1}{c|}{MF-RSD} & \multicolumn{1}{c|}{ACD} & MSMD & \multicolumn{1}{c|}{MF-RSD} & \multicolumn{1}{c|}{ACD} & MSMD & \multicolumn{1}{c|}{MF-RSD} & \multicolumn{1}{c|}{ACD} & MSMD\tabularnewline
\hline 
{\footnotesize{}AAPL} & {\footnotesize{}28888} & {\footnotesize{}35421.60} & {\footnotesize{}35936.97} & {\footnotesize{}30748.48} & {\footnotesize{}-70699.41} & {\footnotesize{}-71812.31} & {\footnotesize{}-61445.60} & {\footnotesize{}1.91} & {\footnotesize{}2.00} & {\footnotesize{}2.91}\tabularnewline
{\footnotesize{}ALXN} & {\footnotesize{}5240} & {\footnotesize{}-704.66} & {\footnotesize{}-318.01} & {\footnotesize{}-1365.19} & {\footnotesize{}1529.22} & {\footnotesize{}687.40} & {\footnotesize{}2773.20} & {\footnotesize{}8.21} & {\footnotesize{}9.38} & {\footnotesize{}9.54}\tabularnewline
{\footnotesize{}AMZN} & {\footnotesize{}7674} & {\footnotesize{}5901.71} & {\footnotesize{}5876.18} & {\footnotesize{}5024.88} & {\footnotesize{}-11678.18} & {\footnotesize{}-11698.69} & {\footnotesize{}-10005.03} & {\footnotesize{}6.95} & {\footnotesize{}7.14} & {\footnotesize{}8.37}\tabularnewline
{\footnotesize{}BIDU} & {\footnotesize{}3819} & {\footnotesize{}-4467.04} & {\footnotesize{}-4130.39} & {\footnotesize{}-4685.14} & {\footnotesize{}9049.55} & {\footnotesize{}8310.27} & {\footnotesize{}9411.52} & {\footnotesize{}12.09} & {\footnotesize{}13.39} & {\footnotesize{}18.05}\tabularnewline
{\footnotesize{}BMRN} & {\footnotesize{}2892} & {\footnotesize{}-2729.22} & {\footnotesize{}-2435.24} & {\footnotesize{}-3127.84} & {\footnotesize{}5570.02} & {\footnotesize{}4918.29} & {\footnotesize{}6295.53} & {\footnotesize{}15.75} & {\footnotesize{}18.96} & {\footnotesize{}21.25}\tabularnewline
{\footnotesize{}CELG} & {\footnotesize{}4717} & {\footnotesize{}-3832.10} & {\footnotesize{}-3420.28} & {\footnotesize{}-3498.79} & {\footnotesize{}7782.63} & {\footnotesize{}6891.31} & {\footnotesize{}7039.87} & {\footnotesize{}8.54} & {\footnotesize{}9.72} & {\footnotesize{}9.81}\tabularnewline
{\footnotesize{}CERN} & {\footnotesize{}2460} & {\footnotesize{}-2424.12} & {\footnotesize{}-2202.68} & {\footnotesize{}-1316.67} & {\footnotesize{}4957.55} & {\footnotesize{}4452.21} & {\footnotesize{}2672.38} & {\footnotesize{}15.50} & {\footnotesize{}19.07} & {\footnotesize{}29.18}\tabularnewline
{\footnotesize{}CMCSA} & {\footnotesize{}5235} & {\footnotesize{}-2031.46} & {\footnotesize{}-1933.45} & {\footnotesize{}-1778.63} & {\footnotesize{}4182.80} & {\footnotesize{}3918.28} & {\footnotesize{}3600.08} & {\footnotesize{}11.11} & {\footnotesize{}12.56} & {\footnotesize{}12.61}\tabularnewline
{\footnotesize{}COST} & {\footnotesize{}4330} & {\footnotesize{}-4394.69} & {\footnotesize{}-4125.27} & {\footnotesize{}-4907.93} & {\footnotesize{}8906.61} & {\footnotesize{}8300.78} & {\footnotesize{}9857.73} & {\footnotesize{}8.96} & {\footnotesize{}10.22} & {\footnotesize{}10.67}\tabularnewline
{\footnotesize{}DISCA} & {\footnotesize{}3055} & {\footnotesize{}-1062.11} & {\footnotesize{}-891.86} & {\footnotesize{}-919.68} & {\footnotesize{}2236.57} & {\footnotesize{}1831.87} & {\footnotesize{}1879.49} & {\footnotesize{}16.66} & {\footnotesize{}16.92} & {\footnotesize{}17.96}\tabularnewline
{\footnotesize{}EBAY} & {\footnotesize{}9697} & {\footnotesize{}1450.65} & {\footnotesize{}2098.67} & {\footnotesize{}-484.82} & {\footnotesize{}-2772.79} & {\footnotesize{}-4142.26} & {\footnotesize{}1015.53} & {\footnotesize{}5.78} & {\footnotesize{}5.59} & {\footnotesize{}6.04}\tabularnewline
{\footnotesize{}FB} & {\footnotesize{}10278} & {\footnotesize{}8222.19} & {\footnotesize{}8579.60} & {\footnotesize{}5865.68} & {\footnotesize{}-16315.06} & {\footnotesize{}-17103.77} & {\footnotesize{}-11685.17} & {\footnotesize{}5.13} & {\footnotesize{}5.46} & {\footnotesize{}7.06}\tabularnewline
{\footnotesize{}GOOG} & {\footnotesize{}6387} & {\footnotesize{}4725.22} & {\footnotesize{}4715.50} & {\footnotesize{}-6186.30} & {\footnotesize{}-9327.78} & {\footnotesize{}-9378.43} & {\footnotesize{}12416.41} & {\footnotesize{}8.13} & {\footnotesize{}8.44} & {\footnotesize{}8.59}\tabularnewline
{\footnotesize{}INTC} & {\footnotesize{}5300} & {\footnotesize{}2149.62} & {\footnotesize{}2144.01} & {\footnotesize{}10.61} & {\footnotesize{}-4179.19} & {\footnotesize{}-4236.57} & {\footnotesize{}21.66} & {\footnotesize{}11.27} & {\footnotesize{}12.20} & {\footnotesize{}12.11}\tabularnewline
{\footnotesize{}ISRG} & {\footnotesize{}1918} & {\footnotesize{}-1128.08} & {\footnotesize{}-1028.33} & {\footnotesize{}-715.34} & {\footnotesize{}2361.99} & {\footnotesize{}2094.46} & {\footnotesize{}1468.48} & {\footnotesize{}25.72} & {\footnotesize{}27.33} & {\footnotesize{}30.48}\tabularnewline
{\footnotesize{}KLAC} & {\footnotesize{}3059} & {\footnotesize{}-3076.43} & {\footnotesize{}-2749.82} & {\footnotesize{}-2536.58} & {\footnotesize{}6265.23} & {\footnotesize{}5547.80} & {\footnotesize{}5113.29} & {\footnotesize{}14.32} & {\footnotesize{}16.17} & {\footnotesize{}26.80}\tabularnewline
{\footnotesize{}MAR} & {\footnotesize{}2404} & {\footnotesize{}-3733.69} & {\footnotesize{}-3430.79} & {\footnotesize{}-3801.85} & {\footnotesize{}7576.38} & {\footnotesize{}6908.28} & {\footnotesize{}7642.62} & {\footnotesize{}21.91} & {\footnotesize{}22.42} & {\footnotesize{}26.41}\tabularnewline
{\footnotesize{}MSFT} & {\footnotesize{}6242} & {\footnotesize{}3472.11} & {\footnotesize{}3354.70} & {\footnotesize{}2887.86} & {\footnotesize{}-6821.88} & {\footnotesize{}-6656.96} & {\footnotesize{}-5732.03} & {\footnotesize{}8.19} & {\footnotesize{}9.57} & {\footnotesize{}9.91}\tabularnewline
{\footnotesize{}NFLX} & {\footnotesize{}3594} & {\footnotesize{}-2158.32} & {\footnotesize{}-1800.04} & {\footnotesize{}-4272.06} & {\footnotesize{}4431.27} & {\footnotesize{}3649.20} & {\footnotesize{}8585.06} & {\footnotesize{}10.98} & {\footnotesize{}12.89} & {\footnotesize{}13.33}\tabularnewline
{\footnotesize{}QCOM} & {\footnotesize{}8726} & {\footnotesize{}-2759.03} & {\footnotesize{}-2061.64} & {\footnotesize{}-6911.35} & {\footnotesize{}5645.10} & {\footnotesize{}4177.72} & {\footnotesize{}13868.07} & {\footnotesize{}5.35} & {\footnotesize{}5.85} & {\footnotesize{}6.04}\tabularnewline
{\footnotesize{}REGN} & {\footnotesize{}3841} & {\footnotesize{}860.04} & {\footnotesize{}999.03} & {\footnotesize{}808.91} & {\footnotesize{}-1604.53} & {\footnotesize{}-1948.53} & {\footnotesize{}-1576.55} & {\footnotesize{}10.89} & {\footnotesize{}12.56} & {\footnotesize{}12.88}\tabularnewline
{\footnotesize{}SBUX} & {\footnotesize{}5740} & {\footnotesize{}-3886.73} & {\footnotesize{}-3290.91} & {\footnotesize{}-4722.18} & {\footnotesize{}7894.64} & {\footnotesize{}6633.76} & {\footnotesize{}9487.64} & {\footnotesize{}7.10} & {\footnotesize{}7.76} & {\footnotesize{}8.64}\tabularnewline
{\footnotesize{}TXN} & {\footnotesize{}5633} & {\footnotesize{}-1794.99} & {\footnotesize{}-1412.53} & {\footnotesize{}-3939.81} & {\footnotesize{}3710.88} & {\footnotesize{}2876.88} & {\footnotesize{}7922.80} & {\footnotesize{}9.38} & {\footnotesize{}11.01} & {\footnotesize{}11.07}\tabularnewline
{\footnotesize{}VOD} & {\footnotesize{}1810} & {\footnotesize{}-2257.49} & {\footnotesize{}-2259.32} & {\footnotesize{}-2312.70} & {\footnotesize{}4619.99} & {\footnotesize{}4563.65} & {\footnotesize{}4662.91} & {\footnotesize{}42.13} & {\footnotesize{}44.50} & {\footnotesize{}45.61}\tabularnewline
{\footnotesize{}YHOO} & {\footnotesize{}3797} & {\footnotesize{}1223.24} & {\footnotesize{}1195.00} & {\footnotesize{}-611.21} & {\footnotesize{}-2331.08} & {\footnotesize{}-4109.70} & {\footnotesize{}1263.63} & {\footnotesize{}16.69} & {\footnotesize{}15.74} & {\footnotesize{}16.48}\tabularnewline
\hline 
\end{tabular}}
\label{table:benchmark}

\textit{\footnotesize{}Note:}{\footnotesize{} BIC=$k*\ln n-2*\ln\hat{L}$, where $\hat{L}$ is the likelihood. 
The lower value BIC is, the better the in-sample fitness. The RMSE is the square root of MSE, and the lower the value is, the better the out-of-sample performance.}{\footnotesize\par}
\end{table}

\subsection{Analysis of regimes}
\begin{figure}[!h]
\centerline{\includegraphics[scale=0.36]{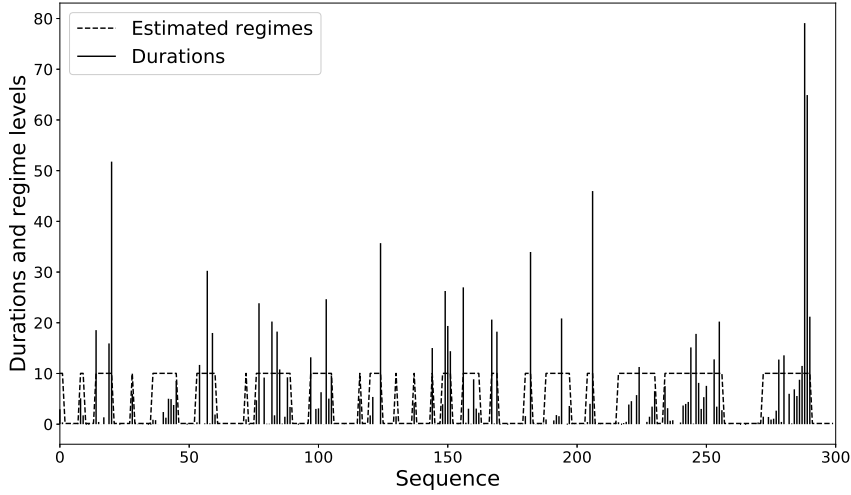}}
\caption{Estimated regime levels for the first 300 MSFT inter-trade durations on January 2, 2013. The estimated
regime level is the mean of the estimated underlying regimes.}
\label{fig:300durations}
\end{figure}

In the next, we want to see the characteristics of the underlying regimes 
that are estimated from our MF-RSD model.
In Figure \ref{fig:300durations}, we plot the first MSFT 300 inter-trade
durations on January 2, 2013, along with their estimated levels of the underlying regimes. 
It shows that the estimated regimes capture the clustering of the long (or short) inter-trade durations, 
and there are significant switchings between the long-duration regime and the short-duration regime.
Given the estimated underlying regimes, we will diagnose the two regimes and
check whether they are related to different channels of profitability for the HFTs, because we think that in
the long-duration regime, HFTs mainly provide liquidity and gain profit from the price spread as market makers, 
while in the short-duration regime, HFTs aggressively take liquidity and earn profit as speculators because of the short-term price movement.

To make things comparable, we divide the trading day into 360 one-minute intervals and calculate the ratio of 
the number of short inter-trade durations to the total number of trades in each interval, which is defined as the HF ratio. 
We categorize the short and long inter-trade durations based on which regime they belong to.
The LF ratio is then defined as the ratio of long inter-trade durations and hence equals to 1-HF ratio. 
Thus, a high HF (LF) ratio means that the short-duration (long-duration) regime is dominated in the interval.
Following \cite{carrion2013very}, we use the permanent price change and 
the realized price spread as the metrics of profit. 
The permanent price change is defined as the absolute percentage change of midpoint price 
in the one-minute interval, which measures the profit of taking liquidity, and the realized price spread 
is the average effective price spread minus the permanent price change, 
which measures the profit of supplying liquidity. 
We also include the order submission rate and the cancellation rate to see the market activity level.  
As the time length of the interval is fixed, we simply use the numbers of submissions and cancellations 
in one minute to represent the rates.

We still use MSFT on January 2, 2013 as an example. For each one-minute interval, 
we calculate the HF ratio, the LF ratio, the order submissions, the order cancellations, 
the permanent price change, and the realized price spread. 
In Table \ref{table:correlation}, we present their correlation matrix, from which 
we can observe that the HF ratio is positively correlated with the order submissions, cancellations, 
and the permanent price change, but negatively correlated with the realized price spread.
The LF ratio is having the opposite relationships with those 4 metrics. Thus, 
the profit of taking liquidity is large when the market in the short-duration regime in which the HFTs are using aggressive trading strategies. While in the period with a high LF ratio, i.e., when HFTs mainly supply liquidity, 
the profit of supplying liquidity is larger as the realized price spread is positively correlated with the LF ratio.

\begin{table}[!h]
\caption{Pearson correlation matrix for the HF ratio, the LF ratio, the order submissions, 
cancellations, the permanent price change and the realized price spread.}
\medskip{}
\centerline{
\begin{tabular}{lllllll}
\hline 
 & \textbf{\small{}HF ratio} & \textbf{\small{}LF ratio} & \textbf{\small{}Cancel} & \textbf{\small{}Submit} & \textbf{\small{}Price} & \textbf{\small{}Realized}\tabularnewline
 &  &  &  &  & \textbf{\small{}change} & \textbf{\small{}spread}\tabularnewline
\hline 
\textbf{\small{}HF ratio} & {\small{}1} &  &  &  &  & \tabularnewline
\textbf{\small{}LF ratio} & {\small{}-1} & {\small{}1} &  &  &  & \tabularnewline
\textbf{\small{}Cancel} & {\small{}0.472**} & {\small{}-0.472**} & {\small{}1} &  &  & \tabularnewline
\textbf{\small{}Submit} & {\small{}0.474**} & {\small{}-0.474**} & {\small{}0.998**} & {\small{}1} &  & \tabularnewline
\textbf{\small{}Price change} & {\small{}0.366**} & {\small{}-0.366**} & {\small{}0.457**} & {\small{}0.463**} & {\small{}1} & \tabularnewline
\textbf{\small{}Realized spread} & {\small{}-0.298**} & {\small{}0.298**} & {\small{}-0.330**} & {\small{}-0.337**} & {\small{}-0.982**} & {\small{}1}\tabularnewline
\hline 
\end{tabular}
}
\footnotesize{*p<0.05, **p<0.01}
\label{table:correlation}
\end{table}

Furthermore, we implement the below regression to see which regime or 
which regime-switching action is followed by a significant short-term price movement.
\begin{equation}
\Delta P_{i,t(+10s)}= constant +\mathbf{1}_{s_i=1} + \hat{P}_{12,i+1} + \hat{P}_{21,i+1} +\varepsilon_{it}
\label{eq: priceOnRegime}
\end{equation}
In the regression, $\Delta P_{i,t(+10s)}$ is the absolute value of the percentage change of midpoint price 
in the next 10 seconds after $i-$th trade.\footnote{As we mainly care about the HFTs' behaviors, 
we think 10 seconds is long enough at the high-frequency level.} 
$\mathbf{1}_{s_i=1}$ is the indicator which equals to 1 if the $i-$th trade belongs to the 
short-duration regime. $\hat{P}_{12,i+1}$ is the estimated transition probability for the next inter-trade duration 
switching from the short-duration regime to the long-duration regime, and $\hat{P}_{21,i+1}$ is the 
estimated regime transition probability for the next inter-trade duration on the contrary; both of them are calculated by equation (\ref{eq:estimatedP}) according to our MF-RSD model.

\begin{table}[!h]
\caption{The regressions of the short-term price change on the short-duration regime and regime-switching probabilities}
\medskip{}
\centerline{%
\begin{tabular}{c>{\centering}p{1.8cm}>{\centering}p{1.8cm}>{\centering}p{1.8cm}>{\centering}p{1.8cm}}
\hline 
 & {\small{}$\Delta P$} & \textbf{\small{}$\Delta P$} & \textbf{\small{}$\Delta P$} & \textbf{\small{}$\Delta P$}\tabularnewline
\hline 
\textbf{$\mathbf{1}_{s_{i}=1}$} & {\small{}.0032} &  &  & {\small{}.0037}\tabularnewline
 & {\small{}(.001){*}} &  &  & {\small{}(.001){*}{*}}\tabularnewline
\textbf{$\hat{P}_{12,i+1}$} &  & {\small{}-.0131} &  & {\small{}.0027}\tabularnewline
 &  & {\small{}(.005){*}} &  & {\small{}(.006)}\tabularnewline
\textbf{$\hat{P}_{21,i+1}$} &  &  & {\small{}.0124} & {\small{}.0146}\tabularnewline
 &  &  & {\small{}(.003){*}{*}} & {\small{}(.004){*}{*}}\tabularnewline
\textbf{$constant$} & {\small{}.0377} & {\small{}.0431} & {\small{}.0350} & {\small{}.0312}\tabularnewline
 & {\small{}(.001){*}{*}} & {\small{}(.001){*}{*}} & {\small{}(.001){*}{*}} & {\small{}(.003){*}{*}}\tabularnewline
\hline 
\end{tabular}
}
{\scriptsize{}Note: $\Delta P$ is the percentage change of mid-price in the next 10 seconds. 
The robust standard errors are shown in parentheses.
$*p<0.05$ and $**p<0.01$.}{\scriptsize\par}
\label{table:PriceOnRegimes}
\end{table}

The regression results are presented in Table \ref{table:PriceOnRegimes}. From it, we can see that the coefficient for $\mathbf{1}_{s_i=1}$ is positive. Hence the short-duration regime will be followed by a larger price change compared with the long-duration regime.  Nonetheless, from the last column of Table \ref{table:PriceOnRegimes}, we note that the effect for the transition probability from the long-duration regime to the short-duration regime is even stronger, as the coefficient of $\hat{P}_{21,i+1}$ is significantly greater than the coefficient of $\mathbf{1}_{s_i=1}$.
This means that a large price change follows the short-duration regime, while a stronger price movement will occur subsequently if the market is switching from the long-duration regime to the short-duration regime. We think this switching could be predicted by the HFTs, so they would take this opportunity to trade on the signal and to earn a profit as price speculators. However, the effect for $\hat{P}_{12,i+1}$ is not significant, so the price change is relatively small if the HFTs turn from the aggressive trading strategy to the passive one.

\subsection{Analysis of LOB factors} \label{SC: LOB factors}
At last, we have studied the informativeness of LOB factors to the trading activities 
by applying the MF-RSD model to the 25 NASDAQ stocks in our sample for 21 trading days (January 2013). 
For each stock and each factor,
we count the number of days in which the factor coefficient in the regression \eqref{model.4lob} is significant.
The results are presented in Table \ref{table:SigSummary} and the main findings are as follows.

\begin{table}[!h]
\caption{Summary of the results for the significance of 4 LOB factors in the MF-RSD model. }
\medskip{}

\centerline{%
\begin{tabular}{c|cccc|cccc|cccc|cccc}
\hline 
\multirow{3}{*}{{\footnotesize{}Stock}} & \multicolumn{4}{c|}{{\footnotesize{}$DI$}} & \multicolumn{4}{c|}{{\footnotesize{}$PS$}} & \multicolumn{4}{c|}{{\footnotesize{}$TV$}} & \multicolumn{4}{c}{{\footnotesize{}$PM$}}\tabularnewline
\cline{2-17} 
 & \multicolumn{2}{c|}{{\scriptsize{}$\mathrm{P_{12}}$}} & \multicolumn{2}{c|}{{\scriptsize{}$\mathrm{P_{21}}$}} & \multicolumn{2}{c|}{{\scriptsize{}$\mathrm{P_{12}}$}} & \multicolumn{2}{c|}{{\scriptsize{}$\mathrm{P_{21}}$}} & \multicolumn{2}{c|}{{\scriptsize{}$\mathrm{P_{12}}$}} & \multicolumn{2}{c|}{{\scriptsize{}$\mathrm{P_{21}}$}} & \multicolumn{2}{c|}{{\scriptsize{}$\mathrm{P_{12}}$}} & \multicolumn{2}{c}{{\scriptsize{}$\mathrm{P_{21}}$}}\tabularnewline
\cline{2-17} 
 & {\scriptsize{}sig(+)} & {\scriptsize{}sig(-)} & {\scriptsize{}sig(+)} & {\scriptsize{}sig(-)} & {\scriptsize{}sig(+)} & {\scriptsize{}sig(-)} & {\scriptsize{}sig(+)} & {\scriptsize{}sig(-)} & {\scriptsize{}sig(+)} & {\scriptsize{}sig(-)} & {\scriptsize{}sig(+)} & {\scriptsize{}sig(-)} & {\scriptsize{}sig(+)} & {\scriptsize{}sig(-)} & {\scriptsize{}sig(+)} & {\scriptsize{}sig(-)}\tabularnewline
\hline 
{\footnotesize{}AAPL} & {\scriptsize{}0} & {\scriptsize{}13} & {\scriptsize{}1} & {\scriptsize{}2} & {\scriptsize{}21} & {\scriptsize{}0} & {\scriptsize{}0} & {\scriptsize{}21} & {\scriptsize{}0} & {\scriptsize{}11} & {\scriptsize{}21} & {\scriptsize{}0} & {\scriptsize{}21} & {\scriptsize{}0} & {\scriptsize{}9} & {\scriptsize{}0}\tabularnewline
{\footnotesize{}ALXN} & {\scriptsize{}3} & {\scriptsize{}1} & {\scriptsize{}1} & {\scriptsize{}4} & {\scriptsize{}7} & {\scriptsize{}0} & {\scriptsize{}0} & {\scriptsize{}19} & {\scriptsize{}0} & {\scriptsize{}10} & {\scriptsize{}9} & {\scriptsize{}1} & {\scriptsize{}20} & {\scriptsize{}0} & {\scriptsize{}0} & {\scriptsize{}21}\tabularnewline
{\footnotesize{}AMZN} & {\scriptsize{}1} & {\scriptsize{}2} & {\scriptsize{}4} & {\scriptsize{}0} & {\scriptsize{}17} & {\scriptsize{}0} & {\scriptsize{}0} & {\scriptsize{}21} & {\scriptsize{}0} & {\scriptsize{}9} & {\scriptsize{}15} & {\scriptsize{}0} & {\scriptsize{}21} & {\scriptsize{}0} & {\scriptsize{}0} & {\scriptsize{}21}\tabularnewline
{\footnotesize{}BIDU} & {\scriptsize{}1} & {\scriptsize{}0} & {\scriptsize{}1} & {\scriptsize{}0} & {\scriptsize{}13} & {\scriptsize{}0} & {\scriptsize{}0} & {\scriptsize{}19} & {\scriptsize{}0} & {\scriptsize{}4} & {\scriptsize{}16} & {\scriptsize{}0} & {\scriptsize{}21} & {\scriptsize{}0} & {\scriptsize{}0} & {\scriptsize{}21}\tabularnewline
{\footnotesize{}BMRN} & {\scriptsize{}0} & {\scriptsize{}0} & {\scriptsize{}3} & {\scriptsize{}0} & {\scriptsize{}9} & {\scriptsize{}1} & {\scriptsize{}0} & {\scriptsize{}21} & {\scriptsize{}0} & {\scriptsize{}6} & {\scriptsize{}11} & {\scriptsize{}0} & {\scriptsize{}15} & {\scriptsize{}0} & {\scriptsize{}0} & {\scriptsize{}12}\tabularnewline
{\footnotesize{}CELG} & {\scriptsize{}0} & {\scriptsize{}3} & {\scriptsize{}6} & {\scriptsize{}0} & {\scriptsize{}4} & {\scriptsize{}2} & {\scriptsize{}0} & {\scriptsize{}14} & {\scriptsize{}0} & {\scriptsize{}13} & {\scriptsize{}20} & {\scriptsize{}0} & {\scriptsize{}21} & {\scriptsize{}0} & {\scriptsize{}0} & {\scriptsize{}21}\tabularnewline
{\footnotesize{}CERN} & {\scriptsize{}1} & {\scriptsize{}0} & {\scriptsize{}1} & {\scriptsize{}1} & {\scriptsize{}7} & {\scriptsize{}0} & {\scriptsize{}0} & {\scriptsize{}18} & {\scriptsize{}0} & {\scriptsize{}4} & {\scriptsize{}9} & {\scriptsize{}0} & {\scriptsize{}21} & {\scriptsize{}0} & {\scriptsize{}0} & {\scriptsize{}20}\tabularnewline
{\footnotesize{}CMCSA} & {\scriptsize{}0} & {\scriptsize{}17} & {\scriptsize{}19} & {\scriptsize{}0} & {\scriptsize{}1} & {\scriptsize{}12} & {\scriptsize{}0} & {\scriptsize{}7} & {\scriptsize{}0} & {\scriptsize{}16} & {\scriptsize{}21} & {\scriptsize{}0} & {\scriptsize{}11} & {\scriptsize{}0} & {\scriptsize{}6} & {\scriptsize{}0}\tabularnewline
{\footnotesize{}COST} & {\scriptsize{}3} & {\scriptsize{}2} & {\scriptsize{}1} & {\scriptsize{}0} & {\scriptsize{}4} & {\scriptsize{}4} & {\scriptsize{}0} & {\scriptsize{}15} & {\scriptsize{}0} & {\scriptsize{}12} & {\scriptsize{}20} & {\scriptsize{}0} & {\scriptsize{}21} & {\scriptsize{}0} & {\scriptsize{}0} & {\scriptsize{}21}\tabularnewline
{\footnotesize{}DISCA} & {\scriptsize{}0} & {\scriptsize{}2} & {\scriptsize{}1} & {\scriptsize{}0} & {\scriptsize{}2} & {\scriptsize{}1} & {\scriptsize{}0} & {\scriptsize{}17} & {\scriptsize{}0} & {\scriptsize{}8} & {\scriptsize{}17} & {\scriptsize{}0} & {\scriptsize{}19} & {\scriptsize{}0} & {\scriptsize{}0} & {\scriptsize{}19}\tabularnewline
{\footnotesize{}EBAY} & {\scriptsize{}1} & {\scriptsize{}3} & {\scriptsize{}0} & {\scriptsize{}6} & {\scriptsize{}10} & {\scriptsize{}0} & {\scriptsize{}0} & {\scriptsize{}21} & {\scriptsize{}0} & {\scriptsize{}18} & {\scriptsize{}20} & {\scriptsize{}0} & {\scriptsize{}20} & {\scriptsize{}0} & {\scriptsize{}0} & {\scriptsize{}3}\tabularnewline
{\footnotesize{}FB} & {\scriptsize{}0} & {\scriptsize{}12} & {\scriptsize{}2} & {\scriptsize{}2} & {\scriptsize{}14} & {\scriptsize{}3} & {\scriptsize{}0} & {\scriptsize{}21} & {\scriptsize{}0} & {\scriptsize{}16} & {\scriptsize{}21} & {\scriptsize{}0} & {\scriptsize{}20} & {\scriptsize{}0} & {\scriptsize{}9} & {\scriptsize{}0}\tabularnewline
{\footnotesize{}GOOG} & {\scriptsize{}1} & {\scriptsize{}4} & {\scriptsize{}2} & {\scriptsize{}1} & {\scriptsize{}21} & {\scriptsize{}0} & {\scriptsize{}0} & {\scriptsize{}21} & {\scriptsize{}0} & {\scriptsize{}12} & {\scriptsize{}20} & {\scriptsize{}0} & {\scriptsize{}21} & {\scriptsize{}0} & {\scriptsize{}0} & {\scriptsize{}18}\tabularnewline
{\footnotesize{}INTC} & {\scriptsize{}0} & {\scriptsize{}20} & {\scriptsize{}18} & {\scriptsize{}0} & {\scriptsize{}0} & {\scriptsize{}15} & {\scriptsize{}1} & {\scriptsize{}3} & {\scriptsize{}0} & {\scriptsize{}14} & {\scriptsize{}20} & {\scriptsize{}0} & {\scriptsize{}2} & {\scriptsize{}5} & {\scriptsize{}5} & {\scriptsize{}0}\tabularnewline
{\footnotesize{}ISRG} & {\scriptsize{}1} & {\scriptsize{}1} & {\scriptsize{}0} & {\scriptsize{}2} & {\scriptsize{}10} & {\scriptsize{}3} & {\scriptsize{}0} & {\scriptsize{}20} & {\scriptsize{}5} & {\scriptsize{}0} & {\scriptsize{}11} & {\scriptsize{}0} & {\scriptsize{}18} & {\scriptsize{}0} & {\scriptsize{}1} & {\scriptsize{}12}\tabularnewline
{\footnotesize{}KLAC} & {\scriptsize{}0} & {\scriptsize{}6} & {\scriptsize{}2} & {\scriptsize{}1} & {\scriptsize{}5} & {\scriptsize{}3} & {\scriptsize{}0} & {\scriptsize{}17} & {\scriptsize{}0} & {\scriptsize{}11} & {\scriptsize{}15} & {\scriptsize{}0} & {\scriptsize{}12} & {\scriptsize{}0} & {\scriptsize{}0} & {\scriptsize{}11}\tabularnewline
{\footnotesize{}MAR} & {\scriptsize{}0} & {\scriptsize{}10} & {\scriptsize{}4} & {\scriptsize{}0} & {\scriptsize{}1} & {\scriptsize{}4} & {\scriptsize{}0} & {\scriptsize{}11} & {\scriptsize{}0} & {\scriptsize{}9} & {\scriptsize{}16} & {\scriptsize{}0} & {\scriptsize{}10} & {\scriptsize{}0} & {\scriptsize{}2} & {\scriptsize{}4}\tabularnewline
{\footnotesize{}MSFT} & {\scriptsize{}0} & {\scriptsize{}20} & {\scriptsize{}20} & {\scriptsize{}0} & {\scriptsize{}1} & {\scriptsize{}17} & {\scriptsize{}1} & {\scriptsize{}2} & {\scriptsize{}0} & {\scriptsize{}13} & {\scriptsize{}20} & {\scriptsize{}0} & {\scriptsize{}1} & {\scriptsize{}6} & {\scriptsize{}5} & {\scriptsize{}0}\tabularnewline
{\footnotesize{}NFLX} & {\scriptsize{}0} & {\scriptsize{}6} & {\scriptsize{}7} & {\scriptsize{}0} & {\scriptsize{}13} & {\scriptsize{}1} & {\scriptsize{}0} & {\scriptsize{}21} & {\scriptsize{}0} & {\scriptsize{}12} & {\scriptsize{}19} & {\scriptsize{}0} & {\scriptsize{}21} & {\scriptsize{}0} & {\scriptsize{}0} & {\scriptsize{}18}\tabularnewline
{\footnotesize{}QCOM} & {\scriptsize{}0} & {\scriptsize{}11} & {\scriptsize{}4} & {\scriptsize{}0} & {\scriptsize{}5} & {\scriptsize{}1} & {\scriptsize{}0} & {\scriptsize{}21} & {\scriptsize{}0} & {\scriptsize{}20} & {\scriptsize{}21} & {\scriptsize{}0} & {\scriptsize{}15} & {\scriptsize{}0} & {\scriptsize{}4} & {\scriptsize{}6}\tabularnewline
{\footnotesize{}REGN} & {\scriptsize{}0} & {\scriptsize{}12} & {\scriptsize{}9} & {\scriptsize{}0} & {\scriptsize{}17} & {\scriptsize{}1} & {\scriptsize{}0} & {\scriptsize{}21} & {\scriptsize{}1} & {\scriptsize{}0} & {\scriptsize{}9} & {\scriptsize{}0} & {\scriptsize{}21} & {\scriptsize{}0} & {\scriptsize{}0} & {\scriptsize{}19}\tabularnewline
{\footnotesize{}SBUX} & {\scriptsize{}0} & {\scriptsize{}14} & {\scriptsize{}4} & {\scriptsize{}0} & {\scriptsize{}2} & {\scriptsize{}5} & {\scriptsize{}0} & {\scriptsize{}15} & {\scriptsize{}0} & {\scriptsize{}17} & {\scriptsize{}21} & {\scriptsize{}0} & {\scriptsize{}11} & {\scriptsize{}0} & {\scriptsize{}0} & {\scriptsize{}3}\tabularnewline
{\footnotesize{}TXN} & {\scriptsize{}0} & {\scriptsize{}13} & {\scriptsize{}8} & {\scriptsize{}0} & {\scriptsize{}1} & {\scriptsize{}3} & {\scriptsize{}0} & {\scriptsize{}8} & {\scriptsize{}0} & {\scriptsize{}11} & {\scriptsize{}21} & {\scriptsize{}0} & {\scriptsize{}9} & {\scriptsize{}0} & {\scriptsize{}2} & {\scriptsize{}1}\tabularnewline
{\footnotesize{}VOD} & {\scriptsize{}0} & {\scriptsize{}15} & {\scriptsize{}17} & {\scriptsize{}0} & {\scriptsize{}3} & {\scriptsize{}2} & {\scriptsize{}0} & {\scriptsize{}0} & {\scriptsize{}1} & {\scriptsize{}1} & {\scriptsize{}19} & {\scriptsize{}0} & {\scriptsize{}9} & {\scriptsize{}0} & {\scriptsize{}1} & {\scriptsize{}0}\tabularnewline
{\footnotesize{}YHOO} & {\scriptsize{}0} & {\scriptsize{}21} & {\scriptsize{}19} & {\scriptsize{}0} & {\scriptsize{}1} & {\scriptsize{}9} & {\scriptsize{}3} & {\scriptsize{}2} & {\scriptsize{}0} & {\scriptsize{}12} & {\scriptsize{}21} & {\scriptsize{}0} & {\scriptsize{}2} & {\scriptsize{}5} & {\scriptsize{}6} & {\scriptsize{}0}\tabularnewline
\hline 
{\footnotesize{}Sum (\%)} & {\scriptsize{}2.3\%} & {\scriptsize{}39.6\%} & {\scriptsize{}29.3\%} & {\scriptsize{}3.6\%} & {\scriptsize{}36.0\%} & {\scriptsize{}16.6\%} & {\scriptsize{}1.0\%} & {\scriptsize{}71.6\%} & {\scriptsize{}1.3\%} & {\scriptsize{}49.3\%} & {\scriptsize{}82.5\%} & {\scriptsize{}0.2\%} & {\scriptsize{}73.0\%} & {\scriptsize{}3.0\%} & {\scriptsize{}9.5\%} & {\scriptsize{}47.8\%}\tabularnewline
\hline 
\end{tabular}}
\label{table:SigSummary}

\textit{\scriptsize{}Note}{\scriptsize{}: $P_{12}$ represents
the regime switching from the short-duration regime to the long-duration
regime, and $P_{21}$ represents the regime switching from
the long-duration regime to the short-duration regime. Significance is measured at the 5\%
level. Sig(+)/Sig(-) means that the estimated coefficient for
the factor is significantly positive/negative. Each cell shows the counts of significance. 
We have 21 trading days for each stock, so the upper limit for each cell is 21. 
The last row records the percentage (\%) of the times that this factor appears
as significant in all 525 instances (25 stocks times 21 days). }{\scriptsize\par}
\end{table}

\begin{enumerate}
\item The effect of $DI$: in approximately 40\% of total instances, it is significantly negative for $P_{12}$,
and in approximately 30\% of total instances, it is significantly positive for $P_{21}$.
Hence, the effect of depth imbalance between the best ask and the best bid on the regime-switching probabilities
is not very common. Nevertheless, for the stocks which have very tight 
price spread, $DI$ has a common effect. 
For instance, we have summarized the results for the 
stocks whose price spread mostly keep at 0.1, including CMCSA, FB, INTC, MSFT, TXN, VOD, and YHOO,
and find that $DI$ is significantly negative for $P_{12}$ in 80\% of total occurrences
and significant positive for $P_{21}$ in 70\% of total occurrences.
Thus, we think that the impact of $DI$ on the dynamics of inter-trade durations 
is price spread dependent. When price spread is tight, the effect of $DI$ is significant, 
i.e., the greater the depth imbalance is, the inter-trade durations are more likely to stay in (or switch to) the short-duration regime.
\item The effect of $PS$: it is not obvious for $P_{12}$, but it is significantly negative for $P_{21}$ in more than 70\% of total instances.
Thus, the effect of the price spread between the best ask and the best bid
on the inter-trade durations switching from the short-duration regime to the long-duration regime is ambiguous,
while it is commonly significantly negative for the switching from the long-duration regime to the short-duration regime.
This means that the larger the price spread is, it is more likely that the inter-trade durations that were previously in the long-duration
regime keep staying in the long-duration regime.
\item The effect of $TV$: it is significantly negative for $P_{12}$ in nearly 50\% of total instances
and is significantly positive for $P_{21}$ in 82\% of total instances.
Thus, the larger quantity the last trade has, the more likely that the short inter-trade
durations stay in the preceding short-duration regime.
Also, a larger quantity of trade is more likely to lead the inter-trade durations
that were previously in the long-duration regime to switch to the short-duration regime.
\item The effect of $PM$: it is significantly positive for $P_{12}$ in 73\% of total instances
and is significantly negative for $P_{21}$ in nearly 50\% of total instances.
Hence, whether the mid-price has changed in the previous trade has
an important effect on the regime-switching probabilities. It is commonly
significantly positive for the inter-trade durations switching from the short-duration regime to the long-duration regime,
which means that if mid-price has moved,
inter-trade durations are very likely to switch to the long-duration
regime. Also in many cases, it is significantly negative for the inter-trade durations switching 
from the long-duration regime to the short-duration regime, which means that if the mid-price moves, inter-trade durations
are inclined to stay in the preceding long-duration regime.
\end{enumerate}

Some of our findings are in harmony with those of the existing research on market microstructure.
First, many papers \citep{LI2005533,goldstein2017high,van2019high}
suggest that order imbalance is a good predictor
for price movement, and algorithm traders would use
it as a trading indicator. A larger imbalance is a signal for future price movement
and hence encourages traders (in particular HFTs) to actively trade to gain profit (or prevent loss).
This is why, in some cases, the effect of depth imbalance is significantly negative for inter-trade durations switching
from the short-duration regime to the long-duration regime and significantly positive for the switching
from the long-duration regime to the short-duration regime.
Second, according to \citep{ranaldo2004order,carrion2013very,hendershott2013algorithmic}, price spread is an
indicator of liquidity and measures the profit (cost) of providing (taking) liquidity.
A large price spread indicates illiquidity of the asset, and the larger the price spread is, 
the more transaction cost liquidity takers face and meanwhile the more profit liquidity providers have.
Thus, when the price spread becomes larger, people are more reluctant to
initialize market orders and the inter-trade durations are more likely to be long. 
Third, trade volume is indeed a momentum variable
that reflects trading willingness \citep{o1995market,hasbrouck1991measuring}. A large trade volume would not only
keep the market in an active trading period (the short-duration regime)
but also lead the market to switch from a passive trading period
to an active one. Finally, the effect of price movement seems to violate the negative correlation between price movement
and duration length \citep{manganelli2005duration,furfine2007inter}.
However, in our high-frequency data, perhaps the change in mid-price in the last
trade is not a momentum variable for future price movement. Moreover, we find that 
the change in mid-price just occurs because the existing limit orders at best bid
or best ask are eliminated by the incoming trade, and consequently 
the price spread between the best ask and the best bid increases.
Thus, we suspect that its effect on the regime switchings of inter-trade durations is probably resulted from 
another channel, i.e., an increase in the price spread.

In addition, we obtain some new findings. First, the depth imbalance plays an important role mostly when the price spread is
small. A possible explanation is that when the price spread is large, the depth imbalance is less informative as it
is relatively costless for traders to submit orders at the best ask/bid.
Moreover, within the revealed price spread, perhaps there are some
hidden limit orders that make the true depth imbalance between the best ask and the best bid unknowable.
Only when the price spread is tight, the depth imbalance reveals more information and becomes a valid
indicator for the subsequent price movement. Next, a large price spread
is more likely to keep inter-trade durations in the long-duration regime
rather than to reverse inter-trade durations from the short-duration regime to
the long-duration regime, as we have observed that the effect of price spread is commonly negative for $P_{21}$ 
but is not obvious for $P_{12}$. This finding suggests that traders are more sensitive to the size of $PS$ 
when the market is in the long-duration regime, in which HFTs mainly provide liquidity and slow traders consume liquidity.
The slow traders seem to be more concerned with a high transaction cost induced by a large $PS$ when the market is relatively stable.
However, when the market enters the short-duration regime in which HFTs aggressively take liquidity and earn a profit because of 
the price movement, a large price spread would not impede the HFTs' trading aggressiveness 
and play a significant role in the arrival time of trades.
Finally, as the effect of $PM$ is commonly positive for $P_{12}$ and 
$PM=1$ is accompanied by an increase of price spread, whether the price spread has just increased 
could be a driving force that leads inter-trade durations to switch to the long-duration regime.
This finding may affirm the viewpoint in \cite{obizhaeva2013optimal} that the optimal strategy for algorithm
traders depends more on the resilience properties of supply/demand such as the change in bid-ask spread, rather than its static property.

To further verify the mechanism of LOB factors on the dynamics of inter-trade durations and our explanation, 
we implement the below `naive' panel data regression 
to see the relationship between the LOB factors and the price change in the next 10 seconds. 
The regression model is like following:
\begin{equation}
\Delta P_{i,t(+10s)}=constant+DI_{i,t}+PS_{i,t}+TV_{it}+PM_{it}+\mu_{i}+\varepsilon_{it}
\label{eq:PriceOnLOB}
\end{equation}
where $i$ and $t$ indicate the $i-$th stock and the $t-$th trade respectively, 
and $\Delta P$ is the short-term price change, the same as in the equation \ref{eq: priceOnRegime}. 
Here we just use the data of 25 stocks on the same day, i.e., January 2, 2013. 
The panel data is unbalanced because different stocks have different numbers of trades. 

\begin{table}[!h]
\caption{The regressions of the short-term price change on the LOB factors}
\medskip{}
\centerline{%
\begin{tabular}{c>{\centering}p{1.9cm}>{\centering}p{1.9cm}>{\centering}m{2.1cm}>{\centering}p{2.1cm}}
\hline 
 & \multicolumn{2}{c}{\textbf{\footnotesize{}Full Sample}} & \textbf{\footnotesize{}Subsample } & \textbf{\footnotesize{}Subsample }\tabularnewline
 &  &  & \textbf{\footnotesize{}(tight spread)} & \textbf{\footnotesize{}(slack spread)}\tabularnewline
\cline{2-5} 
 & \textbf{\footnotesize{}$\Delta P$} & \textbf{\footnotesize{}$\Delta P$} & \textbf{\footnotesize{}$\Delta P$} & \textbf{\footnotesize{}$\Delta P$}\tabularnewline
\hline 
\textbf{\footnotesize{}DI} & {\footnotesize{}.00192} & {\footnotesize{}.00208} & {\footnotesize{}.00672{*}{*}} & {\footnotesize{}-.000325}\tabularnewline
 & {\footnotesize{}(.00119)} & {\footnotesize{}(.00113)} & {\footnotesize{}(.00217)} & {\footnotesize{}(.000597)}\tabularnewline
\textbf{\footnotesize{}PS} & {\footnotesize{}.00436{*}{*}} &  &  & \tabularnewline
 & {\footnotesize{}(.00100)} &  &  & \tabularnewline
\textbf{\footnotesize{}TV} & {\footnotesize{}.00188{*}{*}} & {\footnotesize{}.00189{*}{*}} & {\footnotesize{}.00186{*}{*}} & {\footnotesize{}.0140{*}}\tabularnewline
 & {\footnotesize{}(.000199)} & {\footnotesize{}(.000201)} & {\footnotesize{}(.000192)} & {\footnotesize{}(.005034)}\tabularnewline
\textbf{\footnotesize{}PM} & {\footnotesize{}-.00192} & {\footnotesize{}-.000890} & {\footnotesize{}.00215{*}} & {\footnotesize{}-.00139{*}}\tabularnewline
 & {\footnotesize{}(.000626)} & {\footnotesize{}(.000481)} & {\footnotesize{}(.000782)} & {\footnotesize{}(.000603)}\tabularnewline
\textbf{\footnotesize{}constant} & {\footnotesize{}.0235{*}{*}} & {\footnotesize{}.0268{*}{*}} & {\footnotesize{}.02242{*}{*}} & {\footnotesize{}.0285{*}{*}}\tabularnewline
 & {\footnotesize{}(.00102)} & {\footnotesize{}(.000542)} & {\footnotesize{}(.00112)} & {\footnotesize{}(.000290)}\tabularnewline
\textbf{\footnotesize{}fixed effect} & {\footnotesize{}Y} & {\footnotesize{}Y} & {\footnotesize{}Y} & {\footnotesize{}Y}\tabularnewline
\textbf{\footnotesize{}number of stocks} & {\footnotesize{}25} & {\footnotesize{}25} & {\footnotesize{}12} & {\footnotesize{}13}\tabularnewline
\textbf{\footnotesize{}observations} & {\footnotesize{}146,736} & {\footnotesize{}146,736} & {\footnotesize{}67,921} & {\footnotesize{}78,815}\tabularnewline
\hline 
\end{tabular}
}
{\scriptsize{}Note: Panel data regression with fixed effect. $\Delta P$
is the percentage change of mid-price in the next 10 seconds. The robust standard errors are shown in parentheses.
$*p<0.05$ and $**p<0.01$.}{\scriptsize\par}
\label{table:LOBeffects}
\end{table}

The regression results are presented in Table \ref{table:LOBeffects}. As we can see, for the full sample, 
the LOB factor $DI$ and $PM$ don't show a significant effect on the price change. 
This finding advocates that the depth imbalance is overall not informative 
and the price movement of last trade is not a momentum variable 
for the future price movement. Nevertheless, a large $TV$ predicts a large price change 
because the $TV$ is significantly positive for the $\Delta P$, 
which is in line with our finding that a large trade volume will induce short inter-trade durations 
as HTFs treat it as a signal for future price movement. The coefficient of $PS$ is significant and positive for $\Delta P$, 
however, we think the $PS$ could be endogenously correlated with the price change as the effective price spread ($PS$) 
consists of the permanent price change ($\Delta P$) and the realized price spread, which compensate liquidity providers 
costs on two aspects,  i.e., the adverse select cost and the `real friction' \citep{carrion2013very}.  
Thus, we exclude $PS$ in the
second regression and the result shows that the other three LOB factors have similar effects.
We further divided the 25 stocks into two subgroups to verify the effect of $DI$. 
One subgroup has stocks with relatively tight price spreads,\footnote{There are 12 stocks in total.
75\% percentile of the $PS$ for each stock is less than or equal to 0.2.} 
the other subgroup has the remaining stocks, whose price spreads are relatively slack. 
From the last two columns in Table \ref{table:LOBeffects}, we can observe
that the $DI$ in the first subgroup is significantly positive, while it is insignificant in the second subgroup. 
The finding strongly supports the conclusion that the depth imbalance is informative and a valid indicator for HFTs 
only when the price spread is small, and a large depth imbalance
in such case is very likely to indicate the following price movement.

Moreover, we have provided robustness tests for the findings on the effects of the LOB factors.
The tests include the analysis of NASDAQ stocks in a different month, the analysis of directional trades,
the analysis of $DI$ conditional on the tightness of $PS$, the analysis with a substitution of $DI$
by the aggregated depth imbalance, and the analysis with a substitution 
of $PM$ by a dummy variable for the increase in price spread. These 
robustness test are presented in Section \ref{SC: robustness} the Appendices, 
and the results are very consistent with our findings.

\section{Conclusion}
In this paper, we have proposed a multifactor regime-switching duration (MF-RSD)
model for the inter-trade durations in the high-frequency limit order market.
The motivation is that we have found a common bimodal distribution 
of inter-trade durations for the stocks listed in the NASDAQ market.
The simulation study of MF-RSD model not only validates our estimation method
but also shows that the model can replicate the most empirical facts of inter-trade durations,
including the newly found bimodal distribution. Furthermore, based on the empirical analysis, 
we find that the underlying regimes are related to 
the endogenous liquidity provision and consumption by the HFTs in the market.
Using the factor analysis, the MF-RSD model is also capable to
identify the types of market information that algorithm traders
learn from and react to in high frequency, which causes the inter-trade durations
switching between the short-duration and the long-duration regime.

The contribution of the MF-RSD model in analyzing the high-frequency inter-trade duration data in the LOB 
can be summarized in the following three aspects.
First, the out-of-sample test shows that the performance of the MF-RSD model in predicting inter-trade durations
is significantly improved compared with the two benchmark duration models.
This is because, through the identified impacts of the LOB factors and the parsimonious structure of the two regimes,
our model is able to predict whether the arrival time of the next trade is relatively long or short.
Second, given the estimated regimes and the estimated regime-switching probabilities by the MF-RSD model,
we find empirical evidence that the switchings between the short-duration regime and the long-duration regime
result from the HFTs' behavior of changing strategies between providing liquidity and taking liquidity, 
driven by the different channels of profitability. 
Specifically, the short-duration regime is followed by a relatively large price movement, 
which corresponds to a high profit of taking liquidity, while the long-duration regime is positively correlated 
with the realized price spread, which measures the profit of providing liquidity. 
Third, the MF-RSD model is also useful in testing economic hypotheses in the market microstructure theory, and our results
are consistent with some findings in the prevalent literature. They are as follows: 1) order book
imbalance is applicable to predict the price movement in the short term, and in some cases,
a larger depth imbalance would lead to a shorter inter-trade duration;
2) a larger price spread tends to maintain inter-trade duration in the
long-duration regime because it reflects higher transaction costs for liquidity demanders; and
3) a larger trade volume is a good momentum variable that reveals a strong trading willingness, which normally
induces a shorter duration for the next trade. Moreover, we have some new findings added to the literature.
They are: 1) the impact of depth imbalance on the regime-switching of inter-trade duration is
spread-dependent and is significant when the stock's price spread is tight;
2) a large price spread is not a reverse signal for inter-trade durations switching from
the short-duration regime to the long-duration regime if they were staying in the short-duration regime; and
3) whether the price spread has just increased in the last trade is nonetheless a reverse signal
for the regime switching of inter-trade durations, through our analysis of the impact of the
mid-price movement.

Our work has made progress in analyzing the high-frequency duration data.
Nevertheless, it leaves a great deal to future research. For example, 
our LOB data is rich in terms of having records of order submissions and cancellations, 
we can further investigate the impact of those activities on the arrival time of trades. 
To achieve that, a new econometric model is probably needed.
Moreover, some event studies could be implemented to directly identify the HFTs' behaviors of switching
trading strategies if we can obtain the data that contain exogenous shocks to the market, which helps to
verify our findings on the regime-switching of inter-trade durations.
We can also test whether inter-trade durations in other high-frequency financial markets have similar properties,
in particular, whether the bimodal distribution is a general phenomenon, and a theoretical market microstructure 
model is certainly needed if so. 

\medskip{}

\section*{References}

\bibliography{MF-RSD_for_inter-trade_durations}

\pagebreak

\begin{appendices}

\section{Description of 25 Nasdaq stocks}

\begin{table}[!h]
\caption{A brief description of 25 selected NASDAQ stocks. It includes their
ticker symbols, company names, industry sectors, and market
capitalization, which are measured in 2013.}

\medskip{}

\centerline{%
\begin{tabular}{cccc}
\hline
\textbf{\footnotesize{}Ticker} & \textbf{\footnotesize{}Company name} & \textbf{\footnotesize{}Industry Sector} & \multicolumn{1}{>{\centering}p{2.5cm}}{\textbf{\footnotesize{}Market Cap. }{\scriptsize{}(billion dollars)}}\tabularnewline
\hline 
{\footnotesize{}AAPL} & {\footnotesize{}Apple Inc.} & {\footnotesize{}Technology} & {\footnotesize{}351.75}\tabularnewline
{\footnotesize{}ALXN} & {\footnotesize{}Alexion Pharmaceuticals} & {\footnotesize{}Healthcare} & {\footnotesize{}19.77}\tabularnewline
{\footnotesize{}AMZN} & {\footnotesize{}Amazon.com, Inc. } & {\footnotesize{}Consumer Cyclical} & {\footnotesize{}116.5}\tabularnewline
{\footnotesize{}BIDU} & {\footnotesize{}Baidu, Inc.} & {\footnotesize{}Technology} & {\footnotesize{}30.22}\tabularnewline
{\footnotesize{}BMRN} & {\footnotesize{}BioMarin Pharmaceutical, Inc.} & {\footnotesize{}Healthcare} & {\footnotesize{}8.62}\tabularnewline
{\footnotesize{}CELG} & {\footnotesize{}Celgene Corporation } & {\footnotesize{}Healthcare} & {\footnotesize{}49.98}\tabularnewline
{\footnotesize{}CERN} & {\footnotesize{}Cerner Corporation } & {\footnotesize{}Technology} & {\footnotesize{}18.77}\tabularnewline
{\footnotesize{}CMCSA} & {\footnotesize{}Comcast Corporation} & {\footnotesize{}Communication Services} & {\footnotesize{}101.14}\tabularnewline
{\footnotesize{}COST} & {\footnotesize{}Costco Wholesale Corporation } & {\footnotesize{}Consumer Defensive} & {\footnotesize{}40.91}\tabularnewline
{\footnotesize{}DISCA} & {\footnotesize{}Discovery, Inc. } & {\footnotesize{}Consumer Cyclical} & {\footnotesize{}28.52}\tabularnewline
{\footnotesize{}EBAY} & {\footnotesize{}eBay Inc.} & {\footnotesize{}Consumer Cyclical} & {\footnotesize{}28.68}\tabularnewline
{\footnotesize{}FB} & {\footnotesize{}Facebook, Inc.} & {\footnotesize{}Technology} & {\footnotesize{}68.53}\tabularnewline
{\footnotesize{}GOOG} & {\footnotesize{}Alphabet Inc. } & {\footnotesize{}Technology} & {\footnotesize{}265.76}\tabularnewline
{\footnotesize{}INTC} & {\footnotesize{}Intel Corporation} & {\footnotesize{}Technology} & {\footnotesize{}90.35}\tabularnewline
{\footnotesize{}ISRG} & {\footnotesize{}Intuitive Surgical Inc. } & {\footnotesize{}Healthcare} & {\footnotesize{}19.99}\tabularnewline
{\footnotesize{}KLAC} & {\footnotesize{}KLA-Tencor Corporation} & {\footnotesize{}Technology} & {\footnotesize{}6.01}\tabularnewline
{\footnotesize{}MAR} & {\footnotesize{}Marriott International, Inc.} & {\footnotesize{}Consumer Cyclical} & {\footnotesize{}11.86}\tabularnewline
{\footnotesize{}MSFT} & {\footnotesize{}Microsoft Corporation} & {\footnotesize{}Technology} & {\footnotesize{}211.95}\tabularnewline
{\footnotesize{}NFLX} & {\footnotesize{}Netflix, Inc.} & {\footnotesize{}Consumer Cyclical} & {\footnotesize{}9.98}\tabularnewline
{\footnotesize{}QCOM} & {\footnotesize{}Qualcomm, Inc.} & {\footnotesize{}Technology} & {\footnotesize{}97.96}\tabularnewline
{\footnotesize{}REGN} & {\footnotesize{}Regeneron Pharmaceuticals } & {\footnotesize{}Healthcare} & {\footnotesize{}20.59}\tabularnewline
{\footnotesize{}SBUX} & {\footnotesize{}Starbucks Corporation} & {\footnotesize{}Communication Services} & {\footnotesize{}40.32}\tabularnewline
{\footnotesize{}TXN} & {\footnotesize{}Texas Instruments Inc.} & {\footnotesize{}Technology} & {\footnotesize{}33.09}\tabularnewline
{\footnotesize{}VOD} & {\footnotesize{}Vodafone Group plc } & {\footnotesize{}Communication Services} & {\footnotesize{}102.48}\tabularnewline
{\footnotesize{}YHOO} & {\footnotesize{}Yahoo Corporation} & {\footnotesize{}Financial Services} & {\footnotesize{}25.51}\tabularnewline
\hline 
\end{tabular}}
\label{table:25stocks}
\end{table}

\section{Robustness of the bimodal distribution of inter-trade durations}
\begin{figure}[!h]
\centerline{\includegraphics[scale=0.32]{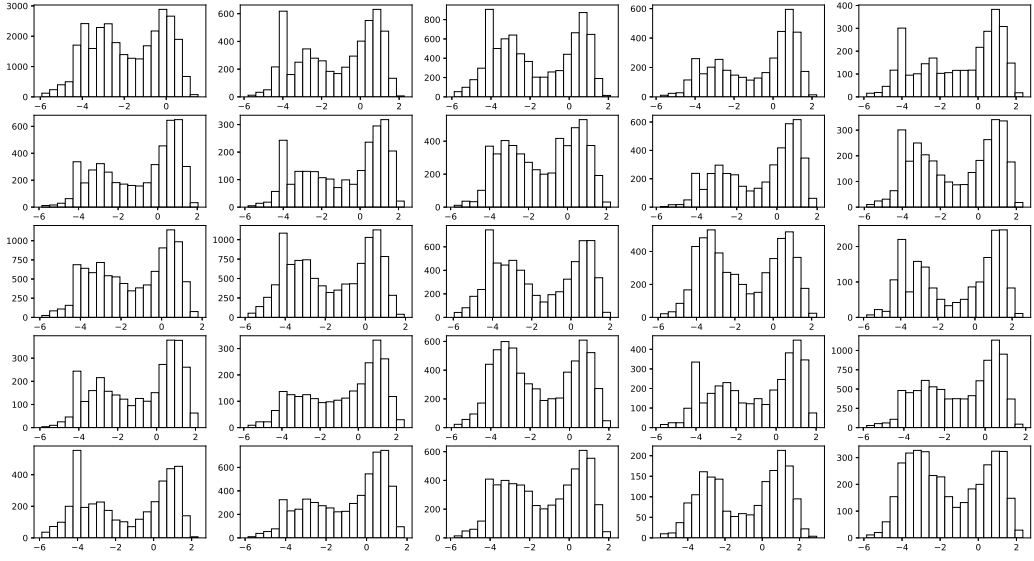}}
\caption{Histograms of the common logarithms of the raw inter-trade durations for 25 NASDAQ stocks on January 2nd, 2013. }
\label{fig:Rawbimodal}
\end{figure}

\begin{figure}[!h]
\centerline{\includegraphics[scale=0.32]{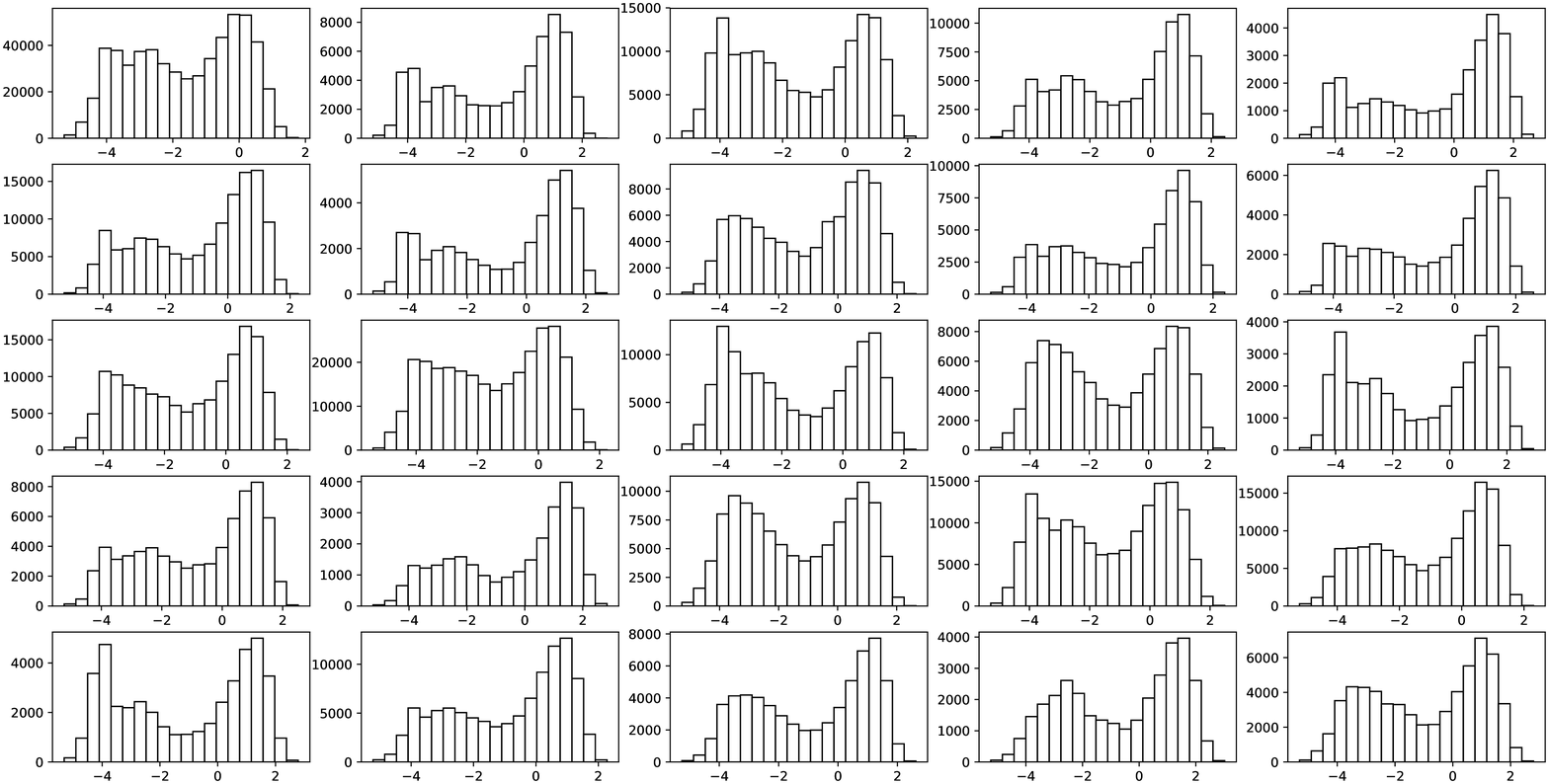}}
\caption{Histograms of the common logarithms of the aggregated inter-trade durations in the whole January of 2013 for 25 NASDAQ stocks.}
\label{fig:Aggbimodal}
\end{figure}

We supplement the robustness check of the bimodal distribution. In Figure \ref{fig:Rawbimodal} we plot the histograms of the common logarithms of the raw inter-trade durations that haven't been adjusted for the intra-day calendar effects. And in Figure \ref {fig:Aggbimodal} we plot the histograms of the common logarithms of the aggregated inter-trade durations in a whole month.

\section{Proof of Proposition 1}
We first restate the Theorem 2 in \cite{dovsla2009conditions}. It says that for two unimodal distributions $f_1$ and $f_2$ to be differentiable 
in the interval $[M_1,M_2]$, let the function $\phi(x)=\left|f'_1(x)/f'_2(x)\right|$ to be continuous on $(M_1,M_2)$ and let 
$\lim_{x\to M_1+}\phi(x)=0$ and $\lim_{x\to M_2-}\phi(x)=\infty$. Assume that there exist points $x_1$, $x_2$ such that $M_1<x_1<x_2<M_2$ 
and function $\phi$ is increasing on the interval $(M_1,x_1)$, decreasing on $(x_1,x_2)$, and again increasing on $(x_2,M_2)$. Then a mixture with
the density $g=pf_1+(1-p)f_2$ is bimodal if and only if $p\in(p_1,p_2)$ where $\frac{1}{p_i}=1+\phi(x_i)$ for $i=1,2$.

In our model of the mixture of two inverse gaussian, i.e, $IVG(\mu_1,\lambda_1)$ and $IVG(\mu_2,\lambda_2)$, we have 
\begin{equation*}
f'_k(y)=\left[\frac{\lambda_k}{2\pi}\right]^{\frac{1}{2}}\exp\left[\frac{-\lambda_k(y-\mu_k)^2}{2\mu_k^2y}\right]\cdot y^{-\frac{3}{2}}
\cdot \left[\frac{-\lambda_k y^2-3\mu_k^2 y+\mu_k^2\lambda_k}{2\mu_k^2 y^2}\right] \quad k=1,2.
\end{equation*}
Then in the interval $(M_1,M_2)$ 
\begin{align*}
\phi(y)&=\left|\frac{f'_1(y)}{f'_2(y)}\right|=\left(\frac{\lambda_1}{\lambda_2}\right)^{\frac{3}{2}}\left(\frac{\mu_2}{\mu_1}\right)^{2}
\frac{(y-M_1)(y-N_1)}{(M_2-y)(y-N_2)}
\exp\left[\left(\frac{\lambda_2}{2\mu_2^2}-\frac{\lambda_1}{2\mu_1^2}\right)y +\frac{\lambda_2-\lambda_1}{2y} + C\right],
\end{align*}
where $C= 2\mu_1\mu_2(\lambda_1\mu_2-\lambda_2\mu_1)$. 
Clearly $\lim_{y\to M_1+}\phi(x)=0$ and $\lim_{y\to M_2-}\phi(x)=+\infty$. And if we take derivative with respect to $\phi(y)$ again,
we have 
\begin{align*}
\phi'(y)&=\phi(y)\cdot\left[\frac{\mu_1^2\lambda_2-\mu_2^2\lambda_1}{2\mu_1^2\mu_2^2}-\frac{\lambda_2-\lambda_1}{2y^2}-\frac{1}{y-N_2}
+\frac{1}{y-M_1}+\frac{1}{y-N_1}+\frac{1}{M_2-y}\right]\\
          &=\phi(y)\cdot R(y).
\end{align*}
For the function $R(y)$, we have $\lim_{y\to M_1+}R(x)=+\infty$ and $\lim_{y\to M_2-}R(y)=+\infty$. 
If there is point $y^{\star}$ that makes $R(y^{\star})<0$, then there must exist two roots $y_1$ and $y_2$ for the equation $R(y)=0$
such that $M_1<y_1<y_2<M_2$. So $\phi'(y)>0$ for $(M_1,y_1)$, $\phi'(y)<0$ for $(y_1,y_2)$ and $\phi'(y)>0$ for $(y_2,M_2)$.
Therefore, function $\phi$ is increasing on the interval $(M_1,y_1)$, decreasing on $(y_1,y_2)$, and again increasing on $(y_2,M_2)$.
Then as long as $p\in(p_1,p_2)$ where $\frac{1}{p_k}=1+\phi(y_k)$ for $k=1,2$, the mixture $g$ is bimodal.

\section{Maximization in EM and updating parameters}

\subsection{Estimation for $\boldsymbol{\beta}$}

From equation \ref{eq:Ex-loglike}, by taking first order condition (F.O.C),
we have

\[
\frac{\partial\mathbf{E}[l]}{\partial\boldsymbol{\beta}_{12}}=\stackrel[i=2]{n}{\sum}\left[\mathbf{x}_{i-1}\cdot\mathbb{P}(s_{i}=2,s_{i-1}=1|\mathcal{F})-\frac{e^{\boldsymbol{\beta}_{12}^{T}\mathbf{x}_{i-1}}}{1+e^{\boldsymbol{\beta}_{12}^{T}\mathbf{x}_{i-1}}}\mathbf{x}_{i-1}\cdot\mathbb{P}(s_{i-1}=1|\mathcal{F})\right]=\vec{0}
\]
which is equivalent to:
\begin{equation}
\stackrel[i=2]{n}{\sum}\left[\mathbb{P}(s_{i}=2,s_{i-1}=1|\mathcal{F})-\frac{e^{\boldsymbol{\beta}_{12}^{T}\mathbf{x}_{i-1}}}{1+e^{\boldsymbol{\beta}_{12}^{T}\mathbf{x}_{i-1}}}\cdot\mathbb{P}(s_{i-1}=1|\mathcal{F})\right]\mathbf{x}_{i-1}=\vec{0}
\end{equation}

As the same, from $\frac{\partial\mathbf{E}[l]}{\partial\boldsymbol{\beta}_{21}}=\vec{0}$,
we have
\begin{equation}
\stackrel[i=2]{n}{\sum}\left[\mathbb{P}(s_{i}=1,s_{i-1}=2|\mathcal{F})-\frac{e^{\boldsymbol{\beta}_{21}^{T}\mathbf{x}_{i-1}}}{1+e^{\boldsymbol{\beta}_{21}^{T}\mathbf{x}_{i-1}}}\cdot\mathbb{P}(s_{i-1}=2|\mathcal{F})\right]\mathbf{x}_{i-1}=\vec{0}
\end{equation}

\subsection{Estimation for the distribution parameters}

Conditional on state $k$, inter-trade duration $y_{i}$ follow
inverse Gaussian distribution. So
\begin{align*}
\log f(y_{i}|s_{i}=k) & =\frac{1}{2}\log\Big[\frac{\lambda_{k}}{2\pi}\big(y_{i}\big)^{-3}\Big]-\frac{\lambda_{k}}{2\mu_{k}^{2}}\frac{\big(y_{i}-\mu_{k})^{2}}{\big(y_{i}\big)}\\
 & =\frac{1}{2}\log\big(\lambda_{k}\big)-\frac{1}{2}\log(2\pi)-\frac{3}{2}\log\big(y_{i}\big)-\frac{\lambda_{k}}{2\mu_{k}^{2}}\frac{(y_{i}-\mu_{k})^{2}}{y_{i}},\quad k=1,2
\end{align*}
Thus, for the part of likelihood that include distribution parameters
in \ref{eq:Ex-loglike}, 
\begin{align*}
\mathbf{E}\left(\sum_{i=1}^{n}\log f\Big(y_{i}\Big)\right) & =-\frac{n}{2}\log\big(2\pi\big)-\frac{3}{2}\stackrel[i=1]{n}{\sum}\log(y_{i})
 +\frac{1}{2}\log\lambda_{1}\cdot\left(\stackrel[i=1]{n}{\sum}\mathbb{P}(s_{i}=1|\mathcal{F})\right)+\frac{1}{2}\log\lambda_{2}\cdot\left(\stackrel[i=1]{n}{\sum}\mathbb{P}(s_{i}=2|\mathcal{F})\right)\\
 & -\frac{\lambda_{1}}{2\mu_{1}^{2}}\stackrel[i=1]{n}{\sum}\Big(y_{i}+\frac{\mu_{1}^{2}}{y_{i}}-2\mu_{1}\Big)\cdot\mathbb{P}(s_{i}=1|\mathcal{F})
 -\frac{\lambda_{2}}{2\mu_{2}^{2}}\stackrel[i=1]{n}{\sum}\Big(y_{i}+\frac{\mu_{2}^{2}}{y_{i}}-2\mu_{2}\Big)\cdot\mathbb{P}(s_{i}=2|\mathcal{F}).
\end{align*}
Then by taking the F.O.C w.r.t $\mu_{1},\mu_{2},\lambda_{1},\lambda_{2}$,
we get the solution in \eqref{M-parameters}.

\section{Calendar effects of inter-trade durations} \label{SC:Calendar effects}
\begin{figure}[!h]
\centerline{\includegraphics[scale=0.32]{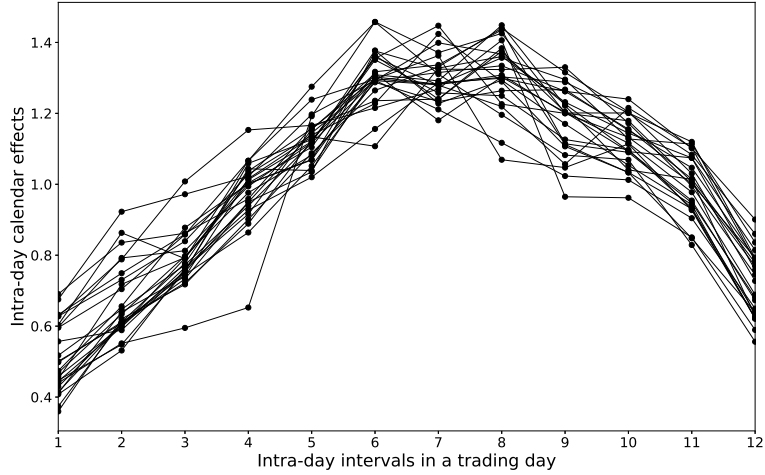}}
\caption{A bundle of 25 stocks' inter-trade duration calendar effects, which
are measured for 12 half hour intervals in a trading day from 9:45am-3:45pm. }
\label{fig:calendarEffects}
\end{figure}

We divide the market operation hour (9:45 am to 3:45 pm) into 12 half-hour blocks
and introduce the a time-of-day variable $z_{ik}$ for the inter-trade duration series.
In particular, $z_{ik}=1$ if the $i$-th inter-trade
duration is in in the $k$-th half-hour time block 
($k=1,\ldots12$), and 0 otherwise. We then regress
$y_i$ on the dummy variables $z_{ik}$, i.e., 
$y_{i}=\sum_{k=1}^{12}a_{k}x_{ik}+\varepsilon_{i}$,
and obtain the coefficients $\hat{a}_{k}$. 
We normalize the coefficients by their average 
$
\tilde{a}_{k}=\frac{\hat{a}_{k}}{\frac{1}{12}\sum_{k=1}^{12}\hat{a}_{k}}
$
$(1\le k \le 12)$
and plot them in Figure \ref{fig:calendarEffects},
which clearly shows a intra-day calendar effect, i.e.,
inter-trade durations are shorter at the beginning and the end of day while longer around the noon.
We then calculate the adjusted inter-trade duration 
by de-trending the calendar effects.

\pagebreak

\section{Robustness test for the effects of LOB factors} \label{SC: robustness}
\subsection{Analysis in different month}
We apply the MF-RSD model \eqref{model.4lob} to the 25 stocks in a different month, i.e., April 2013,
and then summarize the results of the significance of the 4 LOB factors in Table \ref{table:2013April}.
We can see that the results are quite similar to what we have obtained for the sample analysis of January 2013.
Hence, the roles of LOB factors in the regime-switching of inter-trade durations are consistent in different periods.

\begin{table}[!h]
\caption{Summary of the results for the significance of 4 LOB factors in April 2013.}
\medskip{}

\centerline{%
\begin{tabular}{c|cccc|cccc|cccc|cccc}
\hline 
\multirow{3}{*}{{\footnotesize{}Stock}} & \multicolumn{4}{c|}{{\footnotesize{}Depth Imbalance}} & \multicolumn{4}{c|}{{\footnotesize{}Price Spread}} & \multicolumn{4}{c|}{{\footnotesize{}Trade Volume}} & \multicolumn{4}{c}{{\footnotesize{}Price Movement}}\tabularnewline
\cline{2-17} 
 & \multicolumn{2}{c|}{{\scriptsize{}$\mathrm{P_{12}}$}} & \multicolumn{2}{c|}{{\scriptsize{}$\mathrm{P_{21}}$}} & \multicolumn{2}{c|}{{\scriptsize{}$\mathrm{P_{12}}$}} & \multicolumn{2}{c|}{{\scriptsize{}$\mathrm{P_{21}}$}} & \multicolumn{2}{c|}{{\scriptsize{}$\mathrm{P_{12}}$}} & \multicolumn{2}{c|}{{\scriptsize{}$\mathrm{P_{21}}$}} & \multicolumn{2}{c|}{{\scriptsize{}$\mathrm{P_{12}}$}} & \multicolumn{2}{c}{{\scriptsize{}$\mathrm{P_{21}}$}}\tabularnewline
\cline{2-17} 
 & {\scriptsize{}sig(+)} & {\scriptsize{}sig(-)} & {\scriptsize{}sig(+)} & {\scriptsize{}sig(-)} & {\scriptsize{}sig(+)} & {\scriptsize{}sig(-)} & {\scriptsize{}sig(+)} & {\scriptsize{}sig(-)} & {\scriptsize{}sig(+)} & {\scriptsize{}sig(-)} & {\scriptsize{}sig(+)} & {\scriptsize{}sig(-)} & {\scriptsize{}sig(+)} & \multicolumn{1}{c|}{{\scriptsize{}sig(-)}} & {\scriptsize{}sig(+)} & {\scriptsize{}sig(-)}\tabularnewline
\hline 
{\footnotesize{}AAPL} & {\scriptsize{}0} & {\scriptsize{}14} & {\scriptsize{}3} & {\scriptsize{}1} & {\scriptsize{}19} & {\scriptsize{}0} & {\scriptsize{}0} & {\scriptsize{}22} & {\scriptsize{}0} & {\scriptsize{}21} & {\scriptsize{}22} & {\scriptsize{}0} & {\scriptsize{}22} & {\scriptsize{}0} & {\scriptsize{}0} & {\scriptsize{}15}\tabularnewline
{\footnotesize{}ALXN} & {\scriptsize{}1} & {\scriptsize{}2} & {\scriptsize{}1} & {\scriptsize{}1} & {\scriptsize{}8} & {\scriptsize{}5} & {\scriptsize{}0} & {\scriptsize{}20} & {\scriptsize{}0} & {\scriptsize{}10} & {\scriptsize{}18} & {\scriptsize{}0} & {\scriptsize{}22} & {\scriptsize{}0} & {\scriptsize{}0} & {\scriptsize{}22}\tabularnewline
{\footnotesize{}AMZN} & {\scriptsize{}0} & {\scriptsize{}6} & {\scriptsize{}6} & {\scriptsize{}0} & {\scriptsize{}19} & {\scriptsize{}2} & {\scriptsize{}0} & {\scriptsize{}22} & {\scriptsize{}0} & {\scriptsize{}12} & {\scriptsize{}18} & {\scriptsize{}0} & {\scriptsize{}22} & {\scriptsize{}0} & {\scriptsize{}0} & {\scriptsize{}22}\tabularnewline
{\footnotesize{}BIDU} & {\scriptsize{}1} & {\scriptsize{}1} & {\scriptsize{}2} & {\scriptsize{}2} & {\scriptsize{}16} & {\scriptsize{}0} & {\scriptsize{}0} & {\scriptsize{}22} & {\scriptsize{}0} & {\scriptsize{}14} & {\scriptsize{}21} & {\scriptsize{}0} & {\scriptsize{}22} & {\scriptsize{}0} & {\scriptsize{}0} & {\scriptsize{}22}\tabularnewline
{\footnotesize{}BMRN} & {\scriptsize{}0} & {\scriptsize{}3} & {\scriptsize{}0} & {\scriptsize{}0} & {\scriptsize{}11} & {\scriptsize{}1} & {\scriptsize{}0} & {\scriptsize{}21} & {\scriptsize{}0} & {\scriptsize{}7} & {\scriptsize{}9} & {\scriptsize{}0} & {\scriptsize{}18} & {\scriptsize{}0} & {\scriptsize{}0} & {\scriptsize{}22}\tabularnewline
{\footnotesize{}CELG} & {\scriptsize{}1} & {\scriptsize{}7} & {\scriptsize{}3} & {\scriptsize{}1} & {\scriptsize{}13} & {\scriptsize{}2} & {\scriptsize{}0} & {\scriptsize{}22} & {\scriptsize{}0} & {\scriptsize{}11} & {\scriptsize{}20} & {\scriptsize{}0} & {\scriptsize{}22} & {\scriptsize{}0} & {\scriptsize{}0} & {\scriptsize{}21}\tabularnewline
{\footnotesize{}CERN} & {\scriptsize{}2} & {\scriptsize{}1} & {\scriptsize{}1} & {\scriptsize{}1} & {\scriptsize{}7} & {\scriptsize{}1} & {\scriptsize{}0} & {\scriptsize{}17} & {\scriptsize{}0} & {\scriptsize{}2} & {\scriptsize{}5} & {\scriptsize{}0} & {\scriptsize{}21} & {\scriptsize{}0} & {\scriptsize{}0} & {\scriptsize{}21}\tabularnewline
{\footnotesize{}CMCSA} & {\scriptsize{}0} & {\scriptsize{}17} & {\scriptsize{}9} & {\scriptsize{}0} & {\scriptsize{}1} & {\scriptsize{}13} & {\scriptsize{}0} & {\scriptsize{}13} & {\scriptsize{}0} & {\scriptsize{}16} & {\scriptsize{}22} & {\scriptsize{}0} & {\scriptsize{}15} & {\scriptsize{}0} & {\scriptsize{}4} & {\scriptsize{}0}\tabularnewline
{\footnotesize{}COST} & {\scriptsize{}0} & {\scriptsize{}0} & {\scriptsize{}0} & {\scriptsize{}0} & {\scriptsize{}5} & {\scriptsize{}5} & {\scriptsize{}0} & {\scriptsize{}14} & {\scriptsize{}0} & {\scriptsize{}10} & {\scriptsize{}20} & {\scriptsize{}0} & {\scriptsize{}22} & {\scriptsize{}0} & {\scriptsize{}0} & {\scriptsize{}22}\tabularnewline
{\footnotesize{}DISCA} & {\scriptsize{}0} & {\scriptsize{}3} & {\scriptsize{}2} & {\scriptsize{}1} & {\scriptsize{}2} & {\scriptsize{}3} & {\scriptsize{}0} & {\scriptsize{}14} & {\scriptsize{}0} & {\scriptsize{}11} & {\scriptsize{}12} & {\scriptsize{}0} & {\scriptsize{}19} & {\scriptsize{}0} & {\scriptsize{}0} & {\scriptsize{}21}\tabularnewline
{\footnotesize{}EBAY} & {\scriptsize{}0} & {\scriptsize{}13} & {\scriptsize{}5} & {\scriptsize{}1} & {\scriptsize{}2} & {\scriptsize{}8} & {\scriptsize{}0} & {\scriptsize{}19} & {\scriptsize{}0} & {\scriptsize{}20} & {\scriptsize{}22} & {\scriptsize{}0} & {\scriptsize{}17} & {\scriptsize{}0} & {\scriptsize{}1} & {\scriptsize{}3}\tabularnewline
{\footnotesize{}FB} & {\scriptsize{}1} & {\scriptsize{}12} & {\scriptsize{}11} & {\scriptsize{}0} & {\scriptsize{}11} & {\scriptsize{}0} & {\scriptsize{}0} & {\scriptsize{}20} & {\scriptsize{}0} & {\scriptsize{}6} & {\scriptsize{}22} & {\scriptsize{}0} & {\scriptsize{}17} & {\scriptsize{}0} & {\scriptsize{}4} & {\scriptsize{}0}\tabularnewline
{\footnotesize{}GOOG} & {\scriptsize{}0} & {\scriptsize{}6} & {\scriptsize{}3} & {\scriptsize{}0} & {\scriptsize{}21} & {\scriptsize{}0} & {\scriptsize{}0} & {\scriptsize{}22} & {\scriptsize{}0} & {\scriptsize{}13} & {\scriptsize{}21} & {\scriptsize{}0} & {\scriptsize{}22} & {\scriptsize{}0} & {\scriptsize{}0} & {\scriptsize{}21}\tabularnewline
{\footnotesize{}INTC} & {\scriptsize{}0} & {\scriptsize{}22} & {\scriptsize{}21} & {\scriptsize{}0} & {\scriptsize{}1} & {\scriptsize{}14} & {\scriptsize{}1} & {\scriptsize{}1} & {\scriptsize{}0} & {\scriptsize{}19} & {\scriptsize{}22} & {\scriptsize{}0} & {\scriptsize{}6} & {\scriptsize{}5} & {\scriptsize{}4} & {\scriptsize{}0}\tabularnewline
{\footnotesize{}ISRG} & {\scriptsize{}2} & {\scriptsize{}0} & {\scriptsize{}0} & {\scriptsize{}2} & {\scriptsize{}14} & {\scriptsize{}0} & {\scriptsize{}0} & {\scriptsize{}21} & {\scriptsize{}3} & {\scriptsize{}1} & {\scriptsize{}14} & {\scriptsize{}0} & {\scriptsize{}17} & {\scriptsize{}0} & {\scriptsize{}0} & {\scriptsize{}12}\tabularnewline
{\footnotesize{}KLAC} & {\scriptsize{}0} & {\scriptsize{}4} & {\scriptsize{}2} & {\scriptsize{}1} & {\scriptsize{}0} & {\scriptsize{}5} & {\scriptsize{}0} & {\scriptsize{}8} & {\scriptsize{}0} & {\scriptsize{}7} & {\scriptsize{}12} & {\scriptsize{}0} & {\scriptsize{}20} & {\scriptsize{}0} & {\scriptsize{}0} & {\scriptsize{}20}\tabularnewline
{\footnotesize{}MAR} & {\scriptsize{}0} & {\scriptsize{}9} & {\scriptsize{}1} & {\scriptsize{}0} & {\scriptsize{}2} & {\scriptsize{}1} & {\scriptsize{}0} & {\scriptsize{}10} & {\scriptsize{}0} & {\scriptsize{}11} & {\scriptsize{}17} & {\scriptsize{}0} & {\scriptsize{}14} & {\scriptsize{}0} & {\scriptsize{}0} & {\scriptsize{}8}\tabularnewline
{\footnotesize{}MSFT} & {\scriptsize{}0} & {\scriptsize{}22} & {\scriptsize{}22} & {\scriptsize{}0} & {\scriptsize{}0} & {\scriptsize{}21} & {\scriptsize{}3} & {\scriptsize{}5} & {\scriptsize{}0} & {\scriptsize{}17} & {\scriptsize{}22} & {\scriptsize{}0} & {\scriptsize{}10} & {\scriptsize{}0} & {\scriptsize{}8} & {\scriptsize{}0}\tabularnewline
{\footnotesize{}NFLX} & {\scriptsize{}0} & {\scriptsize{}4} & {\scriptsize{}5} & {\scriptsize{}0} & {\scriptsize{}22} & {\scriptsize{}0} & {\scriptsize{}0} & {\scriptsize{}22} & {\scriptsize{}0} & {\scriptsize{}14} & {\scriptsize{}22} & {\scriptsize{}0} & {\scriptsize{}22} & {\scriptsize{}0} & {\scriptsize{}0} & {\scriptsize{}21}\tabularnewline
{\footnotesize{}QCOM} & {\scriptsize{}0} & {\scriptsize{}9} & {\scriptsize{}5} & {\scriptsize{}3} & {\scriptsize{}0} & {\scriptsize{}5} & {\scriptsize{}0} & {\scriptsize{}19} & {\scriptsize{}0} & {\scriptsize{}21} & {\scriptsize{}22} & {\scriptsize{}0} & {\scriptsize{}21} & {\scriptsize{}0} & {\scriptsize{}7} & {\scriptsize{}2}\tabularnewline
{\footnotesize{}REGN} & {\scriptsize{}0} & {\scriptsize{}5} & {\scriptsize{}4} & {\scriptsize{}0} & {\scriptsize{}17} & {\scriptsize{}0} & {\scriptsize{}0} & {\scriptsize{}22} & {\scriptsize{}0} & {\scriptsize{}2} & {\scriptsize{}9} & {\scriptsize{}0} & {\scriptsize{}22} & {\scriptsize{}0} & {\scriptsize{}0} & {\scriptsize{}20}\tabularnewline
{\footnotesize{}SBUX} & {\scriptsize{}0} & {\scriptsize{}8} & {\scriptsize{}3} & {\scriptsize{}0} & {\scriptsize{}1} & {\scriptsize{}5} & {\scriptsize{}0} & {\scriptsize{}10} & {\scriptsize{}0} & {\scriptsize{}16} & {\scriptsize{}22} & {\scriptsize{}0} & {\scriptsize{}19} & {\scriptsize{}0} & {\scriptsize{}0} & {\scriptsize{}11}\tabularnewline
{\footnotesize{}TXN} & {\scriptsize{}0} & {\scriptsize{}16} & {\scriptsize{}9} & {\scriptsize{}1} & {\scriptsize{}0} & {\scriptsize{}9} & {\scriptsize{}0} & {\scriptsize{}12} & {\scriptsize{}0} & {\scriptsize{}10} & {\scriptsize{}22} & {\scriptsize{}0} & {\scriptsize{}21} & {\scriptsize{}0} & {\scriptsize{}1} & {\scriptsize{}0}\tabularnewline
{\footnotesize{}VOD} & {\scriptsize{}1} & {\scriptsize{}13} & {\scriptsize{}13} & {\scriptsize{}0} & {\scriptsize{}7} & {\scriptsize{}1} & {\scriptsize{}0} & {\scriptsize{}12} & {\scriptsize{}1} & {\scriptsize{}6} & {\scriptsize{}21} & {\scriptsize{}0} & {\scriptsize{}10} & {\scriptsize{}1} & {\scriptsize{}2} & {\scriptsize{}0}\tabularnewline
{\footnotesize{}YHOO} & {\scriptsize{}0} & {\scriptsize{}20} & {\scriptsize{}19} & {\scriptsize{}0} & {\scriptsize{}3} & {\scriptsize{}2} & {\scriptsize{}0} & {\scriptsize{}5} & {\scriptsize{}0} & {\scriptsize{}13} & {\scriptsize{}22} & {\scriptsize{}0} & {\scriptsize{}7} & {\scriptsize{}1} & {\scriptsize{}5} & {\scriptsize{}0}\tabularnewline
\hline 
{\footnotesize{}Sum (\%)} & {\scriptsize{}1.6\%} & {\scriptsize{}39.5\%} & {\scriptsize{}27.3\%} & {\scriptsize{}2.7\%} & {\scriptsize{}36.7\%} & {\scriptsize{}18.7\%} & {\scriptsize{}0.7\%} & {\scriptsize{}71.8\%} & {\scriptsize{}0.7\%} & {\scriptsize{}52.7\%} & {\scriptsize{}83.5\%} & {\scriptsize{}0.0\%} & {\scriptsize{}81.8\%} & {\scriptsize{}1.3\%} & {\scriptsize{}6.5\%} & {\scriptsize{}55.6\%}\tabularnewline
\hline 
\end{tabular}}
\label{table:2013April}

\textit{\scriptsize{}Note}{\scriptsize{}: $\mathrm{P_{12}}$ represent
the regime switching from the short-duration regime to the long duration
regime, and $\mathrm{P_{21}}$ represent the regime switching from
the long regime to the short-duration regime. Significance is measured at the 5\%
level. Sig(+)/Sig(-) means that the estimated coefficient for
the factor is significantly positive/negative. Each cell shows the significance. We have 22 trading
days in April 2013, so the upper limit for each cell is 22. The
last row records the percentage (\%) of the times that this factor appears
as significant in all 550 instances (25 stocks times 22 days). }{\scriptsize\par}
\end{table}

\pagebreak

\subsection{Analysis of trades by trading direction}
To test if our finding on the informativeness of $DI$ to the price movement,
we repeat the analysis in the previous section to directional trades.
We divide the trades into buyer-initiated trades and seller-initiated trades
and then construct the series of inter-trade durations between the pure buy trades
and the series of inter-trade durations between the pure sell trades. Moreover,
to gain a clearer insight into the effect of depth imbalance on the
regime-switching probabilities for the directional transactions,
we redefine the order book depth imbalance $DI$ as
\[
DI=\frac{v_{b}^{1}-v_{a}^{1}}{v_{a}^{1}+v_{b}^{1}}.
\]
It is positive if the depth at best bid is greater than the depth
at best ask, and it is negative if $v_{a}^{1}>v_{b}^{1}$.

\begin{table}[!h]
\caption{Summary of the results for the significance of LOB factors for buy trades.}
\medskip{}

\centerline{%
\begin{tabular}{c|cccc|cccc|cccc|cccc}
\hline 
\multirow{3}{*}{{\footnotesize{}Stock}} & \multicolumn{4}{c|}{{\footnotesize{}$DI$}} & \multicolumn{4}{c|}{{\footnotesize{}$PS$}} & \multicolumn{4}{c|}{{\footnotesize{}$TV$}} & \multicolumn{4}{c}{{\footnotesize{}$PM$}}\tabularnewline
\cline{2-17} 
 & \multicolumn{2}{c|}{{\scriptsize{}$\mathrm{P_{12}}$}} & \multicolumn{2}{c|}{{\scriptsize{}$\mathrm{P_{21}}$}} & \multicolumn{2}{c|}{{\scriptsize{}$\mathrm{P_{12}}$}} & \multicolumn{2}{c|}{{\scriptsize{}$\mathrm{P_{21}}$}} & \multicolumn{2}{c|}{{\scriptsize{}$\mathrm{P_{12}}$}} & \multicolumn{2}{c|}{{\scriptsize{}$\mathrm{P_{21}}$}} & \multicolumn{2}{c|}{{\scriptsize{}$\mathrm{P_{12}}$}} & \multicolumn{2}{c}{{\scriptsize{}$\mathrm{P_{21}}$}}\tabularnewline
\cline{2-17} 
 & {\scriptsize{}sig(+)} & {\scriptsize{}sig(-)} & {\scriptsize{}sig(+)} & {\scriptsize{}sig(-)} & {\scriptsize{}sig(+)} & {\scriptsize{}sig(-)} & {\scriptsize{}sig(+)} & {\scriptsize{}sig(-)} & {\scriptsize{}sig(+)} & {\scriptsize{}sig(-)} & {\scriptsize{}sig(+)} & {\scriptsize{}sig(-)} & {\scriptsize{}sig(+)} & {\scriptsize{}sig(-)} & {\scriptsize{}sig(+)} & {\scriptsize{}sig(-)}\tabularnewline
\hline 
{\footnotesize{}AAPL} & {\scriptsize{}1} & {\scriptsize{}11} & {\scriptsize{}12} & {\scriptsize{}0} & {\scriptsize{}16} & {\scriptsize{}0} & {\scriptsize{}0} & {\scriptsize{}21} & {\scriptsize{}0} & {\scriptsize{}10} & {\scriptsize{}21} & {\scriptsize{}0} & {\scriptsize{}21} & {\scriptsize{}0} & {\scriptsize{}3} & {\scriptsize{}0}\tabularnewline
{\footnotesize{}ALXN} & {\scriptsize{}3} & {\scriptsize{}1} & {\scriptsize{}3} & {\scriptsize{}0} & {\scriptsize{}4} & {\scriptsize{}0} & {\scriptsize{}0} & {\scriptsize{}11} & {\scriptsize{}0} & {\scriptsize{}4} & {\scriptsize{}4} & {\scriptsize{}1} & {\scriptsize{}19} & {\scriptsize{}0} & {\scriptsize{}0} & {\scriptsize{}20}\tabularnewline
{\footnotesize{}AMZN} & {\scriptsize{}3} & {\scriptsize{}1} & {\scriptsize{}6} & {\scriptsize{}0} & {\scriptsize{}11} & {\scriptsize{}0} & {\scriptsize{}0} & {\scriptsize{}19} & {\scriptsize{}0} & {\scriptsize{}4} & {\scriptsize{}10} & {\scriptsize{}0} & {\scriptsize{}21} & {\scriptsize{}0} & {\scriptsize{}0} & {\scriptsize{}21}\tabularnewline
{\footnotesize{}BIDU} & {\scriptsize{}1} & {\scriptsize{}2} & {\scriptsize{}4} & {\scriptsize{}0} & {\scriptsize{}6} & {\scriptsize{}1} & {\scriptsize{}0} & {\scriptsize{}15} & {\scriptsize{}0} & {\scriptsize{}2} & {\scriptsize{}10} & {\scriptsize{}0} & {\scriptsize{}20} & {\scriptsize{}0} & {\scriptsize{}0} & {\scriptsize{}21}\tabularnewline
{\footnotesize{}BMRN} & {\scriptsize{}0} & {\scriptsize{}0} & {\scriptsize{}3} & {\scriptsize{}1} & {\scriptsize{}1} & {\scriptsize{}0} & {\scriptsize{}0} & {\scriptsize{}15} & {\scriptsize{}2} & {\scriptsize{}1} & {\scriptsize{}7} & {\scriptsize{}0} & {\scriptsize{}15} & {\scriptsize{}0} & {\scriptsize{}0} & {\scriptsize{}14}\tabularnewline
{\footnotesize{}CELG} & {\scriptsize{}0} & {\scriptsize{}2} & {\scriptsize{}18} & {\scriptsize{}0} & {\scriptsize{}2} & {\scriptsize{}5} & {\scriptsize{}0} & {\scriptsize{}12} & {\scriptsize{}0} & {\scriptsize{}9} & {\scriptsize{}16} & {\scriptsize{}0} & {\scriptsize{}21} & {\scriptsize{}0} & {\scriptsize{}0} & {\scriptsize{}21}\tabularnewline
{\footnotesize{}CERN} & {\scriptsize{}0} & {\scriptsize{}1} & {\scriptsize{}5} & {\scriptsize{}0} & {\scriptsize{}2} & {\scriptsize{}1} & {\scriptsize{}0} & {\scriptsize{}12} & {\scriptsize{}0} & {\scriptsize{}5} & {\scriptsize{}7} & {\scriptsize{}0} & {\scriptsize{}20} & {\scriptsize{}0} & {\scriptsize{}0} & {\scriptsize{}18}\tabularnewline
{\footnotesize{}CMCSA} & {\scriptsize{}0} & {\scriptsize{}11} & {\scriptsize{}19} & {\scriptsize{}0} & {\scriptsize{}6} & {\scriptsize{}1} & {\scriptsize{}4} & {\scriptsize{}0} & {\scriptsize{}3} & {\scriptsize{}2} & {\scriptsize{}19} & {\scriptsize{}0} & {\scriptsize{}11} & {\scriptsize{}0} & {\scriptsize{}0} & {\scriptsize{}4}\tabularnewline
{\footnotesize{}COST} & {\scriptsize{}2} & {\scriptsize{}2} & {\scriptsize{}4} & {\scriptsize{}0} & {\scriptsize{}3} & {\scriptsize{}5} & {\scriptsize{}1} & {\scriptsize{}7} & {\scriptsize{}0} & {\scriptsize{}5} & {\scriptsize{}18} & {\scriptsize{}0} & {\scriptsize{}20} & {\scriptsize{}0} & {\scriptsize{}0} & {\scriptsize{}20}\tabularnewline
{\footnotesize{}DISCA} & {\scriptsize{}1} & {\scriptsize{}1} & {\scriptsize{}2} & {\scriptsize{}0} & {\scriptsize{}0} & {\scriptsize{}2} & {\scriptsize{}0} & {\scriptsize{}11} & {\scriptsize{}0} & {\scriptsize{}3} & {\scriptsize{}8} & {\scriptsize{}1} & {\scriptsize{}20} & {\scriptsize{}0} & {\scriptsize{}0} & {\scriptsize{}20}\tabularnewline
{\footnotesize{}EBAY} & {\scriptsize{}8} & {\scriptsize{}1} & {\scriptsize{}6} & {\scriptsize{}1} & {\scriptsize{}8} & {\scriptsize{}0} & {\scriptsize{}0} & {\scriptsize{}17} & {\scriptsize{}0} & {\scriptsize{}2} & {\scriptsize{}12} & {\scriptsize{}0} & {\scriptsize{}20} & {\scriptsize{}0} & {\scriptsize{}0} & {\scriptsize{}13}\tabularnewline
{\footnotesize{}FB} & {\scriptsize{}0} & {\scriptsize{}5} & {\scriptsize{}10} & {\scriptsize{}0} & {\scriptsize{}8} & {\scriptsize{}1} & {\scriptsize{}0} & {\scriptsize{}11} & {\scriptsize{}2} & {\scriptsize{}3} & {\scriptsize{}21} & {\scriptsize{}0} & {\scriptsize{}21} & {\scriptsize{}0} & {\scriptsize{}0} & {\scriptsize{}8}\tabularnewline
{\footnotesize{}GOOG} & {\scriptsize{}1} & {\scriptsize{}3} & {\scriptsize{}4} & {\scriptsize{}0} & {\scriptsize{}18} & {\scriptsize{}0} & {\scriptsize{}0} & {\scriptsize{}21} & {\scriptsize{}1} & {\scriptsize{}4} & {\scriptsize{}20} & {\scriptsize{}0} & {\scriptsize{}21} & {\scriptsize{}0} & {\scriptsize{}0} & {\scriptsize{}15}\tabularnewline
{\footnotesize{}INTC} & {\scriptsize{}0} & {\scriptsize{}19} & {\scriptsize{}16} & {\scriptsize{}0} & {\scriptsize{}5} & {\scriptsize{}5} & {\scriptsize{}7} & {\scriptsize{}1} & {\scriptsize{}1} & {\scriptsize{}2} & {\scriptsize{}21} & {\scriptsize{}0} & {\scriptsize{}4} & {\scriptsize{}4} & {\scriptsize{}0} & {\scriptsize{}0}\tabularnewline
{\footnotesize{}ISRG} & {\scriptsize{}4} & {\scriptsize{}0} & {\scriptsize{}3} & {\scriptsize{}1} & {\scriptsize{}5} & {\scriptsize{}1} & {\scriptsize{}0} & {\scriptsize{}13} & {\scriptsize{}3} & {\scriptsize{}0} & {\scriptsize{}8} & {\scriptsize{}1} & {\scriptsize{}16} & {\scriptsize{}0} & {\scriptsize{}0} & {\scriptsize{}11}\tabularnewline
{\footnotesize{}KLAC} & {\scriptsize{}1} & {\scriptsize{}2} & {\scriptsize{}6} & {\scriptsize{}0} & {\scriptsize{}7} & {\scriptsize{}3} & {\scriptsize{}0} & {\scriptsize{}11} & {\scriptsize{}4} & {\scriptsize{}4} & {\scriptsize{}9} & {\scriptsize{}0} & {\scriptsize{}15} & {\scriptsize{}0} & {\scriptsize{}0} & {\scriptsize{}14}\tabularnewline
{\footnotesize{}MAR} & {\scriptsize{}0} & {\scriptsize{}3} & {\scriptsize{}9} & {\scriptsize{}0} & {\scriptsize{}3} & {\scriptsize{}1} & {\scriptsize{}0} & {\scriptsize{}3} & {\scriptsize{}1} & {\scriptsize{}1} & {\scriptsize{}8} & {\scriptsize{}0} & {\scriptsize{}9} & {\scriptsize{}0} & {\scriptsize{}0} & {\scriptsize{}4}\tabularnewline
{\footnotesize{}MSFT} & {\scriptsize{}0} & {\scriptsize{}20} & {\scriptsize{}18} & {\scriptsize{}0} & {\scriptsize{}6} & {\scriptsize{}5} & {\scriptsize{}8} & {\scriptsize{}2} & {\scriptsize{}1} & {\scriptsize{}1} & {\scriptsize{}20} & {\scriptsize{}0} & {\scriptsize{}3} & {\scriptsize{}2} & {\scriptsize{}1} & {\scriptsize{}0}\tabularnewline
{\footnotesize{}NFLX} & {\scriptsize{}1} & {\scriptsize{}6} & {\scriptsize{}8} & {\scriptsize{}0} & {\scriptsize{}11} & {\scriptsize{}1} & {\scriptsize{}0} & {\scriptsize{}20} & {\scriptsize{}0} & {\scriptsize{}6} & {\scriptsize{}17} & {\scriptsize{}0} & {\scriptsize{}21} & {\scriptsize{}0} & {\scriptsize{}0} & {\scriptsize{}18}\tabularnewline
{\footnotesize{}QCOM} & {\scriptsize{}5} & {\scriptsize{}1} & {\scriptsize{}10} & {\scriptsize{}0} & {\scriptsize{}10} & {\scriptsize{}1} & {\scriptsize{}0} & {\scriptsize{}14} & {\scriptsize{}0} & {\scriptsize{}4} & {\scriptsize{}19} & {\scriptsize{}0} & {\scriptsize{}16} & {\scriptsize{}0} & {\scriptsize{}0} & {\scriptsize{}12}\tabularnewline
{\footnotesize{}REGN} & {\scriptsize{}0} & {\scriptsize{}11} & {\scriptsize{}14} & {\scriptsize{}0} & {\scriptsize{}9} & {\scriptsize{}0} & {\scriptsize{}0} & {\scriptsize{}20} & {\scriptsize{}1} & {\scriptsize{}0} & {\scriptsize{}9} & {\scriptsize{}0} & {\scriptsize{}20} & {\scriptsize{}0} & {\scriptsize{}0} & {\scriptsize{}18}\tabularnewline
{\footnotesize{}SBUX} & {\scriptsize{}0} & {\scriptsize{}4} & {\scriptsize{}5} & {\scriptsize{}0} & {\scriptsize{}3} & {\scriptsize{}3} & {\scriptsize{}0} & {\scriptsize{}9} & {\scriptsize{}0} & {\scriptsize{}5} & {\scriptsize{}19} & {\scriptsize{}0} & {\scriptsize{}16} & {\scriptsize{}0} & {\scriptsize{}0} & {\scriptsize{}8}\tabularnewline
{\footnotesize{}TXN} & {\scriptsize{}0} & {\scriptsize{}3} & {\scriptsize{}9} & {\scriptsize{}0} & {\scriptsize{}6} & {\scriptsize{}1} & {\scriptsize{}1} & {\scriptsize{}6} & {\scriptsize{}1} & {\scriptsize{}1} & {\scriptsize{}18} & {\scriptsize{}0} & {\scriptsize{}16} & {\scriptsize{}0} & {\scriptsize{}0} & {\scriptsize{}6}\tabularnewline
{\footnotesize{}VOD} & {\scriptsize{}0} & {\scriptsize{}7} & {\scriptsize{}13} & {\scriptsize{}0} & {\scriptsize{}2} & {\scriptsize{}0} & {\scriptsize{}1} & {\scriptsize{}0} & {\scriptsize{}3} & {\scriptsize{}0} & {\scriptsize{}15} & {\scriptsize{}0} & {\scriptsize{}11} & {\scriptsize{}0} & {\scriptsize{}0} & {\scriptsize{}0}\tabularnewline
{\footnotesize{}YHOO} & {\scriptsize{}0} & {\scriptsize{}16} & {\scriptsize{}14} & {\scriptsize{}0} & {\scriptsize{}8} & {\scriptsize{}4} & {\scriptsize{}8} & {\scriptsize{}0} & {\scriptsize{}5} & {\scriptsize{}0} & {\scriptsize{}18} & {\scriptsize{}0} & {\scriptsize{}4} & {\scriptsize{}1} & {\scriptsize{}0} & {\scriptsize{}0}\tabularnewline
\hline 
{\footnotesize{}Sum (\%)} & {\scriptsize{}5.9\%} & {\scriptsize{}25.3\%} & {\scriptsize{}42.1\%} & {\scriptsize{}0.6\%} & {\scriptsize{}30.5\%} & {\scriptsize{}7.8\%} & {\scriptsize{}5.7\%} & {\scriptsize{}51.6\%} & {\scriptsize{}5.3\%} & {\scriptsize{}14.9\%} & {\scriptsize{}67.4\%} & {\scriptsize{}0.6\%} & {\scriptsize{}76.4\%} & {\scriptsize{}1.3\%} & {\scriptsize{}0.8\%} & {\scriptsize{}54.5\%}\tabularnewline
\hline 
\end{tabular}}
\label{table:buyside}

\textit{\scriptsize{}Note}{\scriptsize{}: $\mathrm{P_{12}}$ represent
the regime switching from the short-duration regime to the long duration
regime, and $\mathrm{P_{21}}$ represent the regime switching from
the long regime to the short-duration regime. Significance is measured at the 5\%
level. Sig(+)/Sig(-) means that the estimated coefficient for
the factor is significantly positive/negative. Each cell shows the significance. 
The data is in the month of January 2013. We have 21 trading
days for each stock, so the upper limit for each cell is 21. The
last row records the percentage (\%) of the times that this factor appears
as significant in all 525 instances (25 stocks times 21 days). }{\scriptsize\par}
\end{table}

\pagebreak

The results are summarized in Table \ref{table:buyside} and Table \ref{table:sellside}.
Except for $DI$, the effects of the other 3 LOB factors are very similar to their effects
in the nondirectional trades. Nonetheless, the effect of
$DI$ in sell trades is exactly the opposite of its effect in buy
trades, which is similar to the effect in the nondirectional trades.
It means that, when $DI$ plays a significant role, a
larger and positive $DI$ (when $v_{b}^{1}$ is much higher than $v_{a}^{1}$)
would induce a shorter duration for the next buy trade, while a larger
and negative $DI$ (when $v_{a}^{1}$ is much higher than $v_{b}^{1}$)
is going to prompt a shorter inter-trade duration between sell trades.
This finding supports our explanation before and further elucidates the mechanism of $DI$'s effect.
When the depth at best bid is much higher than the depth at best ask, traders (especially HFTs)
in the market would anticipate that the price will increase,
and hence, they would actively buy securities and build a position
to gain profit because of a potential price increase. Similarly, if the
depth at best ask is much higher than the depth at best bid, traders
would actively sell shares, as they predict that the price is
probably going to subsequently decrease.

\begin{table}[!h]
\caption{Summary of the results for the significance of LOB factors for sell trades.}
\medskip{}

\centerline{%
\begin{tabular}{c|cccc|cccc|cccc|cccc}
\hline 
\multirow{3}{*}{{\footnotesize{}Stock}} & \multicolumn{4}{c|}{{\footnotesize{}$DI$}} & \multicolumn{4}{c|}{{\footnotesize{}$PS$}} & \multicolumn{4}{c|}{{\footnotesize{}$TV$}} & \multicolumn{4}{c}{{\footnotesize{}$PM$}}\tabularnewline
\cline{2-17} 
 & \multicolumn{2}{c|}{{\scriptsize{}$\mathrm{P_{12}}$}} & \multicolumn{2}{c|}{{\scriptsize{}$\mathrm{P_{21}}$}} & \multicolumn{2}{c|}{{\scriptsize{}$\mathrm{P_{12}}$}} & \multicolumn{2}{c|}{{\scriptsize{}$\mathrm{P_{21}}$}} & \multicolumn{2}{c|}{{\scriptsize{}$\mathrm{P_{12}}$}} & \multicolumn{2}{c|}{{\scriptsize{}$\mathrm{P_{21}}$}} & \multicolumn{2}{c|}{{\scriptsize{}$\mathrm{P_{12}}$}} & \multicolumn{2}{c}{{\scriptsize{}$\mathrm{P_{21}}$}}\tabularnewline
\cline{2-17} 
 & {\scriptsize{}sig(+)} & {\scriptsize{}sig(-)} & {\scriptsize{}sig(+)} & {\scriptsize{}sig(-)} & {\scriptsize{}sig(+)} & {\scriptsize{}sig(-)} & {\scriptsize{}sig(+)} & {\scriptsize{}sig(-)} & {\scriptsize{}sig(+)} & {\scriptsize{}sig(-)} & {\scriptsize{}sig(+)} & {\scriptsize{}sig(-)} & {\scriptsize{}sig(+)} & {\scriptsize{}sig(-)} & {\scriptsize{}sig(+)} & {\scriptsize{}sig(-)}\tabularnewline
\hline 
{\footnotesize{}AAPL} & {\scriptsize{}6} & {\scriptsize{}0} & {\scriptsize{}0} & {\scriptsize{}10} & {\scriptsize{}19} & {\scriptsize{}0} & {\scriptsize{}0} & {\scriptsize{}21} & {\scriptsize{}0} & {\scriptsize{}12} & {\scriptsize{}21} & {\scriptsize{}0} & {\scriptsize{}21} & {\scriptsize{}0} & {\scriptsize{}1} & {\scriptsize{}8}\tabularnewline
{\footnotesize{}ALXN} & {\scriptsize{}0} & {\scriptsize{}2} & {\scriptsize{}3} & {\scriptsize{}3} & {\scriptsize{}4} & {\scriptsize{}2} & {\scriptsize{}0} & {\scriptsize{}14} & {\scriptsize{}1} & {\scriptsize{}6} & {\scriptsize{}5} & {\scriptsize{}0} & {\scriptsize{}21} & {\scriptsize{}0} & {\scriptsize{}0} & {\scriptsize{}21}\tabularnewline
{\footnotesize{}AMZN} & {\scriptsize{}2} & {\scriptsize{}1} & {\scriptsize{}0} & {\scriptsize{}6} & {\scriptsize{}11} & {\scriptsize{}0} & {\scriptsize{}0} & {\scriptsize{}20} & {\scriptsize{}1} & {\scriptsize{}5} & {\scriptsize{}12} & {\scriptsize{}0} & {\scriptsize{}21} & {\scriptsize{}0} & {\scriptsize{}0} & {\scriptsize{}21}\tabularnewline
{\footnotesize{}BIDU} & {\scriptsize{}1} & {\scriptsize{}0} & {\scriptsize{}1} & {\scriptsize{}5} & {\scriptsize{}8} & {\scriptsize{}2} & {\scriptsize{}0} & {\scriptsize{}14} & {\scriptsize{}0} & {\scriptsize{}5} & {\scriptsize{}9} & {\scriptsize{}0} & {\scriptsize{}21} & {\scriptsize{}0} & {\scriptsize{}0} & {\scriptsize{}21}\tabularnewline
{\footnotesize{}BMRN} & {\scriptsize{}0} & {\scriptsize{}0} & {\scriptsize{}1} & {\scriptsize{}3} & {\scriptsize{}10} & {\scriptsize{}1} & {\scriptsize{}0} & {\scriptsize{}16} & {\scriptsize{}1} & {\scriptsize{}2} & {\scriptsize{}5} & {\scriptsize{}0} & {\scriptsize{}14} & {\scriptsize{}0} & {\scriptsize{}0} & {\scriptsize{}12}\tabularnewline
{\footnotesize{}CELG} & {\scriptsize{}5} & {\scriptsize{}0} & {\scriptsize{}0} & {\scriptsize{}19} & {\scriptsize{}2} & {\scriptsize{}3} & {\scriptsize{}2} & {\scriptsize{}9} & {\scriptsize{}1} & {\scriptsize{}6} & {\scriptsize{}14} & {\scriptsize{}0} & {\scriptsize{}21} & {\scriptsize{}0} & {\scriptsize{}0} & {\scriptsize{}21}\tabularnewline
{\footnotesize{}CERN} & {\scriptsize{}1} & {\scriptsize{}1} & {\scriptsize{}0} & {\scriptsize{}4} & {\scriptsize{}5} & {\scriptsize{}1} & {\scriptsize{}0} & {\scriptsize{}14} & {\scriptsize{}0} & {\scriptsize{}3} & {\scriptsize{}5} & {\scriptsize{}0} & {\scriptsize{}21} & {\scriptsize{}0} & {\scriptsize{}0} & {\scriptsize{}18}\tabularnewline
{\footnotesize{}CMCSA} & {\scriptsize{}11} & {\scriptsize{}1} & {\scriptsize{}0} & {\scriptsize{}17} & {\scriptsize{}5} & {\scriptsize{}0} & {\scriptsize{}3} & {\scriptsize{}1} & {\scriptsize{}1} & {\scriptsize{}3} & {\scriptsize{}20} & {\scriptsize{}0} & {\scriptsize{}12} & {\scriptsize{}1} & {\scriptsize{}0} & {\scriptsize{}6}\tabularnewline
{\footnotesize{}COST} & {\scriptsize{}1} & {\scriptsize{}1} & {\scriptsize{}0} & {\scriptsize{}2} & {\scriptsize{}1} & {\scriptsize{}2} & {\scriptsize{}1} & {\scriptsize{}10} & {\scriptsize{}1} & {\scriptsize{}6} & {\scriptsize{}18} & {\scriptsize{}0} & {\scriptsize{}20} & {\scriptsize{}0} & {\scriptsize{}0} & {\scriptsize{}19}\tabularnewline
{\footnotesize{}DISCA} & {\scriptsize{}1} & {\scriptsize{}0} & {\scriptsize{}0} & {\scriptsize{}4} & {\scriptsize{}1} & {\scriptsize{}4} & {\scriptsize{}0} & {\scriptsize{}7} & {\scriptsize{}0} & {\scriptsize{}2} & {\scriptsize{}6} & {\scriptsize{}0} & {\scriptsize{}21} & {\scriptsize{}0} & {\scriptsize{}0} & {\scriptsize{}19}\tabularnewline
{\footnotesize{}EBAY} & {\scriptsize{}2} & {\scriptsize{}4} & {\scriptsize{}2} & {\scriptsize{}7} & {\scriptsize{}11} & {\scriptsize{}1} & {\scriptsize{}0} & {\scriptsize{}18} & {\scriptsize{}0} & {\scriptsize{}6} & {\scriptsize{}10} & {\scriptsize{}1} & {\scriptsize{}20} & {\scriptsize{}0} & {\scriptsize{}0} & {\scriptsize{}12}\tabularnewline
{\footnotesize{}FB} & {\scriptsize{}4} & {\scriptsize{}2} & {\scriptsize{}0} & {\scriptsize{}14} & {\scriptsize{}11} & {\scriptsize{}2} & {\scriptsize{}0} & {\scriptsize{}14} & {\scriptsize{}4} & {\scriptsize{}0} & {\scriptsize{}21} & {\scriptsize{}0} & {\scriptsize{}21} & {\scriptsize{}0} & {\scriptsize{}1} & {\scriptsize{}4}\tabularnewline
{\footnotesize{}GOOG} & {\scriptsize{}5} & {\scriptsize{}0} & {\scriptsize{}0} & {\scriptsize{}8} & {\scriptsize{}19} & {\scriptsize{}0} & {\scriptsize{}0} & {\scriptsize{}21} & {\scriptsize{}0} & {\scriptsize{}3} & {\scriptsize{}18} & {\scriptsize{}0} & {\scriptsize{}21} & {\scriptsize{}0} & {\scriptsize{}0} & {\scriptsize{}17}\tabularnewline
{\footnotesize{}INTC} & {\scriptsize{}18} & {\scriptsize{}0} & {\scriptsize{}0} & {\scriptsize{}18} & {\scriptsize{}1} & {\scriptsize{}5} & {\scriptsize{}5} & {\scriptsize{}1} & {\scriptsize{}0} & {\scriptsize{}2} & {\scriptsize{}19} & {\scriptsize{}0} & {\scriptsize{}4} & {\scriptsize{}3} & {\scriptsize{}0} & {\scriptsize{}0}\tabularnewline
{\footnotesize{}ISRG} & {\scriptsize{}0} & {\scriptsize{}3} & {\scriptsize{}1} & {\scriptsize{}3} & {\scriptsize{}11} & {\scriptsize{}2} & {\scriptsize{}0} & {\scriptsize{}17} & {\scriptsize{}2} & {\scriptsize{}0} & {\scriptsize{}5} & {\scriptsize{}0} & {\scriptsize{}17} & {\scriptsize{}0} & {\scriptsize{}0} & {\scriptsize{}5}\tabularnewline
{\footnotesize{}KLAC} & {\scriptsize{}2} & {\scriptsize{}3} & {\scriptsize{}1} & {\scriptsize{}4} & {\scriptsize{}1} & {\scriptsize{}2} & {\scriptsize{}0} & {\scriptsize{}9} & {\scriptsize{}2} & {\scriptsize{}4} & {\scriptsize{}6} & {\scriptsize{}0} & {\scriptsize{}19} & {\scriptsize{}0} & {\scriptsize{}0} & {\scriptsize{}17}\tabularnewline
{\footnotesize{}MAR} & {\scriptsize{}6} & {\scriptsize{}0} & {\scriptsize{}0} & {\scriptsize{}7} & {\scriptsize{}1} & {\scriptsize{}5} & {\scriptsize{}0} & {\scriptsize{}5} & {\scriptsize{}0} & {\scriptsize{}3} & {\scriptsize{}14} & {\scriptsize{}0} & {\scriptsize{}14} & {\scriptsize{}0} & {\scriptsize{}1} & {\scriptsize{}5}\tabularnewline
{\footnotesize{}MSFT} & {\scriptsize{}20} & {\scriptsize{}0} & {\scriptsize{}0} & {\scriptsize{}19} & {\scriptsize{}7} & {\scriptsize{}4} & {\scriptsize{}11} & {\scriptsize{}0} & {\scriptsize{}0} & {\scriptsize{}1} & {\scriptsize{}20} & {\scriptsize{}0} & {\scriptsize{}3} & {\scriptsize{}1} & {\scriptsize{}1} & {\scriptsize{}0}\tabularnewline
{\footnotesize{}NFLX} & {\scriptsize{}7} & {\scriptsize{}0} & {\scriptsize{}0} & {\scriptsize{}11} & {\scriptsize{}12} & {\scriptsize{}1} & {\scriptsize{}0} & {\scriptsize{}19} & {\scriptsize{}0} & {\scriptsize{}10} & {\scriptsize{}15} & {\scriptsize{}0} & {\scriptsize{}21} & {\scriptsize{}0} & {\scriptsize{}0} & {\scriptsize{}19}\tabularnewline
{\footnotesize{}QCOM} & {\scriptsize{}2} & {\scriptsize{}2} & {\scriptsize{}0} & {\scriptsize{}13} & {\scriptsize{}4} & {\scriptsize{}1} & {\scriptsize{}0} & {\scriptsize{}17} & {\scriptsize{}0} & {\scriptsize{}3} & {\scriptsize{}15} & {\scriptsize{}0} & {\scriptsize{}20} & {\scriptsize{}0} & {\scriptsize{}0} & {\scriptsize{}13}\tabularnewline
{\footnotesize{}REGN} & {\scriptsize{}12} & {\scriptsize{}0} & {\scriptsize{}0} & {\scriptsize{}10} & {\scriptsize{}14} & {\scriptsize{}1} & {\scriptsize{}0} & {\scriptsize{}18} & {\scriptsize{}2} & {\scriptsize{}1} & {\scriptsize{}6} & {\scriptsize{}1} & {\scriptsize{}18} & {\scriptsize{}0} & {\scriptsize{}0} & {\scriptsize{}14}\tabularnewline
{\footnotesize{}SBUX} & {\scriptsize{}5} & {\scriptsize{}1} & {\scriptsize{}0} & {\scriptsize{}5} & {\scriptsize{}5} & {\scriptsize{}3} & {\scriptsize{}0} & {\scriptsize{}8} & {\scriptsize{}0} & {\scriptsize{}7} & {\scriptsize{}19} & {\scriptsize{}0} & {\scriptsize{}17} & {\scriptsize{}0} & {\scriptsize{}0} & {\scriptsize{}13}\tabularnewline
{\footnotesize{}TXN} & {\scriptsize{}7} & {\scriptsize{}0} & {\scriptsize{}0} & {\scriptsize{}11} & {\scriptsize{}6} & {\scriptsize{}2} & {\scriptsize{}0} & {\scriptsize{}3} & {\scriptsize{}1} & {\scriptsize{}1} & {\scriptsize{}18} & {\scriptsize{}0} & {\scriptsize{}11} & {\scriptsize{}0} & {\scriptsize{}0} & {\scriptsize{}3}\tabularnewline
{\footnotesize{}VOD} & {\scriptsize{}3} & {\scriptsize{}0} & {\scriptsize{}0} & {\scriptsize{}8} & {\scriptsize{}1} & {\scriptsize{}0} & {\scriptsize{}1} & {\scriptsize{}0} & {\scriptsize{}2} & {\scriptsize{}0} & {\scriptsize{}8} & {\scriptsize{}0} & {\scriptsize{}7} & {\scriptsize{}0} & {\scriptsize{}0} & {\scriptsize{}0}\tabularnewline
{\footnotesize{}YHOO} & {\scriptsize{}15} & {\scriptsize{}0} & {\scriptsize{}0} & {\scriptsize{}16} & {\scriptsize{}7} & {\scriptsize{}1} & {\scriptsize{}5} & {\scriptsize{}0} & {\scriptsize{}1} & {\scriptsize{}1} & {\scriptsize{}18} & {\scriptsize{}0} & {\scriptsize{}3} & {\scriptsize{}2} & {\scriptsize{}0} & {\scriptsize{}0}\tabularnewline
\hline 
{\footnotesize{}Sum (\%)} & {\scriptsize{}25.9\%} & {\scriptsize{}4.0\%} & {\scriptsize{}1.7\%} & {\scriptsize{}43.2\%} & {\scriptsize{}33.7\%} & {\scriptsize{}8.6\%} & {\scriptsize{}5.3\%} & {\scriptsize{}52.6\%} & {\scriptsize{}3.8\%} & {\scriptsize{}17.5\%} & {\scriptsize{}62.3\%} & {\scriptsize{}0.4\%} & {\scriptsize{}77.9\%} & {\scriptsize{}1.3\%} & {\scriptsize{}0.8\%} & {\scriptsize{}54.9\%}\tabularnewline
\hline 
\end{tabular}}
\label{table:sellside}
\textit{\scriptsize{}Note}{\scriptsize{}: $\mathrm{P_{12}}$ represent
the regime switching from the short-duration regime to the long duration
regime, and $\mathrm{P_{21}}$ represent the regime switching from
the long regime to the short-duration regime. Significance is measured at the 5\%
level. Sig(+)/Sig(-) means that the estimated coefficient for
the factor is significantly positive/negative. Each cell shows the significance. 
The data is in the month of January 2013. We have 21 trading
days for each stock, so the upper limit for each cell is 21. The
last row records the percentage (\%) of the times that this factor appears
as significant in all 525 instances (25 stocks times 21 days). }{\scriptsize\par}
\end{table}

\pagebreak

\subsection{Analysis of $DI$ conditional on the tightness of $PS$}
As we have shown in Section 5, $DI$ is a significant factor for the regime-switching
of inter-trade durations mostly for the stocks that have a very tight price spread.
To further test the effect of $DI$ conditional on the tightness of $PS$,
we add one more factor to our regressions for the stocks whose $PS$ is not always tight.
To see it more clearly, we simply define a dummy variable named $IT$ ($Is$ $Tight$),
which equals 1 when $PS\leq 0.2$ (double of the minimum tick) and 0 when $PS>0.2$. Then we 
add one more factor, which is an interaction term $DI\times IT$, 
into the R.H.S. of equation \eqref{model.4lob}. The results are presented in Table \ref{table:DItimesIT},
which omits the report of $TV$ and $PM$, as their effects are almost the same as what they have in Table \ref{table:SigSummary}.
From the regression results, we can see that the factor $DI$ is basically insignificant for the regime switching of inter-trade durations, 
while in more than half of total instances the factor $DI\times IT$ is significantly negative for $P_{12}$ and significantly positive for $P_{21}$.
Hence, it supports our explanation before that the depth imbalance is informative and 
important to the dynamics of inter-trade durations only when the price spread is tight, i.e.,  
the greater the depth imbalance is, inter-trade durations will be more likely to stay in (or switch to) the short-duration regime.

\begin{table}[!h]
\caption{Summary of results for the significance of LOB factors in MF-RSD.}
\medskip{}

\centerline{%
\begin{tabular}{c|cccc|cccc|cccc}
\hline 
\multirow{3}{*}{{\footnotesize{}Stock}} & \multicolumn{4}{c|}{{\footnotesize{}$DI$}} & \multicolumn{4}{c|}{{\footnotesize{}$PS$}} & \multicolumn{4}{c}{{\footnotesize{}$DI\times IT$}}\tabularnewline
\cline{2-13} 
 & \multicolumn{2}{c|}{{\scriptsize{}$\mathrm{P_{12}}$}} & \multicolumn{2}{c|}{{\scriptsize{}$\mathrm{P_{21}}$}} & \multicolumn{2}{c|}{{\scriptsize{}$\mathrm{P_{12}}$}} & \multicolumn{2}{c|}{{\scriptsize{}$\mathrm{P_{21}}$}} & \multicolumn{2}{c|}{{\scriptsize{}$\mathrm{P_{12}}$}} & \multicolumn{2}{c}{{\scriptsize{}$\mathrm{P_{21}}$}}\tabularnewline
\cline{2-13} 
 & {\scriptsize{}sig(+)} & {\scriptsize{}sig(-)} & {\scriptsize{}sig(+)} & {\scriptsize{}sig(-)} & {\scriptsize{}sig(+)} & {\scriptsize{}sig(-)} & {\scriptsize{}sig(+)} & {\scriptsize{}sig(-)} & {\scriptsize{}sig(+)} & {\scriptsize{}sig(-)} & {\scriptsize{}sig(+)} & {\scriptsize{}sig(-)}\tabularnewline
\hline 
{\footnotesize{}AAPL} & {\scriptsize{}0} & {\scriptsize{}13} & {\scriptsize{}1} & {\scriptsize{}4} & {\scriptsize{}20} & {\scriptsize{}0} & {\scriptsize{}0} & {\scriptsize{}21} & {\scriptsize{}0} & {\scriptsize{}19} & {\scriptsize{}18} & {\scriptsize{}0}\tabularnewline
{\footnotesize{}ALXN} & {\scriptsize{}4} & {\scriptsize{}0} & {\scriptsize{}1} & {\scriptsize{}5} & {\scriptsize{}5} & {\scriptsize{}1} & {\scriptsize{}0} & {\scriptsize{}15} & {\scriptsize{}0} & {\scriptsize{}15} & {\scriptsize{}15} & {\scriptsize{}1}\tabularnewline
{\footnotesize{}AMZN} & {\scriptsize{}1} & {\scriptsize{}0} & {\scriptsize{}3} & {\scriptsize{}0} & {\scriptsize{}13} & {\scriptsize{}0} & {\scriptsize{}0} & {\scriptsize{}21} & {\scriptsize{}0} & {\scriptsize{}14} & {\scriptsize{}10} & {\scriptsize{}0}\tabularnewline
{\footnotesize{}BIDU} & {\scriptsize{}3} & {\scriptsize{}0} & {\scriptsize{}1} & {\scriptsize{}1} & {\scriptsize{}10} & {\scriptsize{}2} & {\scriptsize{}0} & {\scriptsize{}16} & {\scriptsize{}0} & {\scriptsize{}15} & {\scriptsize{}18} & {\scriptsize{}0}\tabularnewline
{\footnotesize{}BMRN} & {\scriptsize{}0} & {\scriptsize{}0} & {\scriptsize{}1} & {\scriptsize{}2} & {\scriptsize{}6} & {\scriptsize{}1} & {\scriptsize{}0} & {\scriptsize{}17} & {\scriptsize{}2} & {\scriptsize{}7} & {\scriptsize{}12} & {\scriptsize{}0}\tabularnewline
{\footnotesize{}CELG} & {\scriptsize{}1} & {\scriptsize{}2} & {\scriptsize{}3} & {\scriptsize{}0} & {\scriptsize{}0} & {\scriptsize{}2} & {\scriptsize{}0} & {\scriptsize{}12} & {\scriptsize{}1} & {\scriptsize{}8} & {\scriptsize{}5} & {\scriptsize{}3}\tabularnewline
{\footnotesize{}CERN} & {\scriptsize{}1} & {\scriptsize{}0} & {\scriptsize{}0} & {\scriptsize{}2} & {\scriptsize{}6} & {\scriptsize{}2} & {\scriptsize{}0} & {\scriptsize{}9} & {\scriptsize{}1} & {\scriptsize{}10} & {\scriptsize{}14} & {\scriptsize{}0}\tabularnewline
{\footnotesize{}COST} & {\scriptsize{}3} & {\scriptsize{}1} & {\scriptsize{}1} & {\scriptsize{}1} & {\scriptsize{}0} & {\scriptsize{}2} & {\scriptsize{}0} & {\scriptsize{}8} & {\scriptsize{}1} & {\scriptsize{}8} & {\scriptsize{}5} & {\scriptsize{}2}\tabularnewline
{\footnotesize{}DISCA} & {\scriptsize{}0} & {\scriptsize{}0} & {\scriptsize{}0} & {\scriptsize{}1} & {\scriptsize{}1} & {\scriptsize{}2} & {\scriptsize{}0} & {\scriptsize{}12} & {\scriptsize{}1} & {\scriptsize{}7} & {\scriptsize{}6} & {\scriptsize{}5}\tabularnewline
{\footnotesize{}NFLX} & {\scriptsize{}0} & {\scriptsize{}6} & {\scriptsize{}4} & {\scriptsize{}0} & {\scriptsize{}13} & {\scriptsize{}1} & {\scriptsize{}0} & {\scriptsize{}21} & {\scriptsize{}0} & {\scriptsize{}20} & {\scriptsize{}10} & {\scriptsize{}1}\tabularnewline
\hline 
{\footnotesize{}Sum (\%)} & {\scriptsize{}6.2\%} & {\scriptsize{}10.5\%} & {\scriptsize{}7.1\%} & {\scriptsize{}7.6\%} & {\scriptsize{}35.2\%} & {\scriptsize{}6.2\%} & {\scriptsize{}0.0\%} & {\scriptsize{}72.4\%} & {\scriptsize{}2.9\%} & {\scriptsize{}58.6\%} & {\scriptsize{}53.8\%} & {\scriptsize{}5.7\%}\tabularnewline
\hline 
\end{tabular}}
\label{table:DItimesIT}
\textit{\scriptsize{}Note}{\scriptsize{}: $\mathrm{P_{12}}$ represent
the regime switching from the short-duration regime to the long duration
regime, and $\mathrm{P_{21}}$ represent the regime switching from
the long regime to the short-duration regime. Significance is measured at the 5\%
level. Sig(+)/Sig(-) means that the estimated coefficient for
the factor is significantly positive/negative. Each cell shows the significance. 
The data is in the month of January 2013. We have 21 trading
days for each stock, so the upper limit for each cell is 21. The
last row records the percentage (\%) of the times that this factor appears
as significant in total 210 instances (10 stocks times 21 days). 
From the 25 NASDAQ stocks, we have selected 10 stocks whose $PS$ vary from being tight to being slack.
Here we exclude the stocks that mainly have $PS$ less than 0.2, i.e., 
$IT$ equal to 1 for most time, such as CMCSA, EBAY, FB and etc.
And we also exclude the stocks GOOG, ISRG and REGN
 as their $PS$ are basically greater than 0.2 and have  $IT=0$ for most time.}{\scriptsize\par}
\end{table}

\pagebreak

\subsection{Use the aggregated depth imbalance}
We have also used the aggregated depth imbalance between the ask-side and the sell-side instead of
$DI$ between the best ask and the best bid in the regression model \eqref{model.4lob}.
The aggregated depth imbalance is defined as $ADI=\frac{|v_a^{1-5}-v_b^{1-5}|}{|v_a^{1-5}+v_b^{1-5}|}$, where
$v_a^{1-5}$ ($v_b^{1-5}$ ) is the cumulative depth from the first to the fifth ask (bid) price.
Table \ref{table:ADI} shows the regression results.
We can see that, besides the similarity effects of the other 3 factors, 
$ADI$ is less commonly significant for the regime-switching probabilities than $DI$. Even for the stocks which have very tight 
price spread, $ADI$ doesn't show a very common effect on the regime switchings.
Therefore, it demonstrates that $ADI$ is less informative 
and traders may not treat it as an effective indicator for their trading strategies, compared with $DI$.

\begin{table}[!h]
\caption{Summary of results for the significance of LOB factors in MF-RSD.}
\medskip{}

\centerline{%
\begin{tabular}{c|cccc|cccc|cccc|cccc}
\hline 
\multirow{3}{*}{{\footnotesize{}Stock}} & \multicolumn{4}{c|}{{\footnotesize{}$ADI$}} & \multicolumn{4}{c|}{{\footnotesize{}$PS$}} & \multicolumn{4}{c|}{{\footnotesize{}$TV$}} & \multicolumn{4}{c}{{\footnotesize{}$PM$}}\tabularnewline
\cline{2-17} 
 & \multicolumn{2}{c|}{{\scriptsize{}$\mathrm{P_{12}}$}} & \multicolumn{2}{c|}{{\scriptsize{}$\mathrm{P_{21}}$}} & \multicolumn{2}{c|}{{\scriptsize{}$\mathrm{P_{12}}$}} & \multicolumn{2}{c|}{{\scriptsize{}$\mathrm{P_{21}}$}} & \multicolumn{2}{c|}{{\scriptsize{}$\mathrm{P_{12}}$}} & \multicolumn{2}{c|}{{\scriptsize{}$\mathrm{P_{21}}$}} & \multicolumn{2}{c|}{{\scriptsize{}$\mathrm{P_{12}}$}} & \multicolumn{2}{c}{{\scriptsize{}$\mathrm{P_{21}}$}}\tabularnewline
\cline{2-17} 
 & {\scriptsize{}sig(+)} & {\scriptsize{}sig(-)} & {\scriptsize{}sig(+)} & {\scriptsize{}sig(-)} & {\scriptsize{}sig(+)} & {\scriptsize{}sig(-)} & {\scriptsize{}sig(+)} & {\scriptsize{}sig(-)} & {\scriptsize{}sig(+)} & {\scriptsize{}sig(-)} & {\scriptsize{}sig(+)} & {\scriptsize{}sig(-)} & {\scriptsize{}sig(+)} & {\scriptsize{}sig(-)} & {\scriptsize{}sig(+)} & {\scriptsize{}sig(-)}\tabularnewline
\hline 
{\footnotesize{}AAPL} & {\scriptsize{}0} & {\scriptsize{}9} & {\scriptsize{}4} & {\scriptsize{}0} & {\scriptsize{}21} & {\scriptsize{}0} & {\scriptsize{}0} & {\scriptsize{}21} & {\scriptsize{}0} & {\scriptsize{}11} & {\scriptsize{}21} & {\scriptsize{}0} & {\scriptsize{}21} & {\scriptsize{}0} & {\scriptsize{}9} & {\scriptsize{}0}\tabularnewline
{\footnotesize{}ALXN} & {\scriptsize{}3} & {\scriptsize{}2} & {\scriptsize{}0} & {\scriptsize{}1} & {\scriptsize{}6} & {\scriptsize{}0} & {\scriptsize{}0} & {\scriptsize{}20} & {\scriptsize{}0} & {\scriptsize{}10} & {\scriptsize{}10} & {\scriptsize{}1} & {\scriptsize{}20} & {\scriptsize{}0} & {\scriptsize{}0} & {\scriptsize{}21}\tabularnewline
{\footnotesize{}AMZN} & {\scriptsize{}1} & {\scriptsize{}7} & {\scriptsize{}3} & {\scriptsize{}1} & {\scriptsize{}17} & {\scriptsize{}0} & {\scriptsize{}0} & {\scriptsize{}21} & {\scriptsize{}0} & {\scriptsize{}8} & {\scriptsize{}15} & {\scriptsize{}0} & {\scriptsize{}21} & {\scriptsize{}0} & {\scriptsize{}0} & {\scriptsize{}21}\tabularnewline
{\footnotesize{}BIDU} & {\scriptsize{}2} & {\scriptsize{}2} & {\scriptsize{}0} & {\scriptsize{}1} & {\scriptsize{}13} & {\scriptsize{}0} & {\scriptsize{}0} & {\scriptsize{}19} & {\scriptsize{}0} & {\scriptsize{}5} & {\scriptsize{}16} & {\scriptsize{}0} & {\scriptsize{}21} & {\scriptsize{}0} & {\scriptsize{}0} & {\scriptsize{}21}\tabularnewline
{\footnotesize{}BMRN} & {\scriptsize{}1} & {\scriptsize{}2} & {\scriptsize{}1} & {\scriptsize{}2} & {\scriptsize{}8} & {\scriptsize{}1} & {\scriptsize{}0} & {\scriptsize{}21} & {\scriptsize{}0} & {\scriptsize{}6} & {\scriptsize{}12} & {\scriptsize{}0} & {\scriptsize{}14} & {\scriptsize{}0} & {\scriptsize{}0} & {\scriptsize{}12}\tabularnewline
{\footnotesize{}CELG} & {\scriptsize{}1} & {\scriptsize{}0} & {\scriptsize{}4} & {\scriptsize{}0} & {\scriptsize{}4} & {\scriptsize{}2} & {\scriptsize{}0} & {\scriptsize{}15} & {\scriptsize{}0} & {\scriptsize{}12} & {\scriptsize{}20} & {\scriptsize{}0} & {\scriptsize{}21} & {\scriptsize{}0} & {\scriptsize{}0} & {\scriptsize{}21}\tabularnewline
{\footnotesize{}CERN} & {\scriptsize{}2} & {\scriptsize{}0} & {\scriptsize{}0} & {\scriptsize{}0} & {\scriptsize{}8} & {\scriptsize{}0} & {\scriptsize{}0} & {\scriptsize{}17} & {\scriptsize{}0} & {\scriptsize{}4} & {\scriptsize{}9} & {\scriptsize{}0} & {\scriptsize{}21} & {\scriptsize{}0} & {\scriptsize{}0} & {\scriptsize{}20}\tabularnewline
{\footnotesize{}CMCSA} & {\scriptsize{}0} & {\scriptsize{}13} & {\scriptsize{}8} & {\scriptsize{}1} & {\scriptsize{}3} & {\scriptsize{}4} & {\scriptsize{}0} & {\scriptsize{}7} & {\scriptsize{}0} & {\scriptsize{}12} & {\scriptsize{}20} & {\scriptsize{}0} & {\scriptsize{}12} & {\scriptsize{}0} & {\scriptsize{}4} & {\scriptsize{}0}\tabularnewline
{\footnotesize{}COST} & {\scriptsize{}1} & {\scriptsize{}1} & {\scriptsize{}0} & {\scriptsize{}0} & {\scriptsize{}4} & {\scriptsize{}4} & {\scriptsize{}0} & {\scriptsize{}14} & {\scriptsize{}0} & {\scriptsize{}12} & {\scriptsize{}20} & {\scriptsize{}0} & {\scriptsize{}21} & {\scriptsize{}0} & {\scriptsize{}0} & {\scriptsize{}21}\tabularnewline
{\footnotesize{}DISCA} & {\scriptsize{}0} & {\scriptsize{}5} & {\scriptsize{}2} & {\scriptsize{}0} & {\scriptsize{}2} & {\scriptsize{}1} & {\scriptsize{}0} & {\scriptsize{}17} & {\scriptsize{}0} & {\scriptsize{}6} & {\scriptsize{}17} & {\scriptsize{}0} & {\scriptsize{}19} & {\scriptsize{}0} & {\scriptsize{}0} & {\scriptsize{}19}\tabularnewline
{\footnotesize{}EBAY} & {\scriptsize{}0} & {\scriptsize{}10} & {\scriptsize{}0} & {\scriptsize{}3} & {\scriptsize{}11} & {\scriptsize{}0} & {\scriptsize{}0} & {\scriptsize{}21} & {\scriptsize{}0} & {\scriptsize{}16} & {\scriptsize{}20} & {\scriptsize{}0} & {\scriptsize{}20} & {\scriptsize{}0} & {\scriptsize{}0} & {\scriptsize{}3}\tabularnewline
{\footnotesize{}FB} & {\scriptsize{}4} & {\scriptsize{}5} & {\scriptsize{}3} & {\scriptsize{}3} & {\scriptsize{}16} & {\scriptsize{}2} & {\scriptsize{}0} & {\scriptsize{}21} & {\scriptsize{}0} & {\scriptsize{}16} & {\scriptsize{}21} & {\scriptsize{}0} & {\scriptsize{}20} & {\scriptsize{}0} & {\scriptsize{}9} & {\scriptsize{}0}\tabularnewline
{\footnotesize{}GOOG} & {\scriptsize{}0} & {\scriptsize{}3} & {\scriptsize{}4} & {\scriptsize{}0} & {\scriptsize{}21} & {\scriptsize{}0} & {\scriptsize{}0} & {\scriptsize{}21} & {\scriptsize{}0} & {\scriptsize{}12} & {\scriptsize{}20} & {\scriptsize{}0} & {\scriptsize{}21} & {\scriptsize{}0} & {\scriptsize{}0} & {\scriptsize{}18}\tabularnewline
{\footnotesize{}INTC} & {\scriptsize{}0} & {\scriptsize{}12} & {\scriptsize{}6} & {\scriptsize{}0} & {\scriptsize{}3} & {\scriptsize{}7} & {\scriptsize{}1} & {\scriptsize{}5} & {\scriptsize{}0} & {\scriptsize{}11} & {\scriptsize{}20} & {\scriptsize{}0} & {\scriptsize{}3} & {\scriptsize{}5} & {\scriptsize{}5} & {\scriptsize{}0}\tabularnewline
{\footnotesize{}ISRG} & {\scriptsize{}2} & {\scriptsize{}2} & {\scriptsize{}3} & {\scriptsize{}0} & {\scriptsize{}10} & {\scriptsize{}3} & {\scriptsize{}0} & {\scriptsize{}20} & {\scriptsize{}5} & {\scriptsize{}0} & {\scriptsize{}11} & {\scriptsize{}0} & {\scriptsize{}18} & {\scriptsize{}0} & {\scriptsize{}0} & {\scriptsize{}13}\tabularnewline
{\footnotesize{}KLAC} & {\scriptsize{}1} & {\scriptsize{}2} & {\scriptsize{}1} & {\scriptsize{}0} & {\scriptsize{}5} & {\scriptsize{}3} & {\scriptsize{}0} & {\scriptsize{}17} & {\scriptsize{}0} & {\scriptsize{}8} & {\scriptsize{}15} & {\scriptsize{}0} & {\scriptsize{}12} & {\scriptsize{}0} & {\scriptsize{}0} & {\scriptsize{}11}\tabularnewline
{\footnotesize{}MAR} & {\scriptsize{}1} & {\scriptsize{}3} & {\scriptsize{}1} & {\scriptsize{}0} & {\scriptsize{}2} & {\scriptsize{}4} & {\scriptsize{}0} & {\scriptsize{}12} & {\scriptsize{}0} & {\scriptsize{}6} & {\scriptsize{}17} & {\scriptsize{}0} & {\scriptsize{}11} & {\scriptsize{}0} & {\scriptsize{}1} & {\scriptsize{}5}\tabularnewline
{\footnotesize{}MSFT} & {\scriptsize{}0} & {\scriptsize{}15} & {\scriptsize{}9} & {\scriptsize{}0} & {\scriptsize{}5} & {\scriptsize{}7} & {\scriptsize{}0} & {\scriptsize{}5} & {\scriptsize{}0} & {\scriptsize{}7} & {\scriptsize{}20} & {\scriptsize{}0} & {\scriptsize{}2} & {\scriptsize{}2} & {\scriptsize{}5} & {\scriptsize{}0}\tabularnewline
{\footnotesize{}NFLX} & {\scriptsize{}1} & {\scriptsize{}2} & {\scriptsize{}1} & {\scriptsize{}1} & {\scriptsize{}14} & {\scriptsize{}1} & {\scriptsize{}0} & {\scriptsize{}21} & {\scriptsize{}0} & {\scriptsize{}12} & {\scriptsize{}19} & {\scriptsize{}0} & {\scriptsize{}21} & {\scriptsize{}0} & {\scriptsize{}0} & {\scriptsize{}18}\tabularnewline
{\footnotesize{}QCOM} & {\scriptsize{}0} & {\scriptsize{}8} & {\scriptsize{}1} & {\scriptsize{}2} & {\scriptsize{}6} & {\scriptsize{}1} & {\scriptsize{}0} & {\scriptsize{}20} & {\scriptsize{}0} & {\scriptsize{}17} & {\scriptsize{}21} & {\scriptsize{}0} & {\scriptsize{}16} & {\scriptsize{}0} & {\scriptsize{}3} & {\scriptsize{}6}\tabularnewline
{\footnotesize{}REGN} & {\scriptsize{}1} & {\scriptsize{}2} & {\scriptsize{}2} & {\scriptsize{}1} & {\scriptsize{}17} & {\scriptsize{}0} & {\scriptsize{}0} & {\scriptsize{}21} & {\scriptsize{}3} & {\scriptsize{}0} & {\scriptsize{}8} & {\scriptsize{}0} & {\scriptsize{}21} & {\scriptsize{}0} & {\scriptsize{}0} & {\scriptsize{}19}\tabularnewline
{\footnotesize{}SBUX} & {\scriptsize{}0} & {\scriptsize{}5} & {\scriptsize{}6} & {\scriptsize{}0} & {\scriptsize{}2} & {\scriptsize{}4} & {\scriptsize{}0} & {\scriptsize{}15} & {\scriptsize{}0} & {\scriptsize{}17} & {\scriptsize{}21} & {\scriptsize{}0} & {\scriptsize{}12} & {\scriptsize{}0} & {\scriptsize{}0} & {\scriptsize{}3}\tabularnewline
{\footnotesize{}TXN} & {\scriptsize{}1} & {\scriptsize{}7} & {\scriptsize{}1} & {\scriptsize{}0} & {\scriptsize{}4} & {\scriptsize{}1} & {\scriptsize{}0} & {\scriptsize{}8} & {\scriptsize{}0} & {\scriptsize{}8} & {\scriptsize{}21} & {\scriptsize{}0} & {\scriptsize{}10} & {\scriptsize{}0} & {\scriptsize{}1} & {\scriptsize{}1}\tabularnewline
{\footnotesize{}VOD} & {\scriptsize{}0} & {\scriptsize{}6} & {\scriptsize{}6} & {\scriptsize{}4} & {\scriptsize{}6} & {\scriptsize{}0} & {\scriptsize{}0} & {\scriptsize{}5} & {\scriptsize{}2} & {\scriptsize{}0} & {\scriptsize{}17} & {\scriptsize{}0} & {\scriptsize{}18} & {\scriptsize{}0} & {\scriptsize{}1} & {\scriptsize{}0}\tabularnewline
{\footnotesize{}YHOO} & {\scriptsize{}0} & {\scriptsize{}10} & {\scriptsize{}2} & {\scriptsize{}2} & {\scriptsize{}7} & {\scriptsize{}3} & {\scriptsize{}1} & {\scriptsize{}5} & {\scriptsize{}0} & {\scriptsize{}8} & {\scriptsize{}21} & {\scriptsize{}0} & {\scriptsize{}3} & {\scriptsize{}3} & {\scriptsize{}6} & {\scriptsize{}0}\tabularnewline
\hline 
{\footnotesize{}Sum (\%)} & {\scriptsize{}4.2\%} & {\scriptsize{}25.3\%} & {\scriptsize{}13.0\%} & {\scriptsize{}4.2\%} & {\scriptsize{}41.0\%} & {\scriptsize{}9.1\%} & {\scriptsize{}0.4\%} & {\scriptsize{}74.1\%} & {\scriptsize{}1.9\%} & {\scriptsize{}42.7\%} & {\scriptsize{}82.3\%} & {\scriptsize{}0.2\%} & {\scriptsize{}76.0\%} & {\scriptsize{}1.9\%} & {\scriptsize{}8.4\%} & {\scriptsize{}48.2\%}\tabularnewline
\hline 
\end{tabular}}
\label{table:ADI}
\textit{\scriptsize{}Note}{\scriptsize{}: $\mathrm{P_{12}}$ represent
the regime switching from the short-duration regime to the long duration
regime, and $\mathrm{P_{21}}$ represent the regime switching from
the long regime to the short-duration regime. Significance is measured at the 5\%
level. Sig(+)/Sig(-) means that the estimated coefficient for
the factor is significantly positive/negative. Each cell shows the significance. 
The data is in the month of January 2013. We have 21 trading
days for each stock, so the upper limit for each cell is 21. The
last row records the percentage (\%) of the times that this factor appears
as significant in all 525 instances (25 stocks times 21 days). }{\scriptsize\par}
\end{table}
\pagebreak

\subsection{Use a dummy variable for the increase in price spread}
In this part, we use a dummy variable for the increase in price spread instead of $PM$ in the regression model \eqref{model.4lob}.
We compare the $PS$ at $i-1$ and $i$ to see whether it has increased, and the variable
is defined as $PSI_i=1$ if $PS_i>PS_{i-1}$ and 0 otherwise. It may have some overlap with $PM$,
but some increases may result from order cancellations between trades,
rather than the elimination of the existing best ask or the best bid by an incoming trade, which induces the mid-price movement.
From the results shown in Table \ref{table:PSI}, we find that in more than 70\% of total instances, the effect of $PSI$
is significantly positive for $P_{12}$, which is almost the same as the effect of $PM$.
So it supports our explanation that the increase in $PS$ is a driving force 
for inter-trade duration switching from the short-duration regime to the long-duration regime when $PM=1$.
However, $PSI$ doesn't show a common significant impact on $P_{21}$, which means that when the market is in the 
long-duration regime in which HFTs mainly adopt a passive market making strategy, an increase of price spread seems 
don't have a significant influence to them.

\begin{table}[!h]
\caption{Summary of results for the significance of LOB factors in MF-RSD. }
\medskip{}

\centerline{%
\begin{tabular}{c|cccc|cccc|cccc|cccc}
\hline 
\multirow{3}{*}{{\footnotesize{}Stock}} & \multicolumn{4}{c|}{{\footnotesize{}$DI$}} & \multicolumn{4}{c|}{{\footnotesize{}$PS$}} & \multicolumn{4}{c|}{{\footnotesize{}$TV$}} & \multicolumn{4}{c}{{\footnotesize{}$PSI$}}\tabularnewline
\cline{2-17} 
 & \multicolumn{2}{c|}{{\scriptsize{}$\mathrm{P_{12}}$}} & \multicolumn{2}{c|}{{\scriptsize{}$\mathrm{P_{21}}$}} & \multicolumn{2}{c|}{{\scriptsize{}$\mathrm{P_{12}}$}} & \multicolumn{2}{c|}{{\scriptsize{}$\mathrm{P_{21}}$}} & \multicolumn{2}{c|}{{\scriptsize{}$\mathrm{P_{12}}$}} & \multicolumn{2}{c|}{{\scriptsize{}$\mathrm{P_{21}}$}} & \multicolumn{2}{c|}{{\scriptsize{}$\mathrm{P_{12}}$}} & \multicolumn{2}{c}{{\scriptsize{}$\mathrm{P_{21}}$}}\tabularnewline
\cline{2-17} 
 & {\scriptsize{}sig(+)} & {\scriptsize{}sig(-)} & {\scriptsize{}sig(+)} & {\scriptsize{}sig(-)} & {\scriptsize{}sig(+)} & {\scriptsize{}sig(-)} & {\scriptsize{}sig(+)} & {\scriptsize{}sig(-)} & {\scriptsize{}sig(+)} & {\scriptsize{}sig(-)} & {\scriptsize{}sig(+)} & {\scriptsize{}sig(-)} & {\scriptsize{}sig(+)} & {\scriptsize{}sig(-)} & {\scriptsize{}sig(+)} & {\scriptsize{}sig(-)}\tabularnewline
\hline 
{\footnotesize{}AAPL} & {\scriptsize{}0} & {\scriptsize{}14} & {\scriptsize{}1} & {\scriptsize{}2} & {\scriptsize{}21} & {\scriptsize{}0} & {\scriptsize{}0} & {\scriptsize{}21} & {\scriptsize{}0} & {\scriptsize{}12} & {\scriptsize{}21} & {\scriptsize{}0} & {\scriptsize{}21} & {\scriptsize{}0} & {\scriptsize{}1} & {\scriptsize{}2}\tabularnewline
{\footnotesize{}ALXN} & {\scriptsize{}3} & {\scriptsize{}1} & {\scriptsize{}2} & {\scriptsize{}6} & {\scriptsize{}5} & {\scriptsize{}0} & {\scriptsize{}0} & {\scriptsize{}19} & {\scriptsize{}0} & {\scriptsize{}11} & {\scriptsize{}13} & {\scriptsize{}0} & {\scriptsize{}21} & {\scriptsize{}0} & {\scriptsize{}0} & {\scriptsize{}16}\tabularnewline
{\footnotesize{}AMZN} & {\scriptsize{}0} & {\scriptsize{}3} & {\scriptsize{}2} & {\scriptsize{}0} & {\scriptsize{}16} & {\scriptsize{}0} & {\scriptsize{}0} & {\scriptsize{}21} & {\scriptsize{}0} & {\scriptsize{}11} & {\scriptsize{}18} & {\scriptsize{}0} & {\scriptsize{}21} & {\scriptsize{}0} & {\scriptsize{}0} & {\scriptsize{}16}\tabularnewline
{\footnotesize{}BIDU} & {\scriptsize{}1} & {\scriptsize{}0} & {\scriptsize{}1} & {\scriptsize{}0} & {\scriptsize{}12} & {\scriptsize{}1} & {\scriptsize{}0} & {\scriptsize{}18} & {\scriptsize{}0} & {\scriptsize{}4} & {\scriptsize{}19} & {\scriptsize{}0} & {\scriptsize{}21} & {\scriptsize{}0} & {\scriptsize{}0} & {\scriptsize{}17}\tabularnewline
{\footnotesize{}BMRN} & {\scriptsize{}0} & {\scriptsize{}0} & {\scriptsize{}3} & {\scriptsize{}0} & {\scriptsize{}8} & {\scriptsize{}1} & {\scriptsize{}0} & {\scriptsize{}20} & {\scriptsize{}0} & {\scriptsize{}6} & {\scriptsize{}13} & {\scriptsize{}0} & {\scriptsize{}17} & {\scriptsize{}0} & {\scriptsize{}0} & {\scriptsize{}5}\tabularnewline
{\footnotesize{}CELG} & {\scriptsize{}0} & {\scriptsize{}4} & {\scriptsize{}6} & {\scriptsize{}0} & {\scriptsize{}3} & {\scriptsize{}3} & {\scriptsize{}0} & {\scriptsize{}17} & {\scriptsize{}0} & {\scriptsize{}13} & {\scriptsize{}21} & {\scriptsize{}0} & {\scriptsize{}21} & {\scriptsize{}0} & {\scriptsize{}0} & {\scriptsize{}10}\tabularnewline
{\footnotesize{}CERN} & {\scriptsize{}1} & {\scriptsize{}0} & {\scriptsize{}1} & {\scriptsize{}1} & {\scriptsize{}7} & {\scriptsize{}0} & {\scriptsize{}0} & {\scriptsize{}19} & {\scriptsize{}0} & {\scriptsize{}4} & {\scriptsize{}12} & {\scriptsize{}0} & {\scriptsize{}21} & {\scriptsize{}0} & {\scriptsize{}0} & {\scriptsize{}9}\tabularnewline
{\footnotesize{}CMCSA} & {\scriptsize{}0} & {\scriptsize{}17} & {\scriptsize{}19} & {\scriptsize{}0} & {\scriptsize{}1} & {\scriptsize{}14} & {\scriptsize{}0} & {\scriptsize{}8} & {\scriptsize{}0} & {\scriptsize{}15} & {\scriptsize{}21} & {\scriptsize{}0} & {\scriptsize{}19} & {\scriptsize{}0} & {\scriptsize{}4} & {\scriptsize{}0}\tabularnewline
{\footnotesize{}COST} & {\scriptsize{}2} & {\scriptsize{}1} & {\scriptsize{}1} & {\scriptsize{}0} & {\scriptsize{}4} & {\scriptsize{}2} & {\scriptsize{}0} & {\scriptsize{}18} & {\scriptsize{}0} & {\scriptsize{}11} & {\scriptsize{}21} & {\scriptsize{}0} & {\scriptsize{}21} & {\scriptsize{}0} & {\scriptsize{}0} & {\scriptsize{}9}\tabularnewline
{\footnotesize{}DISCA} & {\scriptsize{}0} & {\scriptsize{}2} & {\scriptsize{}1} & {\scriptsize{}0} & {\scriptsize{}3} & {\scriptsize{}3} & {\scriptsize{}0} & {\scriptsize{}20} & {\scriptsize{}0} & {\scriptsize{}9} & {\scriptsize{}18} & {\scriptsize{}0} & {\scriptsize{}19} & {\scriptsize{}0} & {\scriptsize{}0} & {\scriptsize{}7}\tabularnewline
{\footnotesize{}EBAY} & {\scriptsize{}1} & {\scriptsize{}3} & {\scriptsize{}0} & {\scriptsize{}5} & {\scriptsize{}9} & {\scriptsize{}1} & {\scriptsize{}0} & {\scriptsize{}21} & {\scriptsize{}0} & {\scriptsize{}16} & {\scriptsize{}20} & {\scriptsize{}0} & {\scriptsize{}21} & {\scriptsize{}0} & {\scriptsize{}2} & {\scriptsize{}0}\tabularnewline
{\footnotesize{}FB} & {\scriptsize{}0} & {\scriptsize{}11} & {\scriptsize{}1} & {\scriptsize{}2} & {\scriptsize{}14} & {\scriptsize{}3} & {\scriptsize{}0} & {\scriptsize{}21} & {\scriptsize{}0} & {\scriptsize{}16} & {\scriptsize{}21} & {\scriptsize{}0} & {\scriptsize{}21} & {\scriptsize{}0} & {\scriptsize{}6} & {\scriptsize{}0}\tabularnewline
{\footnotesize{}GOOG} & {\scriptsize{}2} & {\scriptsize{}6} & {\scriptsize{}2} & {\scriptsize{}1} & {\scriptsize{}21} & {\scriptsize{}0} & {\scriptsize{}0} & {\scriptsize{}21} & {\scriptsize{}0} & {\scriptsize{}13} & {\scriptsize{}20} & {\scriptsize{}0} & {\scriptsize{}21} & {\scriptsize{}0} & {\scriptsize{}0} & {\scriptsize{}13}\tabularnewline
{\footnotesize{}INTC} & {\scriptsize{}0} & {\scriptsize{}21} & {\scriptsize{}18} & {\scriptsize{}0} & {\scriptsize{}0} & {\scriptsize{}17} & {\scriptsize{}2} & {\scriptsize{}2} & {\scriptsize{}0} & {\scriptsize{}13} & {\scriptsize{}21} & {\scriptsize{}0} & {\scriptsize{}4} & {\scriptsize{}2} & {\scriptsize{}3} & {\scriptsize{}0}\tabularnewline
{\footnotesize{}ISRG} & {\scriptsize{}1} & {\scriptsize{}1} & {\scriptsize{}0} & {\scriptsize{}3} & {\scriptsize{}10} & {\scriptsize{}2} & {\scriptsize{}0} & {\scriptsize{}19} & {\scriptsize{}7} & {\scriptsize{}0} & {\scriptsize{}12} & {\scriptsize{}0} & {\scriptsize{}21} & {\scriptsize{}0} & {\scriptsize{}0} & {\scriptsize{}5}\tabularnewline
{\footnotesize{}KLAC} & {\scriptsize{}0} & {\scriptsize{}5} & {\scriptsize{}2} & {\scriptsize{}1} & {\scriptsize{}4} & {\scriptsize{}4} & {\scriptsize{}0} & {\scriptsize{}19} & {\scriptsize{}0} & {\scriptsize{}9} & {\scriptsize{}15} & {\scriptsize{}0} & {\scriptsize{}16} & {\scriptsize{}0} & {\scriptsize{}0} & {\scriptsize{}0}\tabularnewline
{\footnotesize{}MAR} & {\scriptsize{}0} & {\scriptsize{}10} & {\scriptsize{}4} & {\scriptsize{}0} & {\scriptsize{}1} & {\scriptsize{}4} & {\scriptsize{}0} & {\scriptsize{}16} & {\scriptsize{}0} & {\scriptsize{}9} & {\scriptsize{}18} & {\scriptsize{}0} & {\scriptsize{}11} & {\scriptsize{}0} & {\scriptsize{}2} & {\scriptsize{}0}\tabularnewline
{\footnotesize{}MSFT} & {\scriptsize{}0} & {\scriptsize{}20} & {\scriptsize{}20} & {\scriptsize{}0} & {\scriptsize{}1} & {\scriptsize{}17} & {\scriptsize{}2} & {\scriptsize{}3} & {\scriptsize{}0} & {\scriptsize{}13} & {\scriptsize{}20} & {\scriptsize{}0} & {\scriptsize{}6} & {\scriptsize{}0} & {\scriptsize{}3} & {\scriptsize{}0}\tabularnewline
{\footnotesize{}NFLX} & {\scriptsize{}0} & {\scriptsize{}6} & {\scriptsize{}7} & {\scriptsize{}0} & {\scriptsize{}13} & {\scriptsize{}1} & {\scriptsize{}0} & {\scriptsize{}21} & {\scriptsize{}0} & {\scriptsize{}13} & {\scriptsize{}20} & {\scriptsize{}0} & {\scriptsize{}21} & {\scriptsize{}0} & {\scriptsize{}0} & {\scriptsize{}10}\tabularnewline
{\footnotesize{}QCOM} & {\scriptsize{}0} & {\scriptsize{}9} & {\scriptsize{}4} & {\scriptsize{}0} & {\scriptsize{}5} & {\scriptsize{}3} & {\scriptsize{}0} & {\scriptsize{}21} & {\scriptsize{}0} & {\scriptsize{}19} & {\scriptsize{}21} & {\scriptsize{}0} & {\scriptsize{}21} & {\scriptsize{}0} & {\scriptsize{}2} & {\scriptsize{}0}\tabularnewline
{\footnotesize{}REGN} & {\scriptsize{}0} & {\scriptsize{}12} & {\scriptsize{}9} & {\scriptsize{}0} & {\scriptsize{}17} & {\scriptsize{}0} & {\scriptsize{}0} & {\scriptsize{}20} & {\scriptsize{}1} & {\scriptsize{}0} & {\scriptsize{}12} & {\scriptsize{}0} & {\scriptsize{}20} & {\scriptsize{}0} & {\scriptsize{}0} & {\scriptsize{}9}\tabularnewline
{\footnotesize{}SBUX} & {\scriptsize{}0} & {\scriptsize{}13} & {\scriptsize{}4} & {\scriptsize{}0} & {\scriptsize{}1} & {\scriptsize{}6} & {\scriptsize{}0} & {\scriptsize{}18} & {\scriptsize{}0} & {\scriptsize{}17} & {\scriptsize{}21} & {\scriptsize{}0} & {\scriptsize{}17} & {\scriptsize{}0} & {\scriptsize{}0} & {\scriptsize{}0}\tabularnewline
{\footnotesize{}TXN} & {\scriptsize{}0} & {\scriptsize{}12} & {\scriptsize{}8} & {\scriptsize{}0} & {\scriptsize{}0} & {\scriptsize{}3} & {\scriptsize{}0} & {\scriptsize{}13} & {\scriptsize{}0} & {\scriptsize{}11} & {\scriptsize{}21} & {\scriptsize{}0} & {\scriptsize{}12} & {\scriptsize{}0} & {\scriptsize{}1} & {\scriptsize{}0}\tabularnewline
{\footnotesize{}VOD} & {\scriptsize{}0} & {\scriptsize{}17} & {\scriptsize{}17} & {\scriptsize{}0} & {\scriptsize{}1} & {\scriptsize{}3} & {\scriptsize{}0} & {\scriptsize{}2} & {\scriptsize{}1} & {\scriptsize{}1} & {\scriptsize{}19} & {\scriptsize{}0} & {\scriptsize{}16} & {\scriptsize{}0} & {\scriptsize{}1} & {\scriptsize{}0}\tabularnewline
{\footnotesize{}YHOO} & {\scriptsize{}0} & {\scriptsize{}21} & {\scriptsize{}19} & {\scriptsize{}0} & {\scriptsize{}1} & {\scriptsize{}9} & {\scriptsize{}2} & {\scriptsize{}1} & {\scriptsize{}0} & {\scriptsize{}12} & {\scriptsize{}21} & {\scriptsize{}0} & {\scriptsize{}5} & {\scriptsize{}1} & {\scriptsize{}4} & {\scriptsize{}0}\tabularnewline
\hline 
{\footnotesize{}Sum (\%)} & {\scriptsize{}2.1\%} & {\scriptsize{}39.8\%} & {\scriptsize{}29.0\%} & {\scriptsize{}4.0\%} & {\scriptsize{}33.9\%} & {\scriptsize{}18.5\%} & {\scriptsize{}1.1\%} & {\scriptsize{}76.0\%} & {\scriptsize{}1.7\%} & {\scriptsize{}49.1\%} & {\scriptsize{}87.4\%} & {\scriptsize{}0.0\%} & {\scriptsize{}82.9\%} & {\scriptsize{}0.6\%} & {\scriptsize{}5.5\%} & {\scriptsize{}24.4\%}\tabularnewline
\hline 
\end{tabular}}
\label{table:PSI}
\textit{\scriptsize{}Note}{\scriptsize{}: $\mathrm{P_{12}}$ represent
the regime switching from the short-duration regime to the long duration
regime, and $\mathrm{P_{21}}$ represent the regime switching from
the long regime to the short-duration regime. Significance is measured at the 5\%
level. Sig(+)/Sig(-) means that the estimated coefficient for
the factor is significantly positive/negative. Each cell shows the significance. 
The data is in the month of January 2013. We have 21 trading
days for each stock, so the upper limit for each cell is 21. The
last row records the percentage (\%) of the times that this factor appears
as significant in all 525 instances (25 stocks times 21 days).  }{\scriptsize\par}
\end{table}

\end{appendices}

\end{doublespacing}

\end{document}